\documentclass{article}
\usepackage[utf8]{inputenc}
\usepackage{authblk}
\usepackage{url}
\usepackage{float}
\usepackage{setspace}
\usepackage[margin=1.25in]{geometry}
\usepackage{graphicx}
\usepackage[hidelinks]{hyperref}
\graphicspath{ {./figures/} }
\usepackage{subcaption}
\usepackage{amssymb,amsmath,amsthm}
\usepackage{lineno}
\usepackage{multirow}
\usepackage{paralist}
\usepackage[section]{placeins}

\DeclareUnicodeCharacter{0306}{}

\usepackage[style=nejm, 
citestyle=numeric-comp,
sorting=none]{biblatex}
\title{A systematic evaluation of adversarial attacks against \\ speech emotion recognition models}

\author[1]{Nicolas Facchinetti}
\author[2]{Federico Simonetta}
\author[1$\ast$]{Stavros Ntalampiras}

\affil[1]{Dept. of Computer Science - University of Milan}
\affil[2]{GSSI -- Gran Sasso Science Institute}

\affil[$\ast$]{Corresponding author: stavros.ntalampiras@unimi.it}

\date{}

\begin{document}

\maketitle

%
\begin{abstract}
Speech emotion recognition (SER) is constantly gaining attention in recent years due to its potential applications in diverse fields and thanks to the possibility offered by deep learning technologies. However, recent studies have shown that deep learning models can be vulnerable to adversarial attacks. In this paper, we systematically assess this problem by examining the impact of various adversarial white-box and black-box attacks on different languages and genders within the context of SER. We first propose a suitable methodology for audio data processing, feature extraction, and CNN-LSTM architecture. The observed outcomes highlighted the significant vulnerability of CNN-LSTM models to adversarial examples (AEs). In fact, all the considered adversarial attacks are able to significantly reduce the performance of the constructed models. Furthermore, when assessing the efficacy of the attacks, minor differences were noted between the languages analyzed as well as between male and female speech. In summary, this work contributes to the understanding of the robustness of CNN-LSTM models, particularly in SER scenarios, and the impact of AEs. Interestingly, our findings serve as a baseline for 
\begin{inparaenum}[a)]
\item developing more robust algorithms for SER, \item designing more effective attacks, \item investigating possible defenses, \item improved understanding of the vocal differences between different languages and genders, and \item overall, enhancing our comprehension of the SER task.
\end{inparaenum}
\end{abstract}

%

\section{Introduction}
The exploration of automatic emotional state detection from vocal expressions has drawn considerable attention in the contemporary era, primarily due to its potential applicability in a broad spectrum of fields such as human-computer interaction, psychology, entertainment, and education \cite{10167766,Ntalampiras2021,Ntalampiras2020jaes}. The introduction of deep learning techniques has markedly improved the performance of Speech Emotion Recognition (SER) models, fostering the development of numerous applications for public use \cite{Nicolini2024}. However, recent studies have underscored the vulnerability of deep learning models to Adversarial Examples (AEs) -- carefully crafted input samples designed to mislead the model into producing erroneous predictions \cite{9909635}. This susceptibility could trigger serious consequences in an SER context, especially in applications that are integral to safety. Evaluating the robustness of SER models is crucial, given their prospective use in areas such as affective computing, human-robot interaction, and mental health monitoring. By assessing and enhancing the resilience of these models, researchers can aid in the creation of reliable and trustworthy tools for these vital real-world applications.\\

Adversarial attacks represent a substantial menace to SER systems, resulting in the erroneous interpretation of a speaker's emotional condition, which might lead to severe implications. The following are five instances that illustrate the aforementioned risk:
\begin{itemize}
    \item In the context of customer service, an adversarial assault on an SER model might misread a customer's irritation as joy, culminating in an unsuitable response and possibly a discontented customer.
    \item In the setting of mental health diagnosis, an adversarial incursion on an SER model utilized to identify depression could lead it to falsely categorize a patient as being in good health, resulting in a faulty diagnosis and insufficient treatment.
    \item In the realm of entertainment, an adversarial onslaught on an SER model employed to modulate a virtual assistant's tone might result in responses that are incongruous with the user's expectations, leading to bewilderment and annoyance.
    \item In a security-oriented scenario, an adversarial intrusion on an SER model used for detecting deceit during a police inquiry might lead to the misclassification of a suspect's truthful declarations as falsehoods, resulting in unwarranted allegations and potential erroneous detentions.
    \item In the situation of a job interview, an adversarial strike on an SER model employed to assess a candidate's emotional aptitude might lead to the misinterpretation of a candidate's anxiety as hostility, culminating in an incorrect evaluation and potentially overlooking a competent candidate.
\end{itemize}

In recent years a novel branch of scientific research has been studying the impact of AEs on SER tasks. Research in this area primarily aims to gauge the resilience of SER systems against various forms of attacks, and to devise methodologies to enhance model performance in the face of such threats \cite{9844239}. Despite these efforts, the subject matter remains largely unexplored, necessitating further investigation to gain a more comprehensive understanding of how different attack techniques could potentially impact system performance. The outcomes of such research could, for instance, highlight specific types of attacks that are exceptionally proficient at misleading the model, or indicate that the model exhibits greater robustness against attacks that alter certain input data features. Furthermore, the study may reveal that a specific language or gender is more susceptible to these attacks.

Although the impact of adversarial examples (AEs) on image models has been extensively studied, their application to speech emotion recognition (SER) models is lacking research. Furthermore, it is not possible to draw the same conclusions since the input data of image classification and SER models are only superficially similar. In fact, while the success of convolutional neural networks (CNNs) and Transformer-based architectures has been extended to the audio processing domain, the input samples used in the audio domain differ from those in the image domain in two main aspects: a) they often consist of sparse matrices, with most entries close to zero, and b) they are not easily segmentable, meaning that the same sound source (i.e., object) is spread across the matrix and is not contiguous as in the case of image segmentation. We believe that these differences warrant further investigation into the use of AEs in SER tasks.\\

This paper endeavours to address this gap in the existing body of knowledge by scrutinising the effects of multiple adversarial attacks on various languages and genders in the SER context. It is of paramount importance to assess the robustness of emotion recognition models, comprehend their limitations, and create more resilient algorithms. This process could entail evaluating the model's precision in the face of AEs that have been manipulated to trick the model into generating erroneous predictions. Within this framework, the primary contributions of this research are to:\begin{itemize}
    \item conduct an exhaustive analysis of the susceptibility of Convolutional Neural Network-Long Short-Term Memory (CNN-LSTM) models to AE in SER;
    \item compare the performance of diverse attack categories;
    \item investigate potential disparities in the attack across three distinct languages and between male and female vocal samples.
\end{itemize}

Following an initial exploration of the scientific literature pertaining to principal techniques and methodologies in SER and Adversarial Machine Learning, an optimal neural network model was identified to address the problem at hand. 
Given that there is no single model architecture that performs well across multiple languages in the literature \cite{meng2019speech, zhao2019speech}, we designed a model consisting of a fusion of CNN and LSTM, and is trained using log Mel-spectrograms derived from audio samples embodying diverse emotional states. The current paper focuses on multiple languages and scrutinizes the impacts of various adversarial attacks on speech data from both genders. To this end, three distinct datasets are utilized: EmoDB \cite{burkhardt2005database} for German, EMOVO \cite{costantini-etal-2014-emovo} for Italian, and Ravdess \cite{livingstone2018ryerson} for English. A conscious decision was made to design and educate our unique model to ensure maximum flexibility during experimentation, rather than depending on pre-existing pre-trained models that may not yield satisfactory outcomes across all languages. \\


Interestingly, a multitude of attacks were employed with the objective of assessing their influence on the established models. A broad spectrum of varied attack methodologies \cite{akhtar2019brief} was assessed, and, when feasible, diverse parameter sets were employed. The white-box attacks included were Fast Gradient Sign Method (FGSM) \cite{goodfellow2014explaining}, Basic Iterative Method (BIM) \cite{kurakin2018adversarial}, DeepFool \cite{moosavi2016deepfool}, Jacobian-based Saliency Map Attack (JSMA) \cite{papernot2016limitations}, and Carlini \& Wagner (C\&W) \cite{carlini2017towards}. For the black-box attacks, PixelAttack \cite{su2019one, kotyan2019adversarial}, and BoundaryAttack \cite{brendel2017decision} were utilized in our experimentation. Following comprehensive experimentation, we present detailed results that evaluate the efficacy of SER models when subjected to various attacks, taking into account language and gender factors. The execution of all the experiments depicted in this article can be found at \url{https://github.com/LIMUNIMI/thesis_adversarial_ml_audio}.

\subsection{Analysis of the literature}
\label{subsec:literature}
\paragraph{Speech emotion recognition}
SER represents the computational challenge of discerning a speaker's emotional state by examining the acoustic properties of their speech signal \cite{abbaschian2021deep}. Emotions, being a crucial component of human communication, are manifested through various speech facets including pitch, tempo, intensity, and spectral features. Despite the complexity and diversity in emotional expression through speech, recent breakthroughs in machine learning and deep learning methodologies have propelled substantial advancements in this domain, thereby stimulating active research interest in SER.

 
  
   
    

SER is a potent instrument, with its utility extending to a range of fields including virtual assistant development, emotion detection in customer service, and mental health surveillance. This paper provides a concise review of some pioneering studies and their respective application areas. 
Among the earliest researches, Nakatsu et al. \cite{nakatsu1999emotion} explored the application of SER in an interactive movie system. This system not only allowed viewers to watch the narrative but also to engage with it, employing emotion recognition to facilitate spontaneous interactions among computer characters. 
In a different context, Petrushin et al. \cite{petrushin1999emotion} concentrated on identifying emotional states in telephone call center dialogues. Here, understanding the caller's mental state proved beneficial in decision support systems for tasks such as prioritizing voice messages, assigning suitable agents for responses, or categorizing voice mails based on the emotions expressed by the caller. Their study revealed anger as the most identifiable emotion. 
Further, France et al. \cite{france2000acoustical} used acoustical characteristics as markers of depression and suicide risk, aiding therapists in comprehending their patients' concealed emotions and overall mental state. 
Lastly, Schuller et al. \cite{schuller2004speech} proposed a method that combined acoustic features and linguistic information in the automotive industry to enhance car ride safety by monitoring the driver's mental state. Remarkably, this approach could trigger safety measures, potentially preventing accidents. \\

The research paper by \cite{abbaschian2021deep} categorizes datasets for SER into three distinct types, based on the method of sample collection. These are simulated, semi-natural, and natural speech datasets. Among these, the simulated datasets are the most prevalent. They are synthesized by trained speakers who articulate the same text, each time embodying a different emotion. Such datasets typically portray a standard set of emotions and, due to their acted nature, exhibit less noise and realism in comparison to datasets of natural speech~\cite{abbaschian2021deep}. The datasets utilized in the present study are all of the simulated kind.\\
One of the most frequently used datasets for SER tasks is the \emph{Berlin Database of Emotional Speech (EMO-DB)} \cite{burkhardt2005database}. This dataset features ten actors, evenly split by gender, who simulate a range of emotions while uttering ten German sentences of varying lengths. Specifically, seven emotions (neutral, anger, fear, joy, sadness, disgust, and boredom) are represented across approximately 800 sentences, including 700 primary samples and some secondary versions. The recordings were conducted in an anechoic chamber, and the resultant material was subjected to an automated listening test. Each sentence was assessed by a panel of 20 listeners. Additionally, electroglottograms are provided to facilitate more precise extraction of prosodic and voice quality features.\\
The \emph{Ryerson Audio-Visual Database of Emotional Speech and Song (RAVDESS)} \cite{livingstone2018ryerson} is a comprehensive, gender-balanced dataset. It comprises emotional speech and song recordings from 24 professional North American actors (12 female, 12 male), each contributing 104 sentences. This results in a total of 7356 speech and 3036 song samples, including both facial and vocal expressions. The speech component encapsulates a spectrum of emotions such as calm, happiness, sadness, anger, fear, surprise, and disgust, whereas the song component encompasses calm, happiness, sadness, anger, and fear. Each emotional expression is represented at two levels of intensity: normal and strong, complemented by a neutral expression. The validation process for RAVDESS involved two stages, with an initial group of 247 raters from North America followed by another group of 72 participants.\\
The \emph{EMOVO} \cite{costantini-etal-2014-emovo} dataset, on the other hand, is the inaugural emotional corpus tailored for the Italian language. It consists of recordings from six actors (three males and three females), each delivering 14 sentences that simulate seven emotions: disgust, fear, anger, joy, surprise, sadness, and neutral. These recordings, made using professional equipment in the Fondazione Ugo Bordoni laboratories, total 588, with each actor contributing approximately 10 minutes of material. This culminates in an overall database duration of one hour. The validation of the EMOVO dataset involved two distinct groups of 24 individuals, achieving an overall recognition accuracy of 80\%.

In the realm of existing SER models, the Convolutional Neural Network-Long Short-Term Memory (CNN-LSTM) models have, in recent years, consistently exhibited remarkable effectiveness, achieving unparalleled results as evidenced in multiple studies \cite{adieu,zhao2019speech,latif2019direct, etienne2018cnn+, meng2019speech}. These models, characterized by their deep architecture, possess the ability to independently extract high-quality features from the data. Consequently, we have chosen to employ log-Mel spectrograms, a decision informed by their proven compatibility with this particular architecture in previous research \cite{etienne2018cnn+, meng2019speech, zhao2019speech, purwins2019deep, pandey2019deep, ren2020generating, chang2022robust}. 
 
 
 
  

\paragraph{Adversarial Machine Learning}
The concept of \emph{``Adversarial machine learning''} encompasses a collection of methodologies devised for instigating malevolent attacks by manipulating models using accessible information. Typically, Machine Learning (ML) models are constructed based on a particular train/test set derived from an identical statistical distribution \cite{9913341}. However, upon deployment, the model may be subjected to interference from an attacker who manipulates the system's operation by introducing meticulously designed input data. Such input, termed as \emph{adversarial example} (AE) \cite{szegedy2013intriguing}, comprises legitimate inputs modified by the addition of minimal, often undetectable, perturbations. These perturbations are designed to deceive the system, thereby altering the anticipated outcomes by exploiting certain susceptibilities, all while being accurately classified by a human observer.
 
  
   
   
The susceptibility of numerous ML models, including neural networks, to attacks instigated by minor modifications to the model's input during testing, is a significant concern. Biggio et al. \cite{biggio2013evasion} underscored this point by illustrating the dependency of an ML model's success on its robustness against adversarial data. Their exemplar was a malware detection system for PDF files, which relied on a differentiable discriminant function.\\
The research explored two distinct scenarios. The first scenario involved an attacker possessing comprehensive knowledge of the target classifier, including the feature space, the model type, and the trained model. Conversely, in the second scenario, the attacker's knowledge was limited. The employed attack strategy hinged on the gradient descent walk of the classifier's discriminant function $g(x)$, which was assumed to be differentiable, or an approximation thereof. The findings highlighted that both support vector machines (SVM) and neural networks could be successfully evaded, even when the adversary's understanding of the system was minimal.\\
Regarding contemporary state-of-the-art deep neural networks, which demonstrate remarkable generalization in classification tasks, one would anticipate robustness against minor perturbations of the input signal. However, Szegedy et al. \cite{szegedy2013intriguing} discovered that even a negligible yet carefully tailored perturbation of an input image could alter the network's prediction. Intriguingly, the error rate induced by these meticulously crafted examples surpassed that of examples perturbed with Gaussian noise, even though the average distortion was less.

\paragraph{Adversarial ML and SER models}
The pioneering scheme for generating AE in the context of linguistic applications was introduced by \cite{gong2017crafting}. In their work, the authors focused on three paralinguistic tasks, including SER. Rather than applying perturbations to specific acoustic features, they opted to directly manipulate the raw waveform of an audio recording. The dataset utilized for their study was IEMOCAP, and the models of choice were WaveRNN and WaveCNN \cite{gong2017crafting}, both of which are considered to be state-of-the-art. 

In the task of emotion recognition, both models demonstrated similar performance levels: WaveRNN achieved an accuracy rate of 84\%, while WaveCNN slightly surpassed it with an accuracy of 85\%. The authors employed the Fast Gradient Sign Method (FGSM) \cite{goodfellow2014explaining}, detailed further in Section \ref{algos}, as their chosen attack strategy. This method was applied twice, using various values of $\epsilon$. When the perturbation factor was set to 0.015, there was a significant increase in the emotion recognition error rate for both models: from 16\% to 48\% for WaveRNN, and from 15\% to 42.5\% for WaveCNN. Notably, these rates approached an upper bound error rate of 50\% when only two classes were considered. The authors also observed that the AEs generated could be profitably transferred from WaveCNN to WaveRNN.

An important observation made by the authors was that the perturbations introduced through their approach were not only smaller, but also more effective than those achievable through an attack at the Mel Frequency Cepstral Coefficients (MFCC) feature level.

The inaugural black-box adversarial attack on SER systems is put forward by the authors in \cite{latif2018adversarial}. This attack, currently recognized as the \textit{Real-World Noise (RWN)} attack, subtly manipulates the speech signal by incorporating minute, indiscernible noise. Upon experimental evaluation, the classification error rates were found to be 56.87\ and 66.87\ for the FAU-AIBO and IEMOCAP datasets, correspondingly. The authors further explore the potential of adversarial examples in enhancing model security through adversarial training. By integrating adversarially crafted examples into the training set, they managed to decrease the error rate by approximately 10 percentage points. Furthermore, the use of a GAN to generate examples yielded superior results. Specifically, the generator was programmed to alter adversarial examples with the aim of deceiving a network that differentiates between adversarial and genuine examples. Interestingly, they discovered that the inclusion of a random noise layer did not benefit SER, a finding that contrasts with its impact on images.\\

The authors in \cite{ren2020generating} present a methodology for enhancing the resilience of a CNN based SER system against adversarial assaults. They employ the FGSM to generate adversarial instances from log Mel-spectrograms extracted from the DEMoS dataset. The robustness of three distinct models, namely a four-layer CNN, a ResNet model, and a VGG model, is assessed against these adversarial instances. The experimental results reveal a significant decline in the Unweighted Average Recall (UAR) of the models, from 0.8 to 0.2, with an increase in the $\epsilon$ parameter of FGSM.\\

To address this vulnerability, the authors propose three robustness-enhancing strategies. The initial strategy involves a data augmentation approach, which incorporates FGSM-generated instances into the training set. Subsequently, two adversarial training methodologies are suggested: vanilla and similarity-based. The former method proves effective in enhancing performance on authentic data compared to the conventional training approach, whereas the latter demonstrates superior efficacy in defending against adversarial attacks.\\


The necessity for pre-processing steps is evident when dealing with audio-type inputs, as it is crucial to extract certain features \cite{8678825}. Unlike CNN that can directly operate on image pixels, this context requires initial extraction of signal features such as MFCCs or a log Mel-spectrogram \cite{8678825}. The complexity increases when considering white-box attacks like Fast Gradient Sign Method (FGSM), which utilize the gradient of the targeted audio relative to the input to calculate the optimal perturbation. Although the backpropagation method is efficiently applicable in image recognition due to the differentiability of all layers, the scenario becomes intricate for SER systems. The complexity arises from the commonly extracted features, such as the introduction of non-linearity during the computation of MFCCs and the significant non-linearity in the output due to the usage of numerous Long Short-Term Memory (LSTM) units \cite{latif2018adversarial,taori2019targeted,carlini2018audio}. In light of empirical studies within this context, iterative methods have demonstrated superior effectiveness compared to single-step approaches \cite{carlini2018audio}.\\

To the best of our knowledge, this constitutes the inaugural comprehensive examination of AEs impact on SER systems, encompassing a broad spectrum of elements, from the nature of the attacks to the spoken language and the gender of the speaker. In the spirit of ensuring complete reproducibility of our methodology and findings, the implementation has been made publicly accessible at \url{https://github.com/LIMUNIMI/thesis_adversarial_ml_audio}.

\section{Materials and Methods}
This section delineates the methodologies employed within the proposed framework for audio pattern recognition, which is dedicated to processing audio data and training SER models. It also provides a brief overview of the attacks utilized in the study.

\subsection{Experimental design}
\label{sec:overview}

In order to assess adversarial attacks for Speech Emotion Recognition (SER) models, a specific pipeline was developed. The objective was to observe the impact of different attacks on different languages, while also controlling for the influence of speaker characteristics such as sex. 

Three datasets were used, each representing a different language: RAVDESS (English), EmoDB (German), and EMOVO (Italian). To ensure consistency in the evaluation of attack effectiveness, the same model was kept fixed throughout the experiments. Since there is currently no existing model capable of performing SER in multiple languages, a custom model was developed for this study. 
The data underwent a rigorous cleaning and preprocessing to obtain log-Mel spectrograms. Labels that exhibited high correlation or were heavily under or over-represented in the datasets were removed. To increase the amount of training data, data augmentation techniques were applied. 

A total of seven different CNN-LSTM architectures were designed and evaluated. The first model ($\mathcal{M}0$) was used to determine the most effective normalization procedure, which was found to be simple standardization. This normalization procedure was then applied to all the remaining tests. 
The seven architectures were compared, and the best performing multi-language model structure ($\mathcal{M}1$) was identified. Further fine-tuning was conducted on the number of LSTM units and other hyper-parameters of this model. 

Finally, the obtained model and datasets were utilized to assess the impact of adversarial attacks from the ART library. 

\begin{figure}[H]
    \centering    \includegraphics[width=1\textwidth]{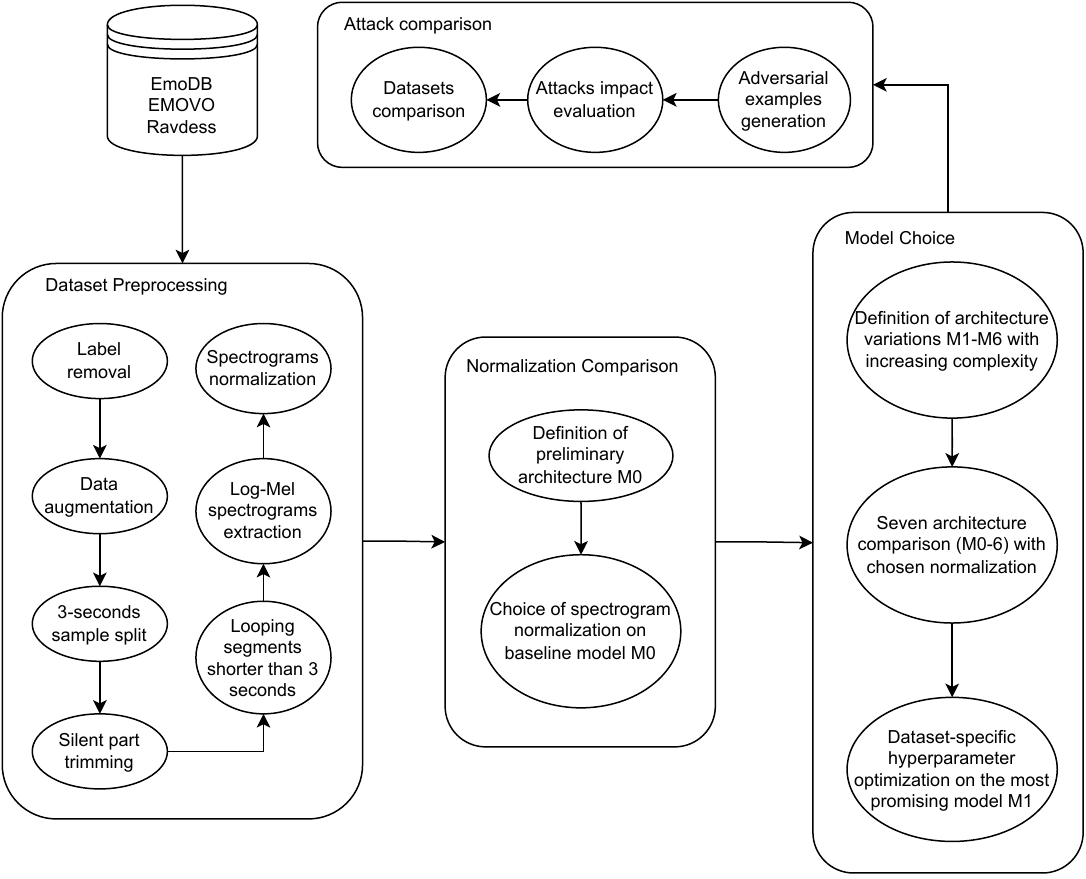}
    \caption{Flowchart of the proposed methodology to conduct the experiment.}
    \label{flowchart}
\end{figure}

Overall, this comprehensive methodology allowed for a thorough evaluation of adversarial attacks on SER models, taking into consideration different languages and controlling for speaker characteristics. The described workflow is shown in Figure~\ref{flowchart}.

\subsection{Datasets}

The EmoDB and EMOVO databases share a similar quantity of samples, in contrast to the substantially larger Ravdess database, which is approximately 2.5 times greater in size. A notable variance is observed in the number of actors across these databases, as previously discussed in Section \ref{subsec:literature}.

In terms of audio duration distributions, the mean and 25, 50, and 75 percentiles demonstrate comparable statistics across the databases. However, significant discrepancies are evident in the maximum duration and inter-file duration. Given our objective to assess the impact of various attacks on SER systems through comparative analysis across different languages and genders, it is essential to maintain uniform training parameterization across all architectures, inclusive of sample duration. When examining the labels, it is observed that EmoDB and EMOVO incorporate 7 emotions, whereas Ravdess includes 8. Further details concerning the extracted metadata can be found in Section \ref{dsproc}.

The datasets under consideration presented several challenges, which are addressed in the subsequent experimental setup. These challenges include:
\begin{inparaenum}[a)]
    \item heterogeneity in the duration and sample rate of audio files,
    \item inconsistency in the number of classes across datasets,
    \item scarcity of data.
\end{inparaenum}

\begin{figure}[h]
    \centering
    \begin{subfigure}{0.49\textwidth}
         \centering
         \includegraphics[width=\textwidth]{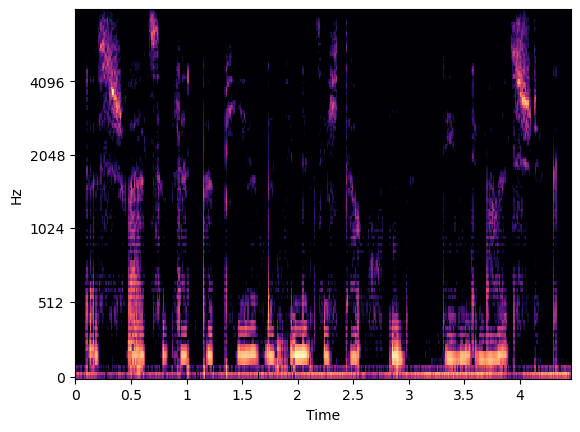}
         \caption{Original}
    \end{subfigure}
    \hfill
    \begin{subfigure}{0.49\textwidth}
         \centering
         \includegraphics[width=\textwidth]{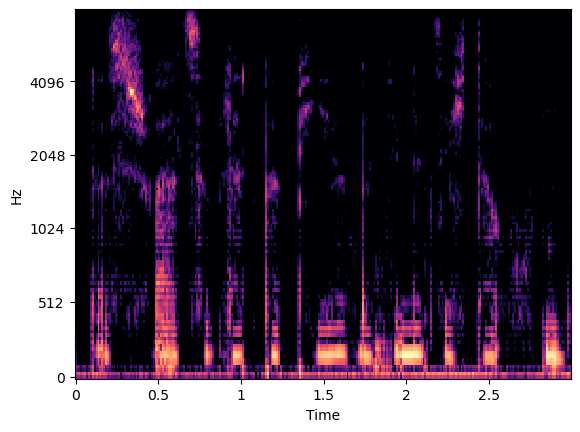}
         \caption{First segment}
    \end{subfigure}
    \vskip\baselineskip
     \begin{subfigure}{0.49\textwidth}
         \centering
         \includegraphics[width=\textwidth]{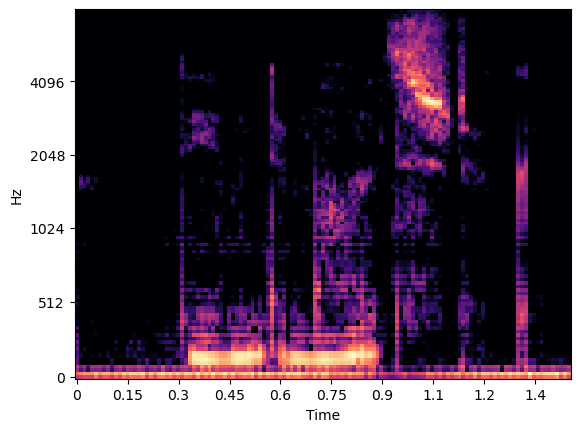}
         \caption{Second segment}
    \end{subfigure}
    \hfill
    \begin{subfigure}{0.49\textwidth}
         \centering
         \includegraphics[width=\textwidth]{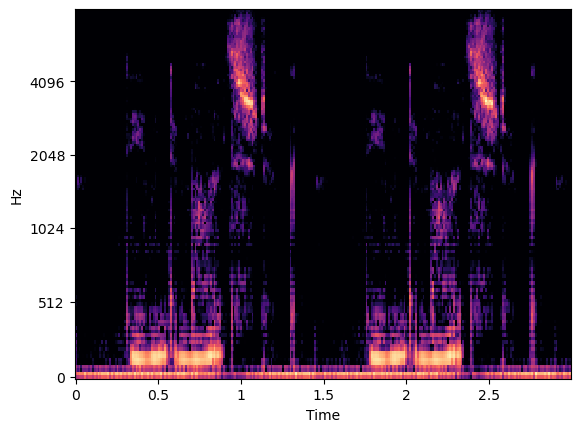}
         \caption{Second segment repeated}
    \end{subfigure}
     \caption{Example of split and repeat process on log Mel-spectrograms. Original log Mel-spectrogram (a), the sliced segments (b) and (c), and segment (c) repeated to 3 seconds (d).}
     \label{example}
\end{figure}

\paragraph{Pre-processing}
In order to obtain homogeneity of the samples, we applied some basic processing to the data. First, the datasets' samples are uniformly resampled at a frequency of 16,000 Hz and the silence at the commencement and termination of each signal is eliminated. Successively, segments with a duration less than 3 seconds are looped, while those exceeding this length are segmented into continuous, non-overlapping 3-second intervals. This process of silence trimming is reiterated on the resulting segments to eliminate superfluous portions. Segments falling short of the 3-second standard are looped until they attain the requisite length. We finally computed the log-Mel spectrograms of the obtained audio excerpts, a choice motivated by prior research~\cite{zhao2019speech}. Log-Mel spectrograms were computed out using 128 Mel bands, a Fast Fourier Transformation window length of 368, and a hop size of 184. These parameters, corresponding to 23 ms and 11.5 ms respectively, were set in line with a sampling rate of 16000 Hz, as suggested by \cite{numberstft}. The spectrogram obtained was subsequently transformed to a logarithmic scale, leading to a 128x261 matrix saved for subsequent analysis. For illustrative purposes, an exemplar is provided in Figure \ref{example}.

Regarding the classes available in the datasets, since we are not interested in finding the optimal SER model, we chose to retain only five labels per dataset. We first computed log-Mel spectrograms of the audio excerpts obtained after segmentation, then, we used PCA and T-SNE to spot classes that were highly correlated. Moreover, other labels were discarded to improve the balance of the datasets. Specifically, the \emph{fear} and \emph{angry} labels were omitted from EmoDB. As detailed in Section \ref{dsproc}, the \emph{angry} label was the most prevalent, and its removal facilitated a more balanced dataset. For EMOVO, the excluded labels were \emph{sad} and \emph{angry}, while for Ravdess, the \emph{calm}, \emph{neutral}, and \emph{angry} labels were discarded. Similarly to the \emph{angry} label in EmoDB, the \emph{neutral} label in Ravdess was disproportionately represented and its exclusion rectified this imbalance.



We approached the problem of the scarcity of data using data augmentation methods.
While GANs have shown encouraging results in SER tasks \cite{chatziagapi2019data,sahu2018enhancing,latif2020augmenting}, our approach is grounded in the application of less computationally intensive techniques. Data augmentation has the potential to significantly enhance the precision of a classifier and facilitate better model generalization to unobserved data, as the model becomes more resilient to the deformations applied \cite{salamon2017deep}. According to \cite{salamon2017deep}, viable augmentation options encompass \textit{time stretching} and \textit{pitch shifting}. We employed an acceleration factor of $[0.75, 0.9, 1.1, 1.25]$ for time stretching. For pitch shifting, each step was configured to correspond to a semitone, with the values $[-3, -1.5, +1.5, +3]$ being considered.


The entire datasets are subjected to these processing, yielding eight supplementary augmented samples. To maintain a uniform duration of 3 seconds across all samples, identical procedures of division and repetition are subsequently applied to these newly generated samples.\\

\paragraph{Normalization} Once we obtained the augmented datasets, we were ready for continuing with subsequent phase as described in Section~\ref{sec:overview}. However, one remaining step for preparing the data for neural models was the normalization of the log-Mel spectrograms. In literature, various methods are adopted without a single methodology being more successful than the others. We were therefore interested in testing them. The objective of the normalization procedure is to transform the input values into a range suitable for neural learning, typically either $[0,1]$ or $[-1,+1]$.

In the initial stages, four distinct transformations of the features are contemplated:
\begin{itemize}
    \item \textbf{Original}: Utilizes the log Mel-spectrogram matrices devoid of any normalization, functioning as a baseline for assessing the practical benefits of the subsequent transformations.
    \item \textbf{NormSum}: Implements normalization by dividing each element of the matrix by the aggregate of all elements, thereby ensuring uniform sound energy across all samples. However, the values of each cell are significantly small and approach zero.
    \item \textbf{NormMaxGlobal}: This method normalizes by dividing each matrix element by the maximum value across the dataset, ensuring all values fall within the $[0,1]$ range while preserving the original proportions between the cells across varying spectrograms.
    \item \textbf{NormMaxLocal}: Normalization is achieved by scaling each element in a matrix through division by the maximum value within that specific spectrogram. This process ensures each matrix has a minimum value of 0 and a maximum value of 1, but the original proportions between cells are not preserved.
\end{itemize}

Beyond the aforementioned transformations, we opted to standardize each previously delineated version. This decision was informed by preliminary experiments that indicated a notable instability in the model's learning process, evidenced by significant fluctuations in the loss function across epochs. In this standardization phase, we initially transformed the matrices into arrays, subsequently standardized these, and then reshaped the data to its original form.

For assessing the best normalization strategy, we developed a rudimentary CNN-LSTM, designated as $\mathcal{M}0$, which demonstrated notable potential in preliminary investigations -- see Section~\ref{sec:models}.  The accuracy and its corresponding standard deviation, derived from each processing variant and dataset, are presented in Table \ref{preacc} as per the definition provided in Section \ref{sec:results}.

The assessment indicates that the use of \emph{NormMaxGlobal} in the context of EmoDB and \emph{Original Standardized} for EMOVO and Ravdess, leads to enhanced accuracies as reflected in Table~\ref{preacc}. Consistently, the standardized variant yields more reliable results, corroborating the preliminary experimental observations. Additionally, the loss function exhibits fewer temporal variations, fostering a steadier learning trajectory.

Although \emph{NormMaxGlobal} exhibited superior performance with respect to accuracy and standard deviation on the EmoDB dataset, we opted to utilize \emph{Original Standardized} consistently across all three instances, which maximizes the accuracy across the datasets on average. The standardization was applied to the entire datasets after the above-described pre-processing steps.


\subsection{Models}
\label{sec:models}

\paragraph{Multi-dataset architecture}
We developed 7 model architectures to search for the an optimal model across the three datasets that could be used for a fair assessment of the attacks.

The base model $\mathcal{M}0$ incorporates three \emph{Conv2D} layers, exhibiting an increase in filters (16, 32, 64) and a decrease in square kernel size, in a manner akin to the models presented in \cite{szegedy2015going, simonyan2014very}. The activation function employed is \emph{ReLU}, as suggested by \cite{simonyan2014very}. Following each convolutional layer is a \emph{MaxPooling} layer, with a pool size that varies ((from (4,4) initially to (2,2) subsequently)) and strides (initially 2, later 1), mirroring the approach in \cite{szegedy2015going, simonyan2014very}. To mitigate overfitting, expedite training, and facilitate the adoption of elevated learning rates, a \emph{BatchNormalization} layer is positioned post the initial convolution \cite{ioffe2015batch}. The output from the CNN is flattened and transferred to an \emph{LSTM} layer comprising three internal units, configured with an internal dropout of 0.2. The final output layer is a \emph{Dense} layer with five units, employing a \emph{Softmax} activation function to generate the probability for the five labels present in each dataset. A comprehensive delineation of the architecture is provided in Section \ref{modarchs}.

We then designed other 6 models, each exhibiting a progressive increase in parameter count and complexity. These models bear the nomenclature $\mathcal{M}1$, $\mathcal{M}2$, $\mathcal{M}3$, $\mathcal{M}4$, $\mathcal{M}5$, and $\mathcal{M}6$. The networks are sequentially arranged based on their parameter quantity, with slight variations in both CNN structure and LSTM internal unit count, while maintaining a foundational structure akin to $\mathcal{M}0$. In-depth information regarding the architectures is provided in Section \ref{modarchs}.
Every convolutional layer uniformly employs the same kernel size, with the exception of models $\mathcal{M}3$ and $\mathcal{M}6$ that utilize more intricate architectures. The pooling operations exhibit a similar pattern, save for the reduced pool size in the initial pooling layer of models $\mathcal{M}3$ and $\mathcal{M}4$. In the case of $\mathcal{M}5$, there is a doubling of the filter count for each layer, whereas model $\mathcal{M}6$ opts for a quartet over a trio. As for the LSTM aspect, all models function in a unidirectional manner, barring $\mathcal{M}1$ which operates bidirectionally. Models $\mathcal{M}2$, $\mathcal{M}4$, and $\mathcal{M}6$ incorporate six units, in contrast to the remaining models which utilize three.

The models undergo training with a batch size of 32 across 50 epochs, utilizing the \emph{Adam} optimization algorithm with a learning rate of 0.001, in line with \cite{tan2018application}'s approach to a similar CNN-LSTM architecture. The \emph{categorical\_crossentropy} serves as the loss function. To mitigate overfitting, an \emph{EarlyStopping} callback is introduced with a tolerance of 10 epochs, which observes the validation loss. Failing to observe an improvement over 10 epochs prompts the restoration of the weights corresponding to the optimal validation loss. Moreover, a \emph{ReduceLROnPlateau} callback is implemented to decrease the learning rate upon observing no improvement in validation loss over six epochs. This strategy aims to prevent the model from straying from the ideal solution due to an excessive learning rate, while also promoting convergence by adopting smaller steps towards the cost function's optimal solution with a diminished learning rate. The underlying theory is that as the model nears a suboptimal solution with the current learning rate, it oscillates around the global minimum. Reducing the rate allows for smaller steps towards the cost function's optimal solution. The validation loss serves as the metric for this callback, and the learning rate is reduced by a factor of 0.1 when there is no improvement in the validation loss for six epochs.

In the realm of SER tasks, a prevalent evaluation strategy is the Leave-One-Speaker-Out
(LOSO) cross-validation method, which designates each unique speaker as a test set.
Nonetheless, this research deviates from the norm, opting for a conventional
train/validation/test partition due to the inconsistency in the number of actors across
datasets, which inevitably yields diverse test sets. Such disparity could potentially skew performance assessments when juxtaposing attacks on varying languages and genders. 
In particular, the data was apportioned into three splits, with proportions of 64\%, 16\%, and 20\% respectively. To ensure a more reliable estimation of the performance metrics, this procedure was reiterated thrice, each time employing a distinct random seed.

 
  
   
   



The architectures and datasets under consideration were evaluated using identical train/validation/test splits, consistent with the methodology employed in the preceding stage. The mean accuracy and corresponding standard deviation for each split, across all architectures and datasets, are tabulated in Table \ref{macc}.

The $\mathcal{M}1$ architecture emerges as the most potent, demonstrating satisfactory performance across all three datasets. Compared to $\mathcal{M}0$, the accuracy exhibits a substantial enhancement: a gain of +0.12 points for EmoDB, +0.11 for EMOVO, and +0.23 for Ravdess. Concurrently, a notable reduction in the standard deviation signifies a robust generalization capability. The bidirectional configuration's efficacy is also evident, as it surpasses $\mathcal{M}2$—an architecture with identical CNN and output shape post-LSTM, but with six unidirectional units.

Furthermore, the convergence across all splits for each dataset is a shared characteristic of both $\mathcal{M}1$ and $\mathcal{M}2$, a trait not observed in other configurations. A noteworthy observation is the inverse relationship between model complexity and accuracy, implying an optimal parameter count in $\mathcal{M}1$ for the given training set size. Intriguingly, the application of more intricate or deeper networks does not confer any advantage for Ravdess, despite its larger number of examples. Once more, EmoDB consistently outperforms, reinforcing the notion that it represents a less complex task.

Consequently, it is cogent to persist in the optimization of hyperparameters for $\mathcal{M}1$, with its architecture succinctly revisited in Section \ref{modarchs}.

\paragraph{Hyper-parameter optimization}
The final stage in defining the model architecture involves hyperparameter tuning to augment the model's performance on the respective datasets. The hyper-parameter optimization was split into two parts: first, we optimized the number of LSTM units, then the remaining hyper-parameters that did not impact the overall model architecture. For this, we used the Hyperband tuner~\cite{li2017hyperband} with categorical cross-entropy loss.

Prior studies indicated an enhancement in performance upon augmenting the LSTM layer's output quantity, as evidenced by the 3 units in $\mathcal{M}0$ and the 6 units in both $\mathcal{M}1$ and $\mathcal{M}2$. Consequently, additional investigations were undertaken on the $\mathcal{M}1$ model to ascertain the ideal quantity of LSTM units. In particular, we examined multiple bidirectional configurations, encompassing 4, 8, 16, 32, 64, 128, 256, 512, and 1024 units.

Typically, the model's accuracy is enhanced by augmenting the quantity of LSTM units, although this association does not consistently apply to the final four values. To illustrate, the configurations yielding the most superior outcomes, in descending order, are as follows:\begin{itemize}
    \item For EmoDB, the top four configurations encompass 512, 256, 128, and 1024 units;
    \item For EMOVO, the top four configurations comprise 512, 256, 1024, and 128 units;
    \item For Ravdess, the top four configurations consist of 1024, 256, 512, and 128 units.
\end{itemize}
Consequently, the LSTM layer was equipped with 256 units due to the following reasons:\begin{enumerate}
    \item It consistently delivered the second highest performance across all three scenarios;
    \item The performance disparity between the top configuration and this one is negligible;
    \item It engenders a more streamlined model, thereby mitigating the potential for overfitting.
\end{enumerate}
For additional insights pertaining to the tuning of model $\mathcal{M}1$, please refer to section \ref{modperf}.

\begin{figure}[H]
    \centering
    \includegraphics[width=1\textwidth]{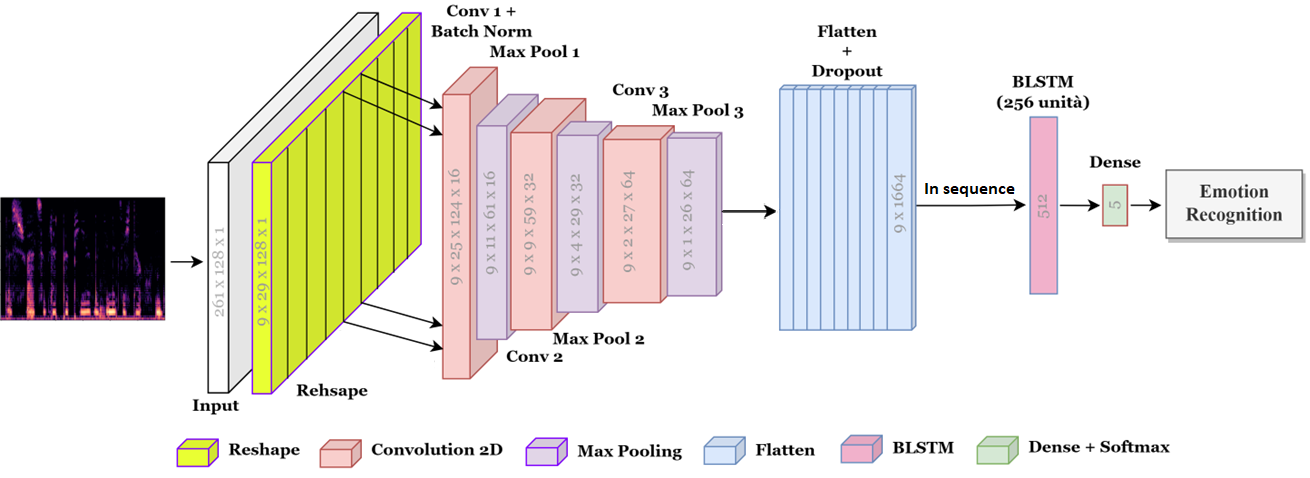}
    \caption{Architecture of the optimized CNN-LSTM model.}
    \label{finalarch}
\end{figure}

Since the number of internal units in the LSTM layer has considerably increased, an additional Dropout layer was added after the Flatten layer to prevent possible overfitting. Figure \ref{finalarch} summarizes the final architecture of the model.\\

The second part of the hyper-parameter optimization consisted of the fine-tuning of the dropout layer's probability, the internal dropout probability of the BLSTM layer, the initial learning rate for the Adam optimizer, and the batch size. 

The tuning of the latter two parameters, namely the learning rate and batch size, is of paramount importance to avert overfitting. These parameters significantly influence the model's performance, convergence, and stability. They are intricately linked to the problem at hand, the input data, and the model's architecture. 

The hyperparameter tuning process has notably enhanced the performance, particularly for larger datasets like EMOVO and Ravdess. For instance, the loss function showed significant reductions:
\begin{itemize}
    \item In the EmoDB dataset, the loss minimally decreased from 0.321270 to 0.312610, which is a reduction of 0.00866 or 2.69\%;
    \item In the EMOVO dataset, the loss substantially decreased from 0.428224 to 0.344954, which is a reduction of 0.08327 or 19.44\%;
    \item In the Ravdess dataset, the loss remarkably decreased from 0.366705 to 0.281702, which is a reduction of 0.08501 or 23.18\%.
\end{itemize}
This enhancement in performance through hyperparameter tuning has particularly allowed for a more effective utilization of datasets with a larger volume of data. Additional details regarding the optimization process can be referred to in Section \ref{modperf}.

The final accuracies of the models are presented in Table~\ref{origacc}.

\subsection{Attack algorithms} \label{algos}
Our experimental study encompasses seven non-targeted algorithms: five white-box attacks and two black-box attacks. The white-box attacks under consideration are Fast Gradient Sign Method (FGSM), Basic Iterative Method (BIM), DeepFool (DF), Jacobian-based Saliency Map Attack (JSMA), and Carlini\&Wagner (C\&W). On the other hand, the black-box attacks include PixelAttack (PA) and BoundaryAttack (BA).

\paragraph{Fast Gradient Sign Method (FGSM)}  \cite{goodfellow2014explaining}, a well-known and straightforward adversarial threat generation technique, primarily exploits the $L_\infty$ distance metric. The method manipulates machine learning models by introducing minimal perturbations to the input data, thereby inducing the model to generate erroneous predictions. Specifically designed to exploit the learning mechanism of neural networks, FGSM utilizes the gradient of the loss with respect to the input data to augment the input data in a way that maximizes the loss. This is achieved by adding a noise vector, derived from the sign of the gradient, to the input data \cite{DBLP:journals/corr/SimonyanZ14a}. The following equation typically illustrates the generation of an adversarial example via this method:
\begin{equation}
    X_{adv} = X + \epsilon * sign(\nabla_{X}J(X,y))
\end{equation}
Given an input $X$ and its corresponding label $y$, the loss function $J(\cdot)$, and a noise magnitude parameter $\epsilon$, the authors demonstrated the potential for successful adversarial attacks. The parameter $\epsilon$ is carefully chosen to balance two conflicting requirements: it must be sufficiently small to render the perturbations imperceptible to humans, yet large enough to mislead the model into making erroneous predictions. The efficacy of such attacks is contingent upon the model's gradient strength, the magnitude of the added noise, and the model's complexity.\\
The authors' experiments on the ImageNet dataset using the GoogLeNet CNN \cite{goodfellow2014explaining}, revealed the ease with which highly effective adversarial examples could be generated. The AEs, created using the fast method, maintained a similar accuracy level until $\epsilon = 32$. Beyond this point, the accuracy gradually declined to nearly zero as $\epsilon$ increased to 128 \cite{kurakin2016adversarial}. This phenomenon can be attributed to the fact that the Fast Gradient Sign Method (FGSM) adds noise scaled by $\epsilon$ to each image. Consequently, utilizing higher $\epsilon$ values effectively obliterates the image content, rendering it unrecognizable to humans.

\paragraph{Basic Iterative Method}  (BIM) \cite{kurakin2018adversarial}. This method, an extension of FGSM utilizing the $L_\infty$ distance metric, employs multiple iterations with a minimal step size. The BIM attack commences with an initial input image, following which the adversary calculates the gradient of the model's loss function relative to this image. Subsequently, the image is incrementally adjusted in the direction of this gradient. This iterative procedure continues either for a predetermined number of iterations or until the attainment of the desired output. Through continuous input modifications based on the model's gradient, the adversary can gradually manipulate the image to induce an erroneous prediction from the model. Specifically, during each iteration, the pixel values of the input image are confined to ensure their location within the $\epsilon$-neighbourhood of the original image.\\
The recursive function used to produce an AE from an input image $X$ is the following:
\begin{equation}
    X_{0}^{adv} = X, X_{N+1}^{adv} = Clip_{X,\epsilon}(X_{N}^{adv} + \alpha * sign(\nabla_{x} J(X_{N}^{adv},y)))
\end{equation}
The function $Clip_{X,\epsilon}(X')$ performs per-pixel clipping on the image $X'$, resulting in an L$\infty$ $\epsilon$-neighbourhood of the original image $X$. The label associated with $X$ is represented by $y$, while the loss function and step size are denoted by $J()$ and $\epsilon$ respectively. The authors empirically determined optimal values, setting $\alpha = 1$ which modifies each pixel value by 1 at every step. The number of iterations was chosen heuristically as $min(\epsilon + 4, 1.25\epsilon)$, a balance ensuring that the AE would reach the boundary of the L$\infty$ $\epsilon$-neighbourhood while maintaining a manageable computational cost for experiments.\\
The experimental results \cite{kurakin2018adversarial} reveal that the iterative method induces subtler perturbations compared to Fast Gradient Sign Method (FGSM), maintaining the integrity of the image even at high $\epsilon$ values. Concurrently, it confounds the classifier at a higher rate. Specifically, the Basic Iterative Method (BIM) generates superior AE for $\epsilon < 48$. Beyond this threshold, however, its performance plateaus and no further improvements are observed. 

\paragraph{DeepFool} (DF) \cite{moosavi2016deepfool} is an adversarial attack method that crafts AEs by iteratively computing the minimal $L_2$ perturbations. This iterative process seeks to ascertain the shortest distance from the original input to the decision boundary of the threat model. The underlying premise of the DeepFool technique is the local approximation of highly nonlinear deep neural networks (NNs) by linear decision boundaries. This assumption allows the authors to analytically formulate the optimal solution to this simplified problem and subsequently construct the AE. Given that NNs are not strictly linear, the algorithm incrementally moves towards the derived solution, repeating the process until a genuine AE is discovered. The DeepFool algorithm utilizes the resulting gradient to determine the optimal direction and magnitude of the perturbations necessary to induce a misclassification by the model. For any differentiable classifier, DeepFool presumes that $f$ is linear around $x'_t$ and iteratively computes the perturbation $r_t$:\begin{equation}
    \underset{r_t}{\arg\min}||r_t||_2 \text{ subject to } f(x'_t) + \nabla(x'_t)^T r_t = 0
\end{equation} 
The algorithm, in every iteration, computes the gradient of the decision function relative to the input data. Subsequently, it determines the least perturbation necessary to transition the input data point beyond the decision boundary into the subsequent class. This procedure is iteratively executed until the model misclassifies the input data point. \\
It has been demonstrated by the authors that the proposed technique effectively deceives advanced image recognition systems. Furthermore, the perturbation instigated by DeepFool is found to be less than that of FGSM across multiple benchmark datasets \cite{moosavi2016deepfool}.

\paragraph{Jacobian Saliency Map Attack} (JSMA) \cite{papernot2016limitations} is an effective adversarial methodology employing a greedy algorithm. This approach leverages $L_0$ distances to generate AEs by iteratively modifying individual pixels. The gradient of the loss, with respect to every input component, is exploited to identify key pixels and their corresponding perturbations. This is facilitated by a saliency map, which pinpoints the input features of significance to the adversary's objectives. The saliency map is derived from the forward derivative (Jacobian) of the function that a Deep Neural Network (DNN) has learned. The procedure of the JSMA attack can be encapsulated as follows:\begin{itemize}
        \item Calculate the Jacobian matrix pertaining to the model's output with respect to the input image;
        \item Generate the saliency map;
        \item Select a target class, denoted as $l$, for the AE;
        \item From the saliency map, pinpoint the most influential features that augment the likelihood of class $l$ while simultaneously diminishing the likelihood of the original class;
        \item Adjust the aforementioned features by a specified parameter $\theta$, to create an AE that is erroneously classified as $l$;
        \item Iterate over steps 2-5 until the AE is successfully produced.
\end{itemize}
The JSMA is a notably potent technique for the creation of AEs with minimal perturbations, which are often challenging to identify. The original work by Papernot et al. \cite{papernot2016limitations} provides a comprehensive explanation of the precise formulation employed, which we recommend for interested readers. 

The authors demonstrate that this algorithm can consistently generate samples that, while perceived as correctly classified by human subjects, are misclassified by a Deep Neural Network (DNN) towards specific targets. This is achieved with an impressive adversarial success rate of 97\% and an average modification of merely 4.02\% of the input features per sample \cite{papernot2016limitations}. 

Despite its effectiveness, the JSMA technique may not always be the most efficient choice due to its computational cost. Furthermore, its performance may vary compared to other attack methods under certain circumstances.

\paragraph{Optimization-Based Attacks by Carlini and Wagner} (C\&W) \cite{carlini2017towards}. This robust method, capable of generating AE measured in $L_0$, $L_2$, and $L_\infty$ norms, is an optimization-based approach. It modifies the objective and the primary constraint of the AE generation optimization problem, as initially proposed in \cite{szegedy2013intriguing}. However, the constraint under consideration possesses a highly non-linear nature, which prompts the authors to recast it in a form more conducive to optimization. Consequently, the optimization problem is reformulated by integrating the constraint within the objective, as shown below:\begin{equation}
\label{newproblem}
     \min ||\delta||_p + c * f(x + \delta)
\end{equation}
In the context of the chosen distance metric $||.||_p$ with $L_p$ norm and a suitably selected constant parameter $c > 0$, while maintaining the second constraint from \cite{szegedy2013intriguing} unaltered, equation \ref{newproblem} yields more potent AEs when tackled with gradient descent in comparison to the FGSM.\\
The implementation nuances diverge based on the employed metric as follows:\begin{itemize}
    \item The $L_2$ norm implementation incorporates multiple initial points for the gradient descent to mitigate the chances of the algorithm landing in unfavorable local minima.
    \item As stated by Carlini and Wagner \cite{carlini2017towards}, the $L_0$ metric is non-differentiable, necessitating the use of an iterative algorithm. This algorithm identifies the least significant pixels in each iteration (utilizing the $L_2$ attack) and subsequently fixes them. Upon identifying a minimal subset of pixels, it is employed to generate an AE. This approach bears resemblance to the JSMA, with the exception that while JSMA expands a set of alterable pixels, the C\&W $L_2$ method reduces the pixel set.
    \item In line with Carlini and Wagner \cite{carlini2017towards}, the $L_\infty$ metric lacks full differentiability and conventional gradient descent yields subpar results. This issue is circumvented with the use of an iterative algorithm, replacing the $||\delta||_p$ term in the objective formulation with a penalty term $\tau=1$. This revised objective is re-evaluated at each iteration and provided $\delta < \tau$, the latter is reduced by a factor of 0.9, allowing transition to the subsequent iteration; otherwise, the search is terminated.
\end{itemize}
The full details of the implementation are pretty complex and can be found in \cite{carlini2017towards}.\\
The experimental results presented by the authors demonstrate that each distance metric employed in the attacks yields AEs that are closer in comparison to those obtained from previous state-of-the-art attacks \cite{carlini2017towards}. The $L_0$ and $L_2$ attacks produce AEs that exhibit 2× to 10× lower distortion compared to the best attacks reported in the literature, boasting a success probability of 100\%. Although the $L_\infty$ attacks yield AEs of similar quality to other studies, they outperform in terms of successful attack rates. Moreover, the effectiveness of the proposed techniques is such that their performance improves with increasing task complexity, a condition under which other methods typically deteriorate. The C\&W attacks consistently achieve a 100\% success rate on naturally trained DNNs across various datasets including MNIST, CIFAR-10, and ImageNet. Additionally, these attacks can successfully compromise defensive distilled models, a feat that DeepFool fails to accomplish in its search for adversarial samples \cite{ren2020adversarial}.

\paragraph{Pixel Attack} (PA) \cite{su2019one, kotyan2019adversarial}. This method,
predicated on differential evolution (DE), fabricates AE by perturbing a single \cite{su2019one} or a limited number \cite{kotyan2019adversarial} of pixels, utilizing either the $L_0$ or $L_\infty$ metric. The attack delineated in these studies concentrates on a small number of pixels without restricting the intensity of modification. This attack aims to derive an AE $x'$ from an original sample $x$ via the computation of a minimal perturbation $\delta$. This results in $x'= x+\delta, f(x) \neq f(x')$, where $f()$ represents the model's output. Consequently, the goal of the associated optimization problem is \cite{kotyan2019adversarial}:\begin{equation}
    \underset{\delta}{\min}\text{ }g(x+\delta)_c  \text{ subject to } ||\delta|| \leq th
\end{equation}
where $th$ denotes a pre-specified threshold parameter that governs the maximum count of alterable pixels, while $g(\cdot)_c$ signifies the confidence associated with the correct class $c$, such that $f(x) = \arg\max g(x)$. The perturbations are encapsulated within arrays, referred to as candidate solutions, and are subjected to optimization through the process of differential evolution \cite{su2019one}. 

Each candidate solution comprises a constant number of perturbations where every perturbation alters a single pixel. This alteration is represented as a quintuple that includes the $(x,y)$ coordinates and the RGB values associated with the perturbation. Upon generation, each candidate solution is pitted against its respective parent, based on the population index, and the victor persists into the subsequent iteration. 

The potential evolution strategies include differential evolution and Covariance Matrix Adaptation. As previously discussed, the attack methodology is designed around two distance metrics \cite{kotyan2019adversarial}, which are detailed as follows:
\begin{itemize}
    \item Threshold Attack: Leveraging the $L_\infty$ metric, this attack can enact slight perturbations across all pixels. It is constrained by the optimization of $||\delta||_{\infty} \leq th$, which allows the algorithm to search within the same space as the input, given that the variables can be any variation of the input, with a maximum limit of $th$.
    
    \item Few-Pixel Attack: Utilizing the $L_0$ metric, this attack can strongly perturb selected pixels. It is a variation of the original One-Pixel Attack \cite{su2019one}, and it optimizes the constraint $||\delta||_{0} \leq th$. In this scenario, the search space for the variables is reduced, as it is a combination of pixel values (dependent on channels 'c' in the image) and position (two values X, Y) for all of the $th$ pixels.
\end{itemize}

The experimental findings reveal a significant disparity in robustness when faced with $L_0$ and $L_\infty$ norm attacks. The attack's effectiveness is remarkably high, even with exceedingly low threshold $th$ values, requiring only a slight perturbation to successfully generate AE \cite{kotyan2019adversarial}.

\paragraph{Boundary Attack} (BA) \cite{brendel2017decision}. The BA is an iterative adversarial attack that operates without a gradient. It commences from a substantial adversarial perturbation and subsequently endeavors to minimize the $L_2$ distance between the original and perturbed examples, while maintaining the adversarial nature. Specifically, the attack begins with an image of the target class and alternates steps between moving the image along the decision boundary (maintaining its adversarial status) and steps moving towards the original image to discover incrementally smaller perturbations. The direction of the steepest ascent on the boundary surface, which is the direction where the model output alters most rapidly, is identified at each iteration by the attacker. The fundamental concept of the algorithm revolves around executing a rejection sampling with an appropriate proposal distribution $P$ to identify incrementally smaller perturbations. During the $k$-th step, the aim is to draw a perturbation $\eta^k$ from a maximum entropy distribution while adhering to the subsequent constraints:\begin{enumerate}
        \item The perturbed instance, denoted as $\tilde{o}$, is confined within the original input domain: $\tilde{o}_i^{k-1} + \eta_i^k \in \text{original domain}$ 
        \item The relative magnitude of the perturbation is represented by $\delta$: $||\delta^k||_2 = \delta * d(o, \tilde{o}^{k-1})$
        \item The perturbation diminishes the Euclidean distance between the perturbed instance and the original input by a relative proportion $\epsilon$: $d(o, \tilde{o}^{k-1}) - d(o,  \tilde{o}^{k-1} + \eta^k) = \epsilon * d(o, \tilde{o}^{k-1})$
    \end{enumerate} Here, $d(o, \tilde{o}) = ||o - \tilde{o}||_2^2$ signifies the Euclidean distance between the original and the perturbed instances, and $o$ denotes the original instance.\\
The original formulation presents substantial complexity due to the challenges in sampling from the given distribution. In light of this, a more straightforward heuristic is employed as follows:\begin{itemize}
    \item Utilization of a Gaussian distribution $N(0,1)$, which is independent and identically distributed;
    \item Perturbed samples undergo rescaling and clipping to ensure the satisfaction of constraints (1) and (2);
    \item The parameter $\eta^k$ is projected onto a sphere centred at $o$, such that $d(o, \tilde{o}^{k-1} + \eta^k) = d(o, \tilde{o}^{k-1}$, thereby maintaining constraint (1). This is referred to as the orthogonal perturbation step;
    \item A modest progression is made towards the original image, ensuring that constraints (1) and (3) are upheld.
\end{itemize}
The algorithm, as elucidated earlier, pivots on two significant parameters: the cumulative perturbation length $\delta$ and the step length $\epsilon$ in the direction of $o$. The intricacies of parameter modification are complex, hence, for a comprehensive understanding, the reader is directed to the original study \cite{brendel2017decision}.\\
Empirical evidence substantiates the efficiency of the Boundary attack as a robust black-box attack. Its capability to locate AEs with minimal perturbations, which are challenging for human detection, underscores its potency. Moreover, its superiority over other gradient-based attacks is evidenced by its ability to bypass the computationally intensive task of gradient computation with respect to the input.

\section{Results and Discussion}

\label{sec:results}
The ensuing discourse delineates the outcomes of the attacks under consideration, as previously introduced in Section \ref{algos}. The ART library \cite{art2018} provided the implementations for all the methodologies under scrutiny, facilitating the examination of a multitude of configuration parameters to comprehensively probe the
potential of each technique. To maintain conciseness, the specifics of each attack's
results are relegated to the \nameref{sec:supplementary}, where the experiments with diverse parameter values are documented. The primary focus of this section, however, remains the comparative analysis of the optimized attacks.\\

The evaluation metrics employed include the model's accuracy over the manipulated samples, the average perturbation introduced to the AEs vis-a-vis the original samples, and the time required for processing. The first two metrics, also utilized in all ART library\footnote{\url{https://adversarial-robustness-toolbox.readthedocs.io/en/latest/guide/notebooks.html}} examples, serve as the primary measures for gauging the performance disparities between attacks and thus, were chosen as the principal metrics for assessing the efficacy of the methods implemented.\\

Accuracy serves to evaluate the performance of classification models. It is informally delineated as the ratio of correct predictions made by our model, specifically for multiclass classification, it is expressed as $\frac{\text{Number of Correctly Classified Samples}}{\text{Total Number of Samples}}$.\\
The mean perturbation introduced is computed as $\frac{1}{|T|} \sum_{s \in T} \frac{\sum_{f \in F_s}|adv^{s}{f} - s{f}|}{|F_s|}$. This denotes the average disparity between the original and altered features for each individual sample. Here, $T$ symbolizes the test set, $F_s$ signifies the feature set of sample $s$, $adv^s$ is the AE derived from sample $s$, and $s_f$ indicates the value of feature $f$ in sample $s$.\\

The training and evaluation of the models were conducted using an identical distribution of data for each split as mentioned in the preceding section, specifically 64/16/20\% for training, validation, and testing, respectively. Detailed insights regarding the structure of the test for each dataset can be referred to in Section \ref{resultsfinal}. The performance of these models, evaluated in terms of accuracy on the comprehensive test set, as well as on the subsets of male and female emotional speech, is presented in Table \ref{origacc}.\\

Initially, a comparative evaluation is conducted predicated on the metrics under consideration. Subsequently, a comprehensive analysis is carried out, factoring in the inherent properties of the diverse attack techniques.\\

In our data presentation, we focus solely on the attack configuration that results in the lowest accuracy parameter, indicative of optimal performance within an adversarial context. This approach is maintained even in light of performance variances across genders under different configurations. In instances where multiple configurations yield identical outcomes, our selection is guided by the configuration that produces the minimum average perturbation.\\
Table \ref{paramrecap} encapsulates the optimal configurations for each dataset and attack, serving as a comparative reference in subsequent chapters.

In general, the optimal configuration remains consistent across the entire test set, for both male and female samples. However, a few exceptions have been noted. For the EMOVO dataset, the FGSM attack outperforms others on the entire data and female samples when $eps=0.5$. Yet, for male samples, equivalent accuracy is achieved when $eps=0.25$, albeit with reduced perturbation. Similarly, the JSMA attack on the EMOVO dataset provides superior results on the entire data and female samples when $theta=+1$. However, male samples exhibit improved accuracy with $theta=-1$. Differently, for the Ravdess dataset, the JSMA attack proves most effective on the entire data and male samples when $theta=+1$, while female samples show enhanced accuracy with $theta=-0.5$.

\subsection{Performance comparison}
In this section, we evaluate the outcomes, taking into account each performance metric separately.

\begin{figure}[H]
    \centering
    \includegraphics[width=1\textwidth]{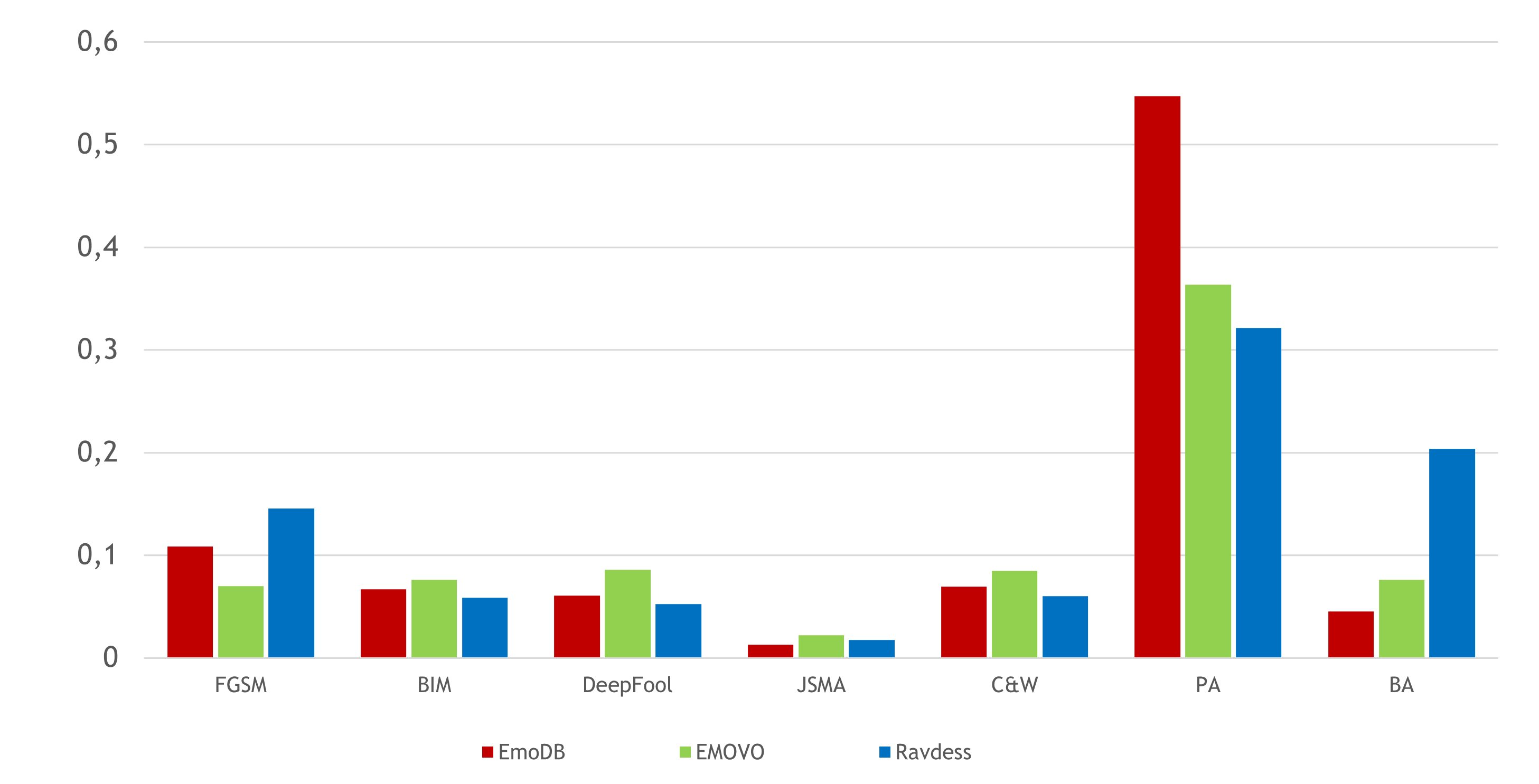}
    \caption{Accuracy obtained by the most effective configuration of each attack. Additional data can be found in Section \ref{resultsfinal}.}
    \label{accalll}
\end{figure}

\paragraph{Accuracy}\label{accpar}
In the experiments, we computed the accuracy as $\frac{\text{Number of Correctly Classified Samples}}{\text{Total Number of Samples}}$.
The precision of the models is assessed using the AEs derived from the instigated attacks, as detailed in Table \ref{paramrecap}. The resultant accuracy metrics are presented in Table~\ref{accresumed} and illustrated in Figure~\ref{accalll}.

The analysis of the results reveals a substantial impact of all attack variants, including the rudimentary FGSM, on the models' performance. Among these, PixelAttack exhibits a distinct efficacy, setting itself apart from the rest. Its effectiveness is particularly pronounced for EMOVO and Ravdess, whereas it shows less impact on EmoDB. Despite this discrepancy, the minimal perturbations it introduces to the samples, as elaborated in the succeeding section, underscore its impressive performance. This idiosyncratic behavior may be attributed to the specific configurations employed, characterized by low thresholds denoted by $th$.
Across all three datasets, the JSMA attack emerges as the most potent, reducing the accuracy to approximately 1-2\%.

The performance of all attacks on the EmoDB dataset is reasonable. The simple yet effective FGSM reduces the accuracy to approximately 10\%, while BIM, DeepFool, and C\&W yield similar results at around 6.5\%. Interestingly, BoundaryAttack, a black-box attack operating with less information, outperforms the latter, suggesting that a successful attack can be executed with minimal or no information about the target.\\
In the EMOVO dataset, FGSM exhibits an unusual efficiency, ranking second and surpassing its iterative BIM variant. The BIM-DeepFool-C\&W trio performs similarly, achieving around 8\% accuracy, on par with BoundaryAttack.\\
In the case of Ravdess, BIM-DeepFool-C\&W again generates similar results, this time around 5.6\%, whereas BoundaryAttack does not maintain the same level of performance as observed in the other datasets.\\

In the comparative analysis of the three languages, the aggregate results exhibit remarkable similarity. The arithmetic mean of accuracies derived from white-box attacks by AEs is the lowest for EmoDB, registering at 0.063, followed by Ravdess at 0.067, and lastly, EMOVO at 0.068. This allows us to infer with confidence that German, Italian, and English demonstrate equivalent susceptibility in the context of a SER task. Further exploration of this subject will be undertaken in the succeeding discourse.\\

Turning our attention to the variance between male and female samples, white-box attacks proved to be more successful on male subjects in 9 out of 15 instances. In the case of EMOVO, this trend is discernible in 4 out of 5 instances, a noteworthy observation given the higher accuracy for males compared to females in the original dataset. Likewise, for Ravdess, male samples proved more susceptible in 3 instances. Conversely, for EmoDB, women were more impacted by 3 out of 5 attacks. The disparity is rather pronounced in certain attacks, with the accuracy differing by approximately 0.04 (excluding PixelAttack), while in other instances, the variation is around 0.01.\\

In summary, JSMA emerges as the most potent assault, substantially undermining the performance of models across all languages. The trio of BIM, DeepFool, and C\&W exhibits a consistent performance across all cases. Even with its unassuming complexity, FGSM has demonstrated its efficacy across all three datasets, with a notable impact on EMOVO. Moreover, despite its black-box characteristics, BoundaryAttack achieves commendable results in two out of the three cases. We have also evidenced how PixelAttack, by altering merely a handful of pixels, can induce significant deviations in a model's behavior.\\

The susceptibility of diverse languages to white-box attacks exhibits no marked disparities. The languages under scrutiny, namely German, Italian, and English, all demonstrate susceptibility to adversarial incursions. Furthermore, the analysis reveals that AEs derived from male audio samples typically yield higher efficacy, notably within the context of the EMOVO and Ravdess datasets.\\

The augmentation of datasets was accomplished through the use of pitch shifting and time stretching techniques, inducing deformations to the input samples. Despite this, the deformations were not adequate to guarantee a robust defense against all forms of AEs attacks. These observations underscore the pronounced susceptibility of the SER task to such attacks when addressed using a CNN-LSTM model trained on log Mel-spectrograms.

\begin{figure}[H]
    \centering
    \includegraphics[width=1\textwidth]{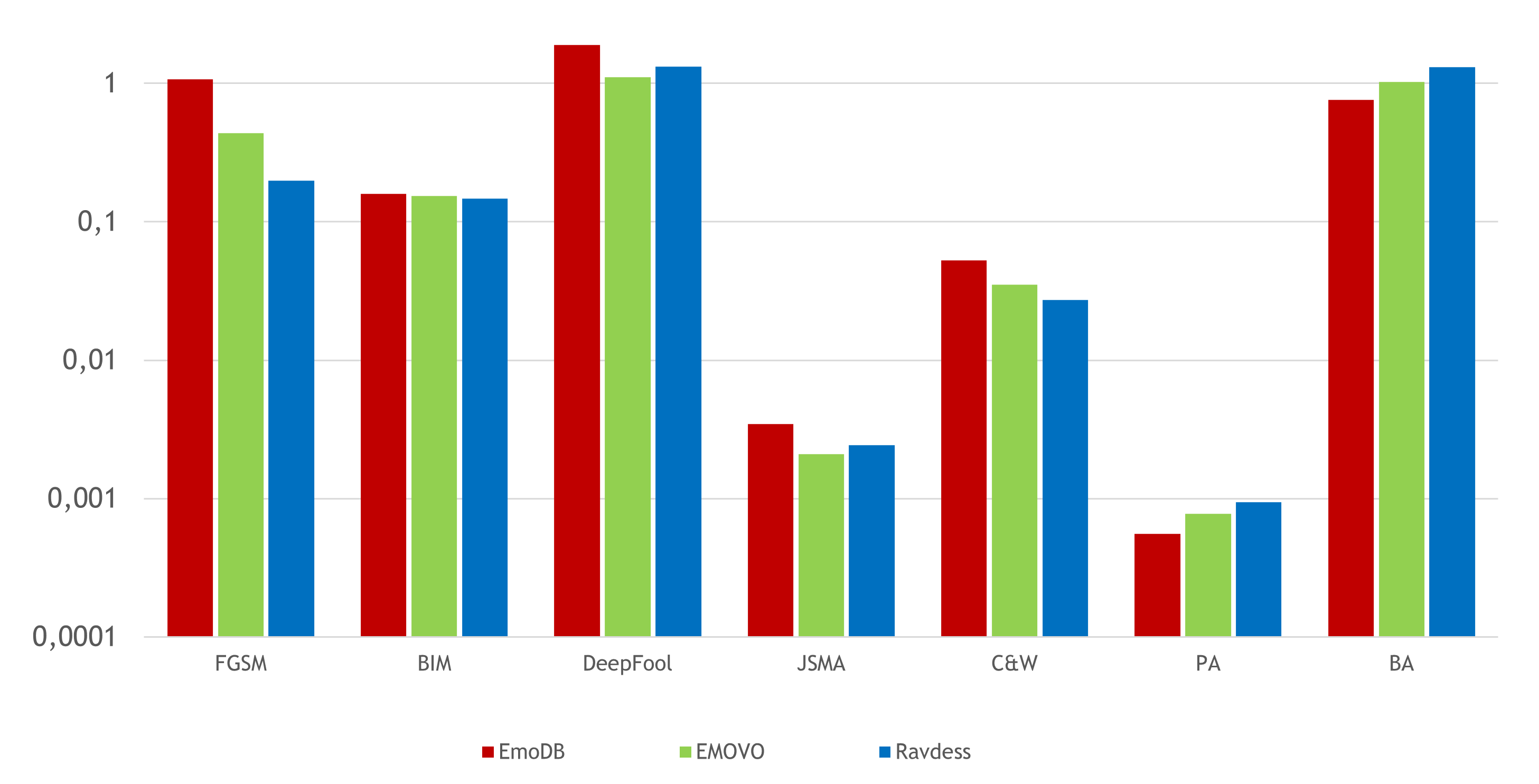}
    \caption{Perturbations injected by the most effective configuration of each attack. Additional data can be found in Section \ref{resultsfinal}}
    \label{pertall}
\end{figure}

\begin{figure}[h]
    \centering
    \begin{subfigure}{0.49\textwidth}
         \centering
         \includegraphics[width=\textwidth]{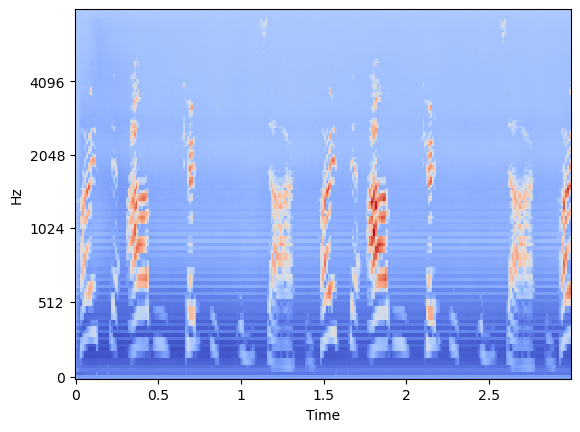}
         \caption{Original}
    \end{subfigure}
    \hfill
    \begin{subfigure}{0.49\textwidth}
         \centering
         \includegraphics[width=\textwidth]{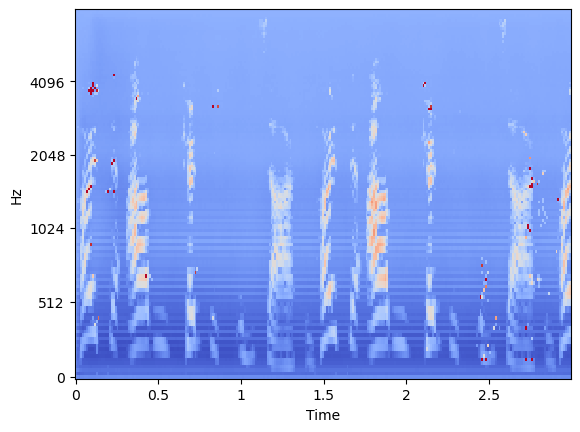}
         \caption{Attacked}
    \end{subfigure}
     \caption{Example of \textbf{male} samples from \textbf{EmoDB}. (a) standardized original sample and (b) its JSMA-attacked version.}
     \label{d1M}
\end{figure}

\begin{figure}[h]
    \centering
    \begin{subfigure}{0.49\textwidth}
         \centering
         \includegraphics[width=\textwidth]{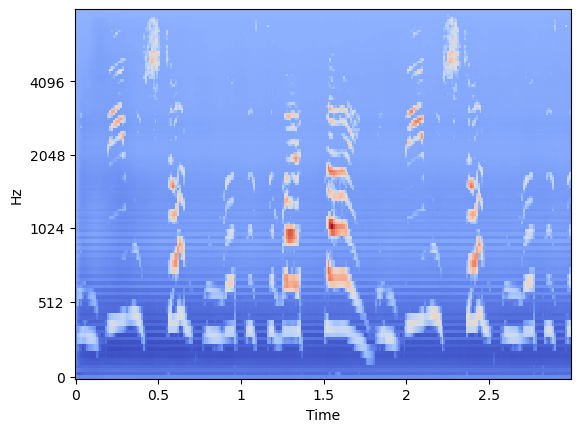}
         \caption{Original}
    \end{subfigure}
    \hfill
    \begin{subfigure}{0.49\textwidth}
         \centering
         \includegraphics[width=\textwidth]{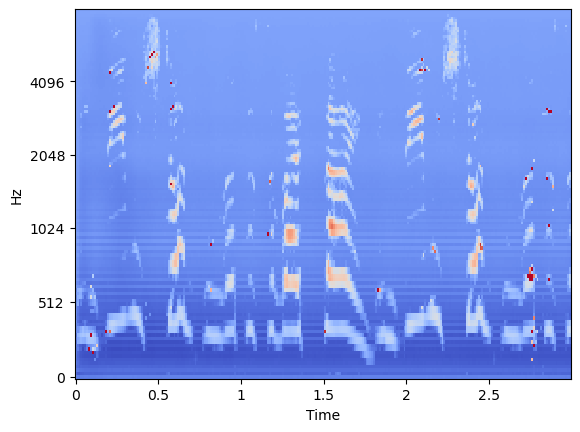}
         \caption{Attacked}
    \end{subfigure}
     \caption{Example of \textbf{female} samples from \textbf{EmoDB}. (a) standardized original sample and (b) its JSMA-attacked version.}
     \label{d1F}
\end{figure}

\begin{figure}[h]
    \centering
    \begin{subfigure}{0.49\textwidth}
         \centering
         \includegraphics[width=\textwidth]{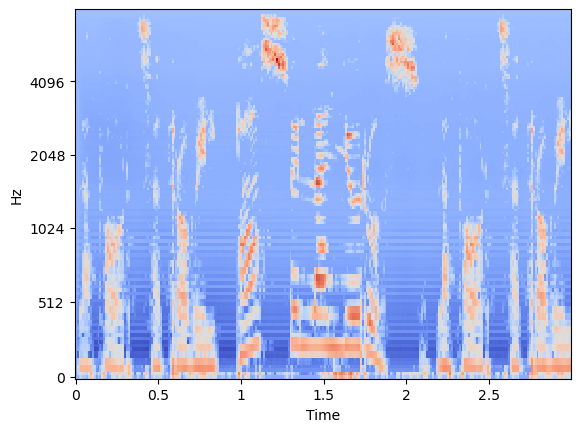}
         \caption{Original}
    \end{subfigure}
    \hfill
    \begin{subfigure}{0.49\textwidth}
         \centering
         \includegraphics[width=\textwidth]{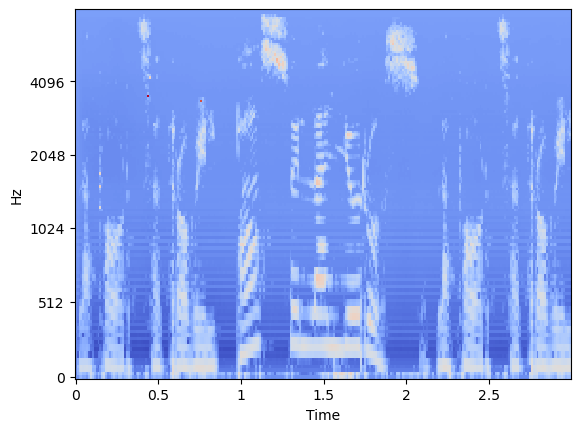}
         \caption{Attacked}
    \end{subfigure}
     \caption{Example of \textbf{male} samples from \textbf{EMOVO}. (a) standardized original sample and (b) its JSMA-attacked version.}
     \label{d2M}
\end{figure}

\begin{figure}[h]
    \centering
    \begin{subfigure}{0.49\textwidth}
         \centering
         \includegraphics[width=\textwidth]{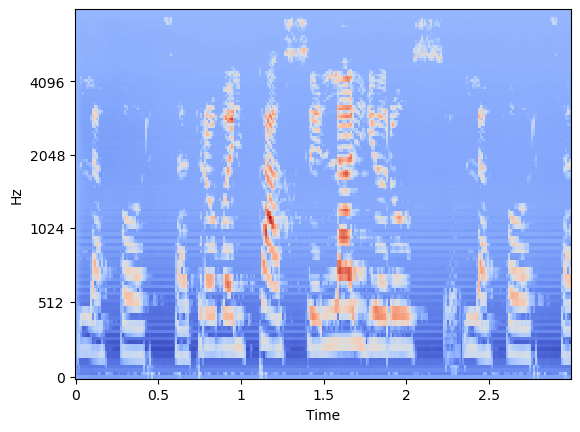}
         \caption{Original}
    \end{subfigure}
    \hfill
    \begin{subfigure}{0.49\textwidth}
         \centering
         \includegraphics[width=\textwidth]{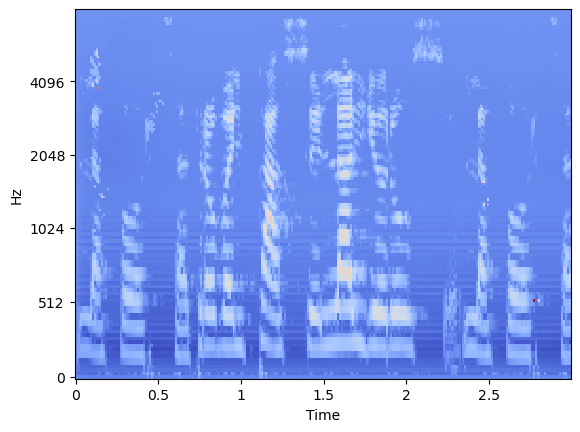}
         \caption{Attacked}
    \end{subfigure}
     \caption{Example of \textbf{female} samples from \textbf{EMOVO}. (a) standardized original sample and (b) its JSMA-attacked version.}
     \label{d2F}
\end{figure}

\begin{figure}[h]
    \centering
    \begin{subfigure}{0.49\textwidth}
         \centering
         \includegraphics[width=\textwidth]{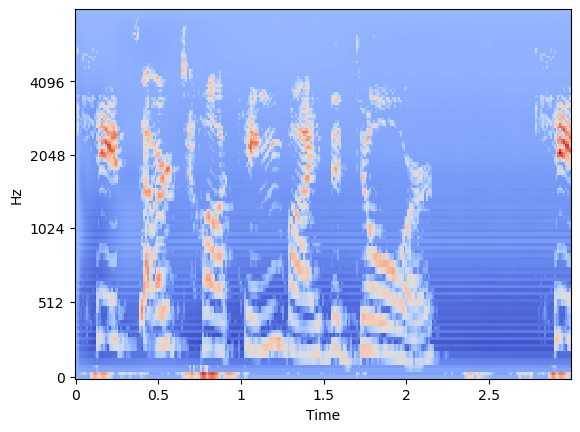}
         \caption{Original}
    \end{subfigure}
    \hfill
    \begin{subfigure}{0.49\textwidth}
         \centering
         \includegraphics[width=\textwidth]{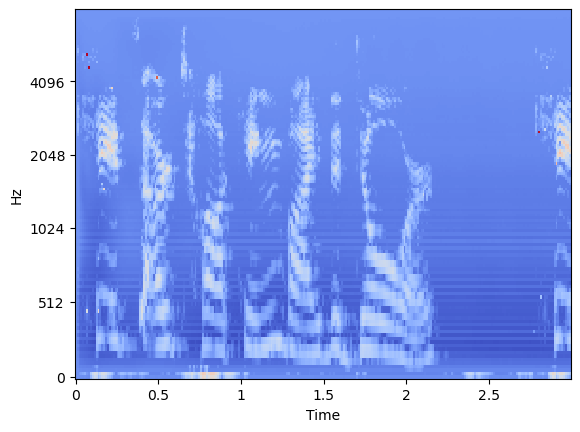}
         \caption{Attacked}
    \end{subfigure}
     \caption{Example of \textbf{male} samples from \textbf{Ravdess}. (a) standardized original sample and (b) its JSMA-attacked version.}
     \label{d3M}
\end{figure}

\begin{figure}[h]
    \centering
    \begin{subfigure}{0.49\textwidth}
         \centering
         \includegraphics[width=\textwidth]{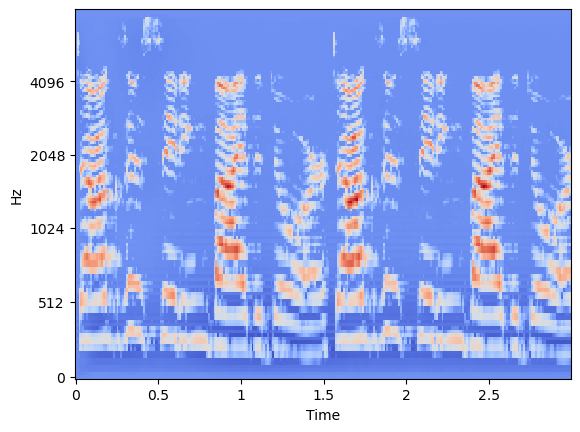}
         \caption{Original}
    \end{subfigure}
    \hfill
    \begin{subfigure}{0.49\textwidth}
         \centering
         \includegraphics[width=\textwidth]{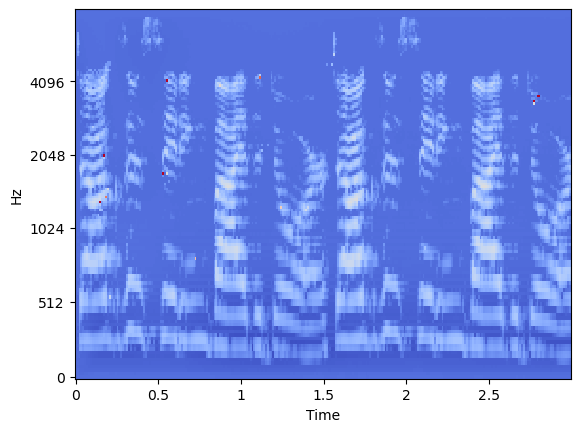}
         \caption{Attacked}
    \end{subfigure}
     \caption{Example of \textbf{female} samples from \textbf{Ravdess}. (a) standardized original sample and (b) its JSMA-attacked version.}
     \label{d3F}
\end{figure}

\paragraph{Perturbation} 
We now turn our attention to the average perturbation induced by the attacks in the creation of the AEs. The data presented in Table~\ref{pertresumed} pertain to the highest accuracy outcomes derived from the parameter configurations encapsulated in Table~\ref{paramrecap} and depicted in Figure~\ref{pertall}. For further inspection, we also provide samples of spectrograms perturbed in Figures~\ref{d1M}-\ref{d3F}.
\\

The preceding section discussed the configuration of PixelAttack, which manipulates only a restricted set of pixels, thereby leading to a markedly diminished mean perturbation. Apart from PixelAttack, the most commendable performance is delivered by JSMA. Intriguingly, the configuration yielding the lowest accuracy for this attack is not the one introducing the minimal perturbation, but rather corresponds to $theta=0.5$. Further details are provided in Section \ref{resultsfinal}, which elucidates that each $theta$ value corresponds to an average perturbation that is lower than that of all instances presented in Table \ref{pertresumed}. In the most unfavorable scenario, i.e., when $theta=-1$, the outcomes are akin to those obtained with C\&W using the L2 distance, albeit surpassing the results of the other attacks.

The triplet comprising BIM, DeepFool, and C\&W, previously discussed for their comparable accuracy, demonstrates substantial discrepancies when it comes to the magnitude of introduced perturbations. Notably, an order of magnitude difference is observed between attacks for each dataset. In ascending order of perturbation magnitude, the attacks are C\&W, BIM, and DeepFool.\\

Upon application to EmoDB, the optimal configuration of DeepFool, irrespective of gender, does not necessarily yield the least amount of noise introduced, albeit the discrepancy is negligible. In the case of Ravdess, the least perturbation is attained with a solitary iteration of the algorithm, the difference being a mere 0.02. Comprehensive details for both instances are available in Section \ref{resultsfinal}. Remarkably, despite its nature as a black-box attack, BoundaryAttack exhibits performance on par with, and occasionally surpassing that of DeepFool. \\

The outcomes of the evaluated algorithms are generally consistent across various languages, with the exception of FGSM. This divergence can be attributed to performance fluctuations corresponding to different $eps$ values, which dictate the attack step size. As anticipated, a decrease in $eps$ values results in reduced perturbation levels. Based on the configurations outlined in Table \ref{paramrecap}, the perturbation levels span from optimal to suboptimal for Ravdess, EMOVO, and EmoDB respectively.\\

The influence of the speaker's gender on all adversarial attacks is generally insignificant, with the notable exception of DeepFool, which exhibits the most substantial disparities.\\

To summarize, the Jacobian-based Saliency Map Attack  (JSMA) emerges as the most efficient attack mechanism, given its proficiency in inducing perturbations in the sample data. Its effectiveness is further underscored by the resultant decrease in accuracies, thereby solidifying its position as the preeminent attack strategy.\\

As evidenced in tables \ref{dfacc} and \ref{dfpert}, DeepFool is characterized by the introduction of substantial noise. However, neither the accuracy nor the noise level exhibits noticeable enhancement with an escalation in the iteration count. The PixelAttack method, on the other hand, substantiates the feasibility of misleading the model through the alteration of a minimal number of pixels in the log Mel-spectrogram.

As Figures~\ref{d1M}-\ref{d3F} exemplify, the resulting spectrograms are mainly degraded because of an almost uniform decrease of the amplitude of the spectrograms, resulting in sparse outlier points. When listening to such samples, the user is usually able to detect the attacked audio due to a noticeable decrease in volume. The GitHub repository contains examples of such audio.

\begin{figure}[H]
    \centering
    \includegraphics[width=0.8\textwidth]{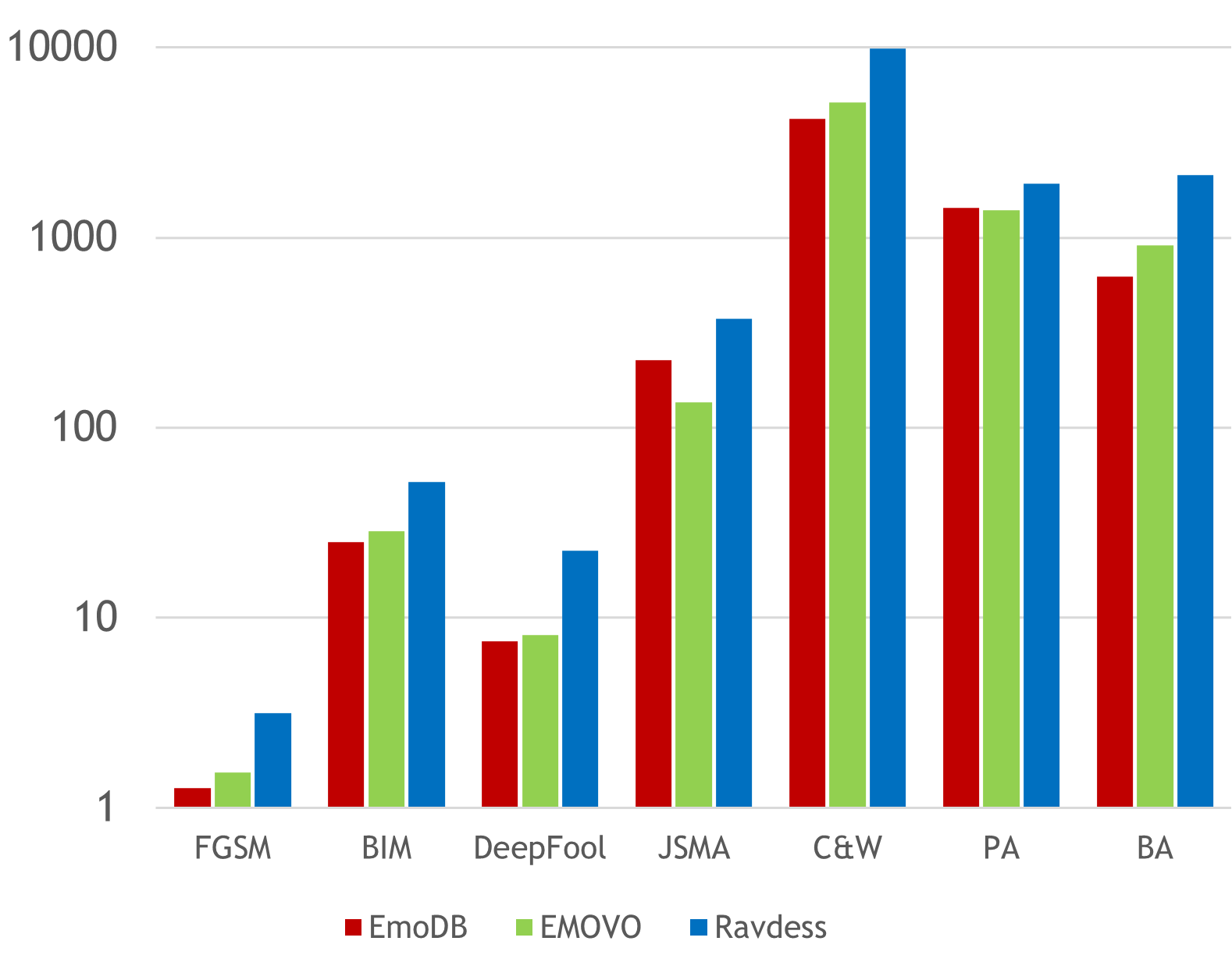}
    \caption{Time (s) required to generate the AEs for the various attacks and datasets for the best-performing configuration. Additional data can be found in Section \ref{resultsfinal}.}
    \label{timealll}
\end{figure}

\paragraph{Execution Time} In line with the evaluation carried out for precision and average disturbance, our attention now turns to the duration required by the under-consideration attack algorithms for the creation of AEs. 
The trial runs were conducted on a workstation of the HP Z4 G4 series, equipped with an i9-9820X CPU, a Nvidia TITAN V GPU possessing 12 GB of RAM, and a CPU RAM of 64 GB. The findings depicted in Figure \ref{timealll} derive from the most effective precision outcomes garnered from the parameter setups delineated in Table \ref{paramrecap}. \\

The FGSM, as anticipated, outperforms all other attacks in terms of speed, even while introducing substantial noise. This makes it an effective and efficient approach for generating AEs rapidly. Although none of the selected configurations prove to be the fastest in terms of execution time, the disparities across various experiments are trivial, merely amounting to fractional seconds.\\

Focusing on the BIM-DeepFool-C\&W trio, which we reiterate achieves comparable accuracy outcomes, noticeable variations are evident in terms of execution time aside from perturbation.\\
The C\&W attack is the most time-consuming and requires a substantial duration to yield results. The scenario remains unchanged even when the $L_2$ distance is taken into account: the ensuing durations are akin to those of PixelAttack with $th=1$, as highlighted in Section \ref{resultsfinal}.\\

Contrarily, DeepFool is capable of generating AEs promptly, although the perturbation induced is substantially high, as previously discussed.

BIM necessitates marginally extended durations compared to DeepFool, yet significantly less than C\&W, positioning it as an optimal choice for creating high-quality AEs with minimal noise and in a reasonable timeframe. Although the configurations utilized are not the quickest, the disparities are inconsequential, akin to FGSM.\\

As previously noted, JSMA presents a high degree of effectiveness in impairing performance and introducing perturbation, albeit it requires a longer duration to generate samples in comparison to FGSM, BIM, and DeepFool. Nevertheless, its execution time remains considerably less than that of C\&W. It is important to highlight that the parameter configurations selected to minimize accuracy concurrently result in reduced execution times. Despite the outcome for EMOVO not being strictly superior, the difference in execution time is a mere second.\\

Contrary to the majority of white-box attacks, the two black-box attacks necessitate a longer duration, except for C\&W which is the slowest technique overall. This is anticipated due to their limited insight into the internal workings of the target model.\\

For PixelAttack, it is pertinent to mention, as elaborated further in Section \ref{resultsfinal}, that increasing the number of modifiable pixels reduces the computation time. As previously stated, it is important to consider that experimenting with larger values may lead to a decrease in both the model's accuracy and the time required.

\subsection{Comparison between characteristics}
In this section, a critical examination of the metrics under consideration is undertaken, with subsequent conclusions drawn from the intrinsic properties of each methodology and potential variances within the data.

\paragraph{Variations in Distance Metrics} It is imperative to understand that diverse attack types strive to minimize the disparity between the original sample, employing a range of distance metrics. A summary of the distance metrics utilized by the implemented ART is presented in Table \ref{disttried}.

Drawing definitive conclusions about potential disparities among distances and attacks utilizing the same distance metric is a complex task. Notably, JSMA and PixelAttack employ the $L_0$ distance, yet their effectiveness varies significantly. JSMA proves to be the most efficient, whereas PixelAttack is less effective. The gathered data indicates that these attacks introduce minimal perturbations, as they aim to reduce the quantity of altered pixels, a characteristic inherent to the $L_0$ distance.\\
Considering the $L_2$ category, it encompasses DeepFool and one variant of C\&W. The interpretation of results becomes intricate here as the two attacks yield considerably divergent outcomes. Despite C\&W achieving superior accuracy (tripling the score on EmoDB), the perturbation it introduces is remarkably lower (by three orders of magnitude). However, the generation of AEs via C\&W is significantly more time-consuming, taking thousands of times longer than DeepFool. Consequently, it is challenging to extract consistent patterns from the executed experiments. The only conclusive remark is that the outcomes derived using the $L_2$ distance are profoundly influenced by the attack's intrinsic logic.\\
Lastly, the $L_\infty$ category, which encompasses the most substantial number of attacks, is considered. Despite all attacks striving to minimize the maximum discrepancy between the original and manipulated examples, the results exhibit significant variations. The accuracy is comparable in all instances (with the exception of FGSM and BoundaryAttack on Ravdess, which demonstrate notably superior results), yet the average perturbation and execution times differ considerably among various cases. Hence, the overall behavior of these techniques is primarily determined by their internal mechanisms.\\

The task of discerning a universal pattern through the juxtaposition of dissimilarities among attack groups utilizing identical metrics presents a considerable challenge. The primary source of variability in algorithms arises from the inherent methodology, rather than the minimized distance. The complexity is further heightened when attempting to compare attacks employing disparate metrics.\\
A preliminary inference drawn from the acquired results suggests that $L_0$ attacks appear to induce fewer perturbations compared to attacks that deploy other distance metrics. Yet, this inference warrants additional scrutiny, considering the fact that our testing was confined to merely two algorithms within this category, and PixelAttack was set up differently compared to the rest of the techniques.

\paragraph{Iterative and normal versions}
BIM and FGSM, two techniques with a comparative relationship, given BIM is an extension of FGSM, exhibit differing performance under varying conditions. Upon examination of their accuracy results in light of all parameters considered, as presented in Tables \ref{fgsmacc} and \ref{bimacc} in Section \ref{resultsfinal}, it is notable that the $eps$ parameter appears to exert no influence on BIM's performance. This is evidenced by the consistent accuracy across different $eps$ values. Conversely, FGSM's accuracy fluctuates with the parameter, presenting a unique trend for each language. This suggests that BIM, with its independence from identifying the optimal configuration for deceiving the model, may be more advantageous.\\
The choice of datasets also influences the performance, as demonstrated in Table \ref{accresumed}. BIM was observed to be more proficient in diminishing accuracy in the EmoDB and Ravdess datasets, while FGSM demonstrated slightly superior performance in the EMOVO dataset.\\
With regards to the perturbation introduced, an increase in $eps$ results in heightened noise for both attacks, as illustrated in Tables \ref{fgsmpert} and \ref{bimpert} in Section \ref{resultsfinal}. This is an anticipated outcome, given that the $eps$ parameter signifies the maximum perturbation an attacker can introduce. Nevertheless, BIM generates more refined AEs with lower perturbation values.\\
Although BIM requires a significantly longer duration than FGSM to yield results, the attack time of 51 seconds for the largest dataset, Ravdess, is deemed acceptable, as indicated in Tables \ref{bimtime} and \ref{fgsmtime} in Section \ref{resultsfinal}.\\

\paragraph{White and black box attacks}
In this study, we executed a series of five white-box attacks and two black-box attacks. Despite the disparity in their quantities, a comparative analysis of these two categories is still significant. This significance stems from the distinct configuration of PixelAttack, which leads to considerably dissimilar results compared to other forms of attacks. Moreover, a closer inspection of the internal procedures utilized by these algorithms allows us to perceive this comparison in the context of gradient-based (white-box) and gradient-free (black-box) attacks, and the varying degrees of access to the model's information.\\

Contrary to intuitive expectations, black-box attacks do not necessarily underperform due to their limited access to model information. This assertion is substantiated by the data presented in Table \ref{accresumed}. For instance, the BoundaryAttack surpasses almost all white-box attacks in terms of performance for the EmoDB and EMOVO datasets. Nevertheless, this does not hold true for the Ravdess dataset, where its performance is significantly inferior to all gradient-based attacks.\\
Upon evaluating the accuracy achieved by white-box attacks, it is evident that the results are comparable and consistently effective across different datasets. On the other hand, black-box techniques demonstrate wider variances. This indicates that gradient-based methods could potentially be language-independent, or at the very least, more so than population-based (PixelAttack) or decision-based (BoundaryAttack) methods, and are capable of working efficiently with log Mel-spectrograms.\\

From the average perturbation data presented in Table \ref{pertresumed}, it is evident that there are substantial variances in the performance of BoundaryAttack across different datasets. A similar trend is also discernible in the case of FGSM, where the selection of the $eps$ value for accuracy minimization is of considerable significance, and DeepFool, where the noise level is even more pronounced. In contrast, PixelAttack consistently produces AEs with a comparable, low level of noise, given its configuration to alter only a minimal number of pixels. 

On analyzing the perturbation variations across different languages, it is noted that black-box attacks introduce a lesser degree of perturbation for EmoDB, while the perturbation is more pronounced for white-box attacks. A more detailed discussion on this observation will be presented in the subsequent section.

As indicated in Table \ref{timestab}, the execution time for black-box attacks is typically longer, owing to the limited information available about the targeted model.

To summarize, despite the paucity of information about the victim model, black-box attacks can sometimes outperform their white-box counterparts by generating AEs with superior performance and lower disruption. However, our experimental results suggest that, on average, black-box attacks necessitate a longer execution time.

\paragraph{Differences between languages}
The trained models exhibit proficient performance across the three languages under consideration, as demonstrated in Table \ref{origacc}. Ravdess yields the highest accuracy on the original data, registering at 0.912. This is closely followed by EmoDB and EMOVO, with respective accuracies of 0.909 and 0.872.\\

In section \ref{accpar}, we posited that the vulnerability to attacks across all three languages is relatively uniform. To elucidate this further, Figure \ref{accalll} presents the accuracy across the three datasets for the most effective attack configurations, as detailed in Table \ref{accscores}, applied to the entire test set.\\
While the discrepancies between the achieved values are generally insubstantial, it is noteworthy that black-box attacks exhibit more significant variations. Nevertheless, these observations enable us to derive some intriguing insights.\\

The model trained on the EMOVO dataset, despite exhibiting the lowest accuracy on the original data, outperforms the other models in terms of resistance to AEs. Specifically, it achieves superior performance when subjected to 4 out of 7 attack methods, thus suggesting a diminished impact of the attacks on this model. Consequently, it can be deduced that the model trained on Italian samples exhibits a marginally higher resilience.\\
In contrast, models trained on the Ravdess dataset present a different scenario. These models, while attaining the highest accuracy on the original data, demonstrate a drop in performance when exposed to AEs, thereby making them the least resistant in 4 out of the 7 cases. This is particularly alarming given that the Ravdess model was trained with a larger dataset and over a greater number of epochs, factors that would typically contribute to increased robustness. Crucially, this underscores the fact that the resilience of a model to AEs is not solely contingent on the volume of the training data, but also its quality.\\

An analogous analysis can be conducted on the injected perturbations. The accuracy of the most effective attack configurations across the three datasets, as presented in Table \ref{pertcores}, is depicted in Figure \ref{pertall} for the entire test set.\\

The EmoDB dataset is subjected to the most substantial perturbations in five out of all the attacks, implying that the majority of the implemented attacks have introduced the maximum level of noise. Notwithstanding the elevated perturbation, the attacks executed on EmoDB do not necessarily yield the lowest accuracies among the languages, as corroborated by Table \ref{accscores}. This observation infers that the scrutinized techniques instigate an increased level of noise, which does not unequivocally translate into superior-performing AEs.\\

As previously alluded to, EmoDB exhibits a higher degree of perturbation across all white-box attacks, yet it manifests less interference under black-box attacks in contrast to the other two languages. This could infer that the efficacy of gradient-based methodologies in generating adversarial examples might be diminished when applied to the German language.\\
In terms of average perturbation, both EMOVO and Ravdess demonstrate analogous scores, with the latter predominantly impacted by the outcomes of black-box attacks. By synthesizing the data from tables \ref{accscores} and \ref{pertcores} in Section \ref{resultsfinal}, it can be inferred that the assaults on Ravdess are both effective (evidenced by low accuracy) and efficient (indicated by minimal noise introduction). This insinuates that the English language might be more susceptible to AEs, and reinforces the notion that a model's robustness does not necessarily equate to its resistance against such attacks.\\
EMOVO, on the other hand, registers the lowest average perturbation and, as anticipated, the highest accuracy score.\\

In summary, the present analysis demonstrates that AEs based on log Mel-spectrogram, when fed to a CNN-LSTM, can significantly degrade the performance of a SER model, regardless of the language considered. Although the performance differences among languages are relatively small, the experiments provide valuable insights.\\
Our findings indicate a heightened susceptibility of the English language to the discussed attacks, evidenced by its diminished accuracy notwithstanding its superior performance on the pristine data. Moreover, the observed mean perturbation is relatively insignificant, implying the generation of high-quality AEs.\\
Conversely, the Italian language demonstrates a greater degree of resilience to the same attacks, as inferred from its marginally superior accuracy and diminished perturbation.\\
The German language, however, presents a scenario that lies intermediate to the aforesaid languages. It exhibits an increased vulnerability specifically to gradient-based attacks, given that the perturbation introduced in these instances surpasses that noted for the other languages.

\begin{figure}[H]
    \centering
    \includegraphics[width=1\textwidth]{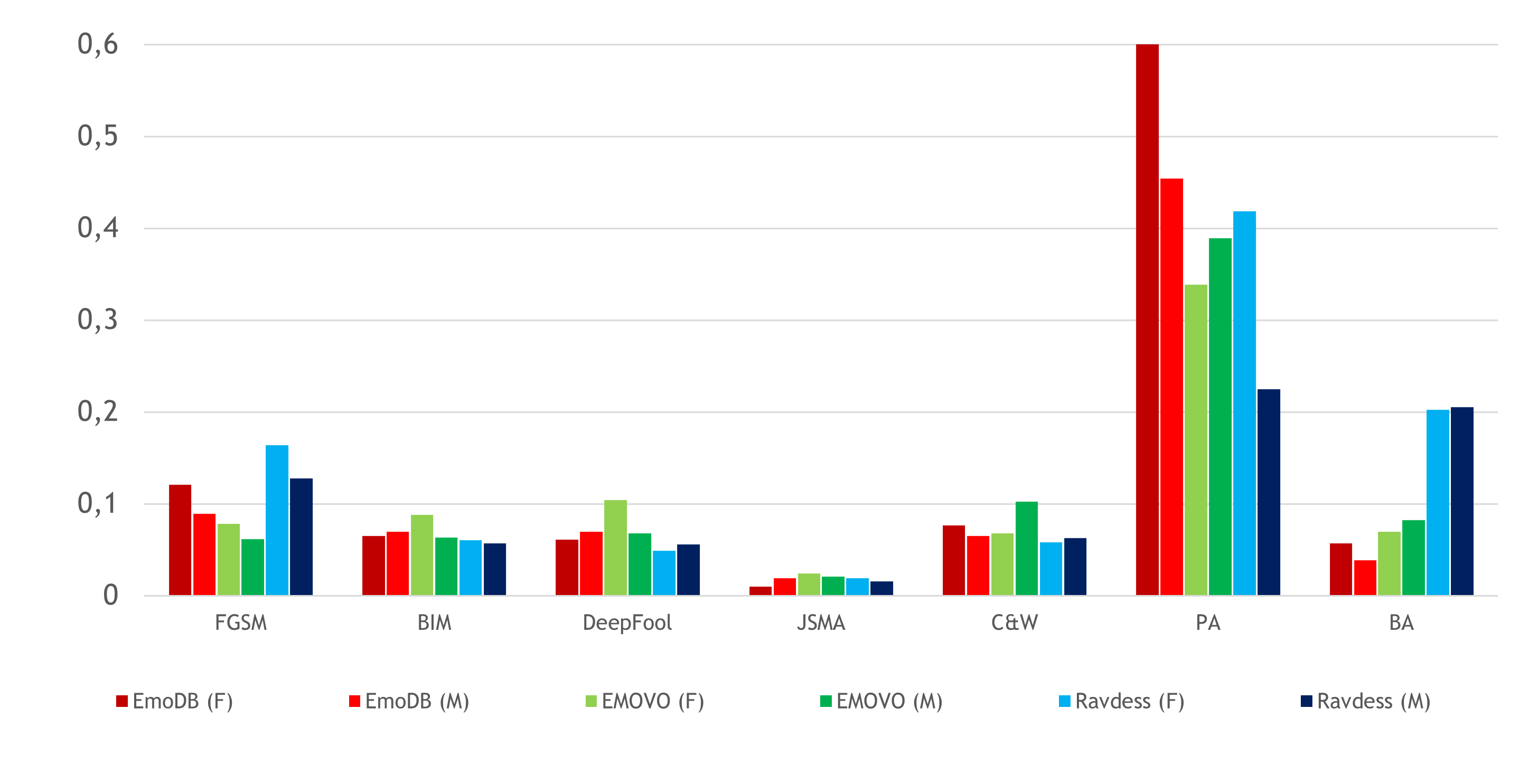}
    \caption{Accuracy obtained by the most effective configuration of each attack for male/female data across datasets.}
    \label{acclang}
\end{figure}

\paragraph{Differences between genders}
Table \ref{origacc} illustrates that the trained models distinguish between male and female samples with negligible variations in the EmoDB original data. However, a pronounced discrepancy is discernible in the EMOVO dataset. Conversely, the models exhibit insignificant fluctuations in the classification of male and female samples in the Ravdess dataset.\\
Although gender disparities are generally nuanced, Table \ref{acclang} provides a more detailed insight by presenting the accuracy exclusively for male/female samples across each dataset. Additional details are elaborated in Section \ref{resultsfinal}.\\

A salient observation is that the utilization of PA yields significant disparities between the two genders. This phenomenon could be attributed to the specific configuration employed. However, it implies that for marginal deviations from the original samples, both German and English languages exhibit increased resilience towards female AEs, conversely, the Italian language demonstrates greater resistance against male AEs.\\

Our preceding analyses, delineated in Section \ref{accpar}, revealed a distinct pattern of white-box attacks exerting a greater impact on male subjects in 9 out of 15 instances -- additional details can be found in \nameref{sec:comparisonAdditional}. This pattern was not exclusive to the aforementioned cases but was also observed in 4 out of 5 instances within the EMOVO dataset, despite the higher initial data accuracy of male subjects compared to their female counterparts. 
In the case of the Ravdess dataset, the AEs proved to be more effective in 3 instances concerning male subjects, who, interestingly, exhibited marginally lower initial data accuracy. However, in the EmoDB dataset, a contrasting trend was observed. Here, 3 out of 5 attacks were more potent on female subjects, who had initially achieved higher accuracy scores on the original data compared to male subjects. 
In summary, the data suggests that the gender with superior initial data accuracy is more susceptible to attacks in two out of the three datasets analyzed. Consequently, this leads to diminished accuracy on AEs relative to the other gender. This indicates that the model's resilience to gradient-based attacks on the best-performing gender cannot be reliably predicted solely based on the performance of the original data. \\inal data accuracy. In contrast, for EmoDB, 3 out of 5 attacks were found to be more influential on female subjects, who had higher accuracy scores on the original data than male subjects.\\
Overall, these results suggest that, for two out of three datasets, the gender with higher original data accuracy is more vulnerable to attacks, resulting in lower accuracy on AEs than the other gender. Thus, relying on the original data's performance may not provide a reliable indicator of the model's susceptibility to gradient-based attacks on the best-performing gender\\

The outcomes of the black-box attack scenario present an equitable distribution, with males outperforming on the EmoDB, while females demonstrate superior attack efficacy on EMOVO. In the case of Ravdess, each gender triumphs in one attack.\\

In addition, the findings corroborate those delineated in Table \ref{accscores}, which pertain to the most effective attacks on the datasets. This consistency in performance is observed even when the dataset is bifurcated into male and female categories. The minor variations in accuracy between the two genders do not significantly impact the overall efficacy of the attacks against the comprehensive AE dataset. Notably, Ravdess comprises the majority of subsets with diminished accuracies, two attacks excel on EmoDB, while EMOVO records a single instance of superior performance with FGSM.\\

\begin{figure}[H]
    \centering
    \includegraphics[width=1\textwidth]{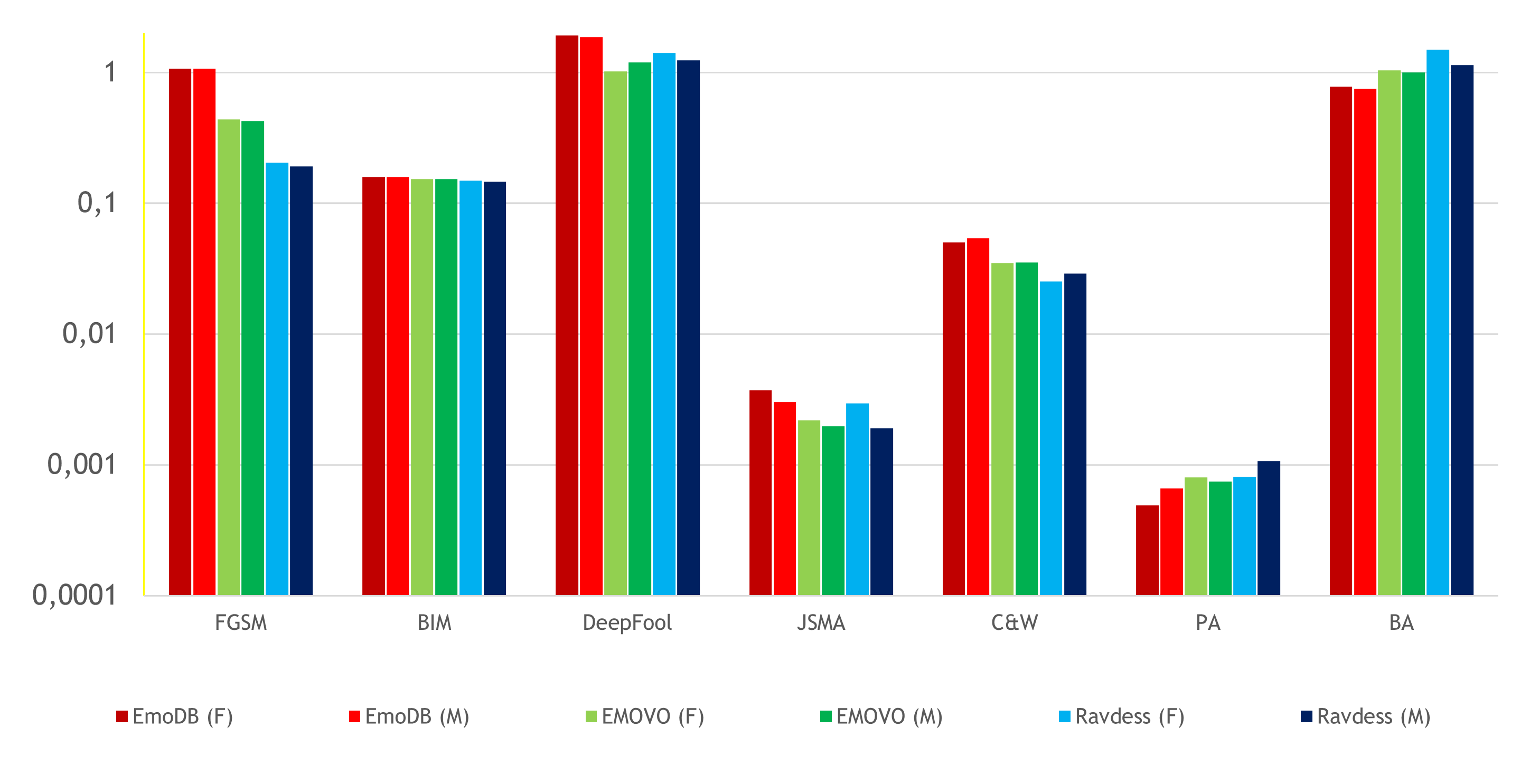}
    \caption{Perturbation injected by the most effective configuration of each attack for male/female data across datasets.}
    \label{pertlang}
\end{figure}

In a similar vein, Figure \ref{pertlang} provides perturbations exclusively for male and female samples within each dataset. Further elaboration on this topic can be located in Section \ref{resultsfinal}.\\

An initial cursory examination suggests an insubstantial distinction between genders. Yet, a more meticulous analysis of the white-box results uncovers that in 11 out of 15 scenarios, male AEs manifest a diminished level of perturbation compared to female AEs. In addition, in 7 out of these 11 instances, males also demonstrate a reduced accuracy rate than females, as illustrated in Table \ref{accsgender} in Section \ref{resultsfinal}. From these observations, it can be deduced that male AEs typically yield superior quality in terms of both the degradation of model performance and the magnitude of induced perturbation.\\ 

In the context of black-box attacks, males registered a lower accuracy in 4 out of 6 situations, although their accuracy was only inferior to females in a single case.\\ 

Upon evaluating individual attacks, it is discernible that male speech consistently manifests diminished perturbation across all instances for the FGSM, BIM, JSMA, and BA attacks. Conversely, for female speech, this phenomenon is solely observed with the C\&W attack.\\ 

In summary, the findings suggest that there are negligible differences in performance between males and females in the majority of instances. Nevertheless, upon a more detailed examination, males seem to hold a minor edge in terms of efficacy (lower accuracy) and quality (reduced induced noise), particularly in the context of white-box attacks.

\section{Conclusion}
The scientific community has been increasingly focusing on adversarial machine learning in recent years. Despite the surge in the development and application of new techniques for SER, the susceptibility of these methods to various forms of attacks has not been sufficiently explored. This paper aims to fill this research gap by evaluating the robustness of SER systems against AEs. Our study scrutinizes three languages—German, Italian, and English—sourced from distinct datasets (EmoDB, EMOVO, and Ravdess, respectively) to discern potential disparities among them. Furthermore, we have incorporated a gender-based perspective into our analysis, investigating the differential impacts of adversarial attacks on male and female speech. Importantly, the implementation of the presented experimental set-up is publicly available at \url{https://github.com/LIMUNIMI/thesis_adversarial_ml_audio} ensuring full reproducibility of the achieved results. \\

We devised a pipeline to standardize the samples across the three languages and extract log Mel-spectrograms. Our methodology involved augmenting the datasets using pitch shifting and time stretching techniques, while maintaining a maximum sample duration of 3 seconds. Specifically, we generated eight distinct versions of the processed data, each differing in the data normalization method applied. The outcomes of our experiments were highly encouraging, demonstrating that the Convolutional Neural Network - Long Short-Term Memory (CNN-LSTM) models performed optimally and consistently when standardized log Mel-spectrograms were used, across all datasets.\\

To address the SER task, we established a uniform CNN-LSTM architecture across all datasets, thereby ensuring methodological consistency for attack comparisons. Through rigorous experimentation with diverse configurations of the neural network, optimal performance was achieved with a modestly sized CNN and 256 bidirectional LSTM units. Subsequent hyperparameter tuning further refined the performance for each dataset. This design strategy yielded high accuracy results on the EmoDB, EMOVO, and Ravdess test sets, with respective accuracies of 90.92\%, 89.52\%, and 91.76\%. These findings underscore the efficacy of employing a CNN-LSTM network trained on log Mel-spectrograms for the SER task while being in line with the state of the art.\\

Upon completion of the model development phase, we assessed the vulnerability of the resultant models to the previously mentioned attacks, under varying parameter configurations. Our empirical investigation revealed a substantial susceptibility of the SER task to AEs. Each examined attack method, including the relatively straightforward FGSM or PixelAttack, which was designed to alter a minimal number of features, successfully misled the network's predictions. In light of these findings, it is evident that the CNN-LSTM model did not exhibit resilience against any of the employed attack techniques. Consequently, we advocate for the exploration of more robust models or alternative training data to enhance system robustness.\\

Our research findings indicate that amongst the multitude of attacks considered, Jacobian-based Saliency Map Attack (JSMA) surfaced as the most potent. The optimal configuration of its parameters led to a significant dip in accuracy rates, resulting in 1.31\, 2.23\, and 1.73\ for EmoDB, EMOVO, and Ravdess datasets, respectively. Furthermore, JSMA introduces only a minuscule degree of perturbation into the AEs. It is second only to PixelAttack, which is specifically tailored to alter a minimal number of pixels per spectrogram, in achieving the least perturbation.\\

The comparative analysis between white-box and black-box methodologies revealed that black-box techniques exhibit superior performance and minimal perturbation in two out of the three cases, specifically with BoundaryAttack. This is notwithstanding their limited access to information about the targeted model. Using BoundaryAttack, we recorded a significant drop in the accuracy of EmoDB, EMOVO, and Ravdess to 4.54\, 7.6\, and 20.38\, respectively. This observation is alarming as it implies that attackers can potentially achieve remarkable results without any understanding of the model's internal operation, simply by scrutinizing its output.\\

When we evaluated the impacts of the attacks across the three languages, no substantial difference in performance was observed. However, the results suggest that English appears to be the most susceptible, while Italian displays the highest resistance.\\

The comparative analysis between male and female samples revealed only negligible variations. A meticulous examination of the results, however, indicates a slight superiority of male samples, particularly in white-box attack scenarios, where they exhibited marginally lesser accuracy and perturbation.

To encapsulate, we have introduced a reliable and efficacious approach for the training of a deep neural network for SER, which has been corroborated on three distinct languages. Our exploratory study on the model's susceptibility to various adversarial attacks has unveiled substantial vulnerabilities to all examined techniques, even revealing critical deficiencies in the face of black-box attacks. The empirical trials showed that the proposed method does not exhibit considerable disparities in attack performance across different languages or gender samples, but only minor variances.\\


This work has significantly advanced the field of SER research through the application of deep learning, offering a comprehensive understanding of the resilience of Convolutional Neural Network-Long Short-Term Memory (CNN-LSTM) models and the influence of AEs on them. The findings presented herein establish a foundation for future studies focused on the creation of sturdier SER techniques, the development of more competent and impactful attacks, the investigation of potential defense mechanisms, the in-depth analysis of vocal variations across diverse languages and genders, and an overall enhanced understanding of the SER task.

\section*{Acknowledgments}
We gratefully acknowledge the support of NVIDIA Corp. with the donation of two Titan V GPUs. 

\subsection*{Author Contributions} 

N. Facchinetti designed, implemented, and conducted the experiments, analysed the experimental results, and prepared the original draft. 
F. Simonetta designed the experiments, analysed the experimental results, reviewed and edited the paper. 
S. Ntalampiras conceived the idea, analysed the experimental results, reviewed and edited the paper.

\subsection*{Conflicts of Interest}
The authors declare that there is no conflict of interest regarding the publication of this article.

\printbibliography
\newpage

\section{Figures and Tables}

\begin{table}[H]
\centering
    \begin{tabular}{|c|c|c|c|c|}
    \hline
    \textbf{Processing} &   \textbf{EmoDB} &  \textbf{EMOVO}   & \textbf{Ravdess} & \textbf{Average} \\ \hline
    \hline
    Original      &  $0.68 \pm 0.04$  &	$0.35 \pm 0.13$ &	$0.45 \pm 0.1 $   &	$0.50 \pm 0.09$     \\ \hline
    NormSum	      &  $0.46 \pm 0.16$  &	$0.29 \pm 0.10$ &	$0.21 \pm 0 	$ &	$0.32 \pm 0.09 $    \\ \hline
    NormMaxGlobal	      &  $0.79 \pm 0   $  &	$0.36 \pm 0.20$ &	$0.21 \pm 0 	$ &	$0.45 \pm 0.07 $    \\ \hline
    NormMaxLocal         &  $0.63 \pm 0.07$  &	$0.48 \pm 0.02$ &	$0.60 \pm 0.03$   &	$0.57 \pm 0.04 $    \\ \hline
    Original standardized  &  $0.71 \pm 0.02$  &	$0.60 \pm 0.07$ &	$0.62 \pm 0.01$   &	$0.64 \pm 0.04$    \\ \hline
    NormSum standardized	  &  $0.71 \pm 0.01$  &	$0.35 \pm 0.19$ &	$0.21 \pm 0 	$ &	$0.42 \pm 0.07 $    \\ \hline
    NormMaxGlobal standardized     &  $0.68 \pm 0.02$  &	$0.24 \pm 0.03$ &	$0.21 \pm 0 	$ &	$0.38 \pm 0.02 $    \\ \hline
    NormMaxLocal standardized     &  $0.68 \pm 0.02$  &	$0.42 \pm 0.15$ &	$0.58 \pm 0.01$   &	$0.56 \pm 0.06 $    \\ \hline
    \end{tabular}
    \caption{Mean accuracy $\pm$ standard deviation over the three splits for each dataset and normalization type.}
    \label{preacc}
\end{table}

\begin{table}[H]
\centering
    \begin{tabular}{|c|c|c|c|c|}
    \hline
    \textbf{Architecture} &   \textbf{EmoDB} &  \textbf{EMOVO}   & \textbf{Ravdess} & \textbf{Mean} \\ \hline
    \hline
    $\mathcal{M}0$ &	$0.71 \pm 0.05$ &	$0.59 \pm 0.03$ &	$0.53 \pm 0.11$ &	$0.61 \pm 0.06$ \\ \hline
    $\mathcal{M}1$ &	$0.83 \pm 0.00$ &	$0.70 \pm 0.03$ &	$0.76 \pm 0.01$ &	$0.76 \pm 0.01$ \\ \hline
    $\mathcal{M}2$ &	$0.78 \pm 0.01$ &	$0.68 \pm 0.03$ &	$0.70 \pm 0.02$ &	$0.72 \pm 0.02$ \\ \hline
    $\mathcal{M}3$ &	$0.68 \pm 0.05$ &	$0.39 \pm 0.14$ &	$0.39 \pm 0.13$ &	$0.49 \pm 0.10$ \\ \hline
    $\mathcal{M}4$ &	$0.76 \pm 0.01$ &	$0.66 \pm 0.03$ &	$0.45 \pm 0.15$ &	$0.62 \pm 0.06$ \\ \hline
    $\mathcal{M}5$ &	$0.58 \pm 0.05$ &	$0.25 \pm 0.03$ &	$0.26 \pm 0.04$ &	$0.36 \pm 0.04$ \\ \hline
    $\mathcal{M}6$ &	$0.32 \pm 0.06$ &	$0.24 \pm 0.03$ &	$0.23 \pm 0.03$ &	$0.26 \pm 0.04$ \\ \hline
    \end{tabular}
    \caption{Mean accuracy and standard deviation over the three splits for each dataset and model architecture.}
    \label{macc}
\end{table}

\begin{table}[H]
    \centering
    \begin{tabular}{|c|c|c|c|c|c|}
        \hline
        \textbf{Data}        & \textbf{EmoDB}     &  \textbf{EMOVO}    & \textbf{Ravdess} \\[0.5ex] \hline
        \hline
        All         & 0.909  & 0.872  & 0.911 \\ \hline
        Male        & 0.895  & 0.893  & 0.911 \\ \hline
        Female      & 0.918  & 0.852  & 0.911 \\ \hline
    \end{tabular}
    \caption{Accuracy on the original test set of the models that will be attacked.}
    \label{origacc}
\end{table}

\begin{table}[H]
\centering
    \begin{tabular}{|c|c|c|c|}
    \hline
    \textbf{Attack}          & \textbf{EmoDB}      & \textbf{EMOVO}        & \textbf{Ravdess}    \\ [0.5ex] \hline
    \hline
    FGSM            & $eps=1.25$ & $eps=0.5$    & $eps=0.25$ \\ \hline
    BIM             & $eps=0.25$ & $eps=0.25$   & $eps=0.25$ \\ \hline
    DeepFool        & $iter=5$   & $iter=5$     & $iter=5$ \\ \hline
    JSMA            & $theta=+1$ & $theta=+1$   & $theta=+1$  \\ \hline
    C\&W            & $metric=L_{\infty}$       & $metric=L_{\infty}$ & $metric=L_{\infty}$ \\ \hline \hline
    PixelAttack     & $th=10$    & $th=10$      & $th=10$ \\ \hline 
    BoundaryAttack  & - & - & - \\ \hline
    \end{tabular}
    \caption{Best accuracy-performing configuration of each attack.}
    \label{paramrecap}
\end{table}

\begin{table}[H]
\centering
    \begin{tabular}{|c|c|c|c|c|}
    \hline
    \textbf{Attack}          & \textbf{Dataset} & \textbf{All}      & \textbf{Female}   & \textbf{Male}    \\ [0.5ex] \hline
    \hline
FGSM        & EmoDB   & 0.109 & 0.121 & 0.089 \\ \hline
BIM         & EmoDB   & 0.067 & 0.065 & 0.070 \\ \hline
DeepFool    & EmoDB   & 0.061 & 0.061 & 0.070 \\ \hline
JSMA        & EmoDB   & \textbf{0.013} & \textbf{0.010} & 0.019 \\ \hline
C\&W        & EmoDB   & 0.069 & 0.076 & 0.065 \\ \hline
PixelAttack & EmoDB   & 0.547 & 0.603 & 0.454 \\ \hline 
BoundaryAttack  & EmoDB & 0.045 & 0.057 & 0.038 \\ \hline 
Original    & EmoDB   & 0.909 & 0.918 & 0.895 \\ \hline \hline
FGSM        & EMOVO  & 0.070 & 0.078 & 0.061 \\ \hline
BIM         & EMOVO  & 0.076 & 0.088 & 0.064 \\ \hline
DeepFool    & EMOVO  & 0.086 & 0.104 & 0.068 \\ \hline
JSMA        & EMOVO  & 0.022 & 0.024 & 0.020 \\ \hline
C\&W        & EMOVO  & 0.085 & 0.068 & 0.102 \\ \hline 
PixelAttack & EMOVO  & 0.364 & 0.339 & 0.389 \\ \hline
BoundaryAttack  & EMOVO & 0.076 & 0.070 & 0.082 \\ \hline 
Original    & EMOVO  & 0.872 & 0.852 & 0.893 \\ \hline \hline
FGSM        & Ravdess& 0.146 & 0.164 & 0.127 \\ \hline
BIM         & Ravdess& 0.059 & 0.060 & 0.057 \\ \hline
DeepFool    & Ravdess& 0.052 & 0.049 & 0.056 \\ \hline
JSMA        & Ravdess& 0.017 & 0.019 & \textbf{0.016} \\ \hline
C\&W        & Ravdess& 0.060 & 0.058 & 0.062 \\ \hline 
PixelAttack & Ravdess& 0.322 & 0.419 & 0.225 \\ \hline
BoundaryAttack  & Ravdess & 0.204 & 0.202 & 0.205 \\ \hline 
Original    & Ravdess& 0.911 & 0.911 & 0.911 \\ \hline
    \end{tabular}
    \caption{Accuracy obtained w.r.t all considered datasets with the original test data and with the best-performing configuration for each attack. The highest accuracies for each gender are highlighted in bold. }
    \label{accresumed}
\end{table}

\begin{table}[H]
\centering
    \begin{tabular}{|c|c|c|c|c|}
    \hline
    \textbf{Attack}   & \textbf{Dataset}       & \textbf{All}      & \textbf{Female}   & \textbf{Male}    \\ [0.5ex] \hline
    \hline
FGSM            & EmoDB   & 1.070 & 1.070 & 1.068 \\ \hline
BIM             & EmoDB   & \textbf{0.159} & \textbf{0.159} & \textbf{0.158} \\ \hline
DeepFool        & EmoDB   & \textbf{1.894} & 1.917 & 1.858 \\ \hline
JSMA            & EmoDB   & 0.003 & 0.004 & 0.003 \\ \hline
C\&W            & EmoDB   & 0.053 & 0.050 & 0.054 \\ \hline
    PixelAttack     & EmoDB     &5.55e-4 & 4.92e-4 & 6-6e-4  \\ \hline 
BoundaryAttack  & EmoDB   & \textbf{0.757} & \textbf{0.775} & \textbf{0.746} \\ \hline \hline
FGSM            & EMOVO   & 0.436 & 0.436 & 0.427 \\ \hline
BIM             & EMOVO   & \textbf{0.153} & \textbf{0.154} & \textbf{0.153} \\ \hline
DeepFool        & EMOVO   & \textbf{1.105} & \textbf{1.020} & \textbf{1.192} \\ \hline
JSMA            & EMOVO   & 0.002 & 0.002 & 0.002 \\ \hline
C\&W            & EMOVO   & 0.035 & 0.035 & 0.035 \\ \hline 
    PixelAttack     & EMOVO     & 7.76e-4 & 8-03e-4 & 7.48e-4  \\ \hline
FGSM            & Ravdess & \textbf{0.198}  & \textbf{0.204} & \textbf{0.192} \\ \hline
BIM             & Ravdess & \textbf{0.147}  & \textbf{0.149} & \textbf{0.146} \\ \hline
DeepFool        & Ravdess & 1.327 & 1.411 & 1.243 \\ \hline
JSMA            & Ravdess & 0.002 & 0.003 & 0.002 \\ \hline
C\&W            & Ravdess & 0.027 & 0.025 & 0.029 \\ \hline \hline
    PixelAttack     & Ravdess   &0.000939 & 0.000813 & 0.001066  \\ \hline
BoundaryAttack  & Ravdess & \textbf{1.314} & \textbf{1.493} & \textbf{1.136} \\ \hline    \end{tabular}
    \caption{Mean perturbation injected by the best-performing configuration for each attack w.r.t all considered datasets. The entries in bold indicate the cases for which the best-performing configuration is also the less perturbing one.}
    \label{pertresumed}
\end{table}

\begin{table}[H]
    \centering
    \begin{tabular}{|c|c|c|c|}
        \hline
        \textbf{Distance Metric}      & \textbf{$L_0$}         & \textbf{$L_2$}         & \textbf{$L_\infty$} \\ [0.5ex] \hline \hline
        FGSM                &               &               & \checkmark\\ \hline
        BIM                 &               &               & \checkmark\\ \hline
        DeepFool            &               & \checkmark    &           \\ \hline
        JSMA                & \checkmark    &               &           \\ \hline
        C\&W                &               & \checkmark    & \checkmark\\ \hline \hline
        PixelAttack         & \checkmark    &               &           \\ \hline
        BoundaryAttack      &               &               & \checkmark\\ \hline
    \end{tabular}
    \caption{Distance metric used by the tested attacks.}
    \label{disttried}
\end{table}

\newpage

\section{Supplementary Material}\label{sec:supplementary}
In this section, all the supplementary materials have been placed to integrate the presented results.
\subsection{Dataset Processing}\label{dsproc}
The summary statistics regarding the duration of the samples are presented in Table \ref{lengthanalysis}, the number of examples for each label in the three datasets is resumed in detail in Table \ref{labelanalysis}, and Table \ref{augtab} summarizes the number of original samples (before all processes, with only labels removed) and the number of samples after data augmentation and split\&repeat for each dataset.
\begin{table}[h]
    \centering
    \begin{tabular}{|c|c|c|c|}
    \hline
                        Statistic &       EmoDB &       EMOVO &      Ravdess \\ [0.5ex] \hline
    \hline
    Mean  &    2.78   &    3.121 &     3.701 \\ \hline
    Std   &    1.028  &    1.357 &     0.337 \\ \hline
    Min   &    1.226  &    1.291 &     2.936 \\ \hline
    25\%   &    2.027 &    2.133 &     3.47 \\ \hline
    50\%   &    2.59  &    2.773 &     3.67 \\ \hline
    75\%   &    3.308 &    3.84  &     3.871 \\ \hline
    Max   &    8.978  &   13.995 &     5.272 \\ \hline
    \end{tabular}
    \caption{Descriptive statistics of duration in seconds about the length of the 3 datasets.}
    \label{lengthanalysis}
\end{table}
\begin{table}[h]
    \centering
    \begin{tabular}{|c|c|c|c|}
    \hline
    Label &       EmoDB &       EMOVO &      Ravdess \\ [0.5ex] \hline
    \hline
    angry     &  127 &   84 &    192 \\ \hline
    bored     &   81 &    NaN &      NaN \\ \hline
    neutral   &   79 &   84 &     96 \\ \hline
    happy     &   71 &   84 &    192 \\ \hline
    fear      &   69 &   84 &    192 \\ \hline
    sad       &   62 &   84 &    192 \\ \hline
    disgust   &   46 &   84 &    192 \\ \hline
    surprised &    NaN &   84 &    192 \\ \hline
    calm      &    NaN &    NaN &    192 \\ \hline
    \end{tabular}
    \caption{Number of examples for each label in the three datasets.}
    \label{labelanalysis}
\end{table}
\begin{table}
\centering
    \begin{tabular}{|c|c|c|c|}
    \hline
    Dataset &   Original &  Augmented   & Factor \\ \hline
    \hline
    EmoDB	& 339 &	4181 &	12.333    \\ \hline
    EMOVO	& 420 &	4931 &	11.74    \\ \hline
    Ravdess	& 960 &	8952 &	9.325     \\ \hline
    \end{tabular}
    \caption{Number of original samples (after removing correlated classes) and after data augmentation step for each dataset. The increase factor is also reported for each case.}
    \label{augtab}
\end{table}

\subsection{Models Architectures}\label{modarchs}
The model's architecture for$\mathcal{M}0$, $\mathcal{M}1$, $\mathcal{M}2$, $\mathcal{M}3$, $\mathcal{M}4$, $\mathcal{M}5$, and $\mathcal{M}6$ are summarized in tables \ref{m0}, \ref{m1}, \ref{m2}, \ref{m3}, \ref{m4}, \ref{m5}, and \ref{m6}. The rows marked with \checkmark in TD column indicate the layers where the \emph{TimeDistributed} wrapper is applied, allowing the CNN layer to be applied to each temporal slice of the input, and the input for these should be in the form $(timesteps, dim1\_size, dim2\_size, n\_channels)$. To achieve this format, a \emph{Reshape} layer is used after the \emph{Input} layer to split the input matrix into timestamps. Each input matrix is divided into 9 non-overlapping vertical slices, each representing one-third of a second, resulting in a shape of \textit{timestamps} x \textit{time frames} x \textit{Mel bands} x \textit{channels}.
\begin{table}
\centering
     \begin{tabular}{|c|c|c|c|c|c|c|} 
     \hline
        \textbf{Layer}              & \textbf{Output shape} & \textbf{Num.} & \textbf{Size} & \textbf{Stride} & \textbf{Function} & \textbf{TD} \\ [0.5ex]  \hline
     \hline
     Input      & (261, 128, 1)     &  &  & & & \\ \hline
     Reshape    & (9, 29, 128, 1)   &  & & & & \\ \hline
     Conv       & (9, 25, 124, 16)  & 16 & (5,5) & 1 & ReLU & \checkmark \\ \hline 
     BatchNorm  &                   &  &  & & & \checkmark \\ \hline
     Pool       & (9, 11, 61, 16)   & & (4,4) & 2 & & \checkmark \\ \hline
     Conv       & (9, 9, 59, 32)    & 32 & (3,3) & 1 & ReLU & \checkmark \\ \hline 
     Pool       & (9, 4, 29, 32)    & & (2,2) & 2 & & \checkmark \\ \hline
     Conv       & (9, 2, 27, 64)    & 64 & (3,3) & 1 & ReLU & \checkmark \\ \hline 
     Pool       & (9, 1, 26, 64)    & & (2,2) & 1 & & \checkmark \\ \hline
     Flatten    & (9, 1664) &       &  & & & \checkmark \\ \hline
     LSTM       & (3) & 3           & \multicolumn{4}{|c|}{Internal dropout = 0.2}  \\ \hline
     Dense      & (5)               & 5  &  & & Softmax &  \\ \hline
     \hline
    \multicolumn{5}{|c|}{Total parameters} & \multicolumn{2}{|c|}{43652} \\ \hline
    \end{tabular}
    \caption{Architecture of the CNN-LSTM base model $\mathcal{M}0$}
    \label{m0}
\end{table}

\begin{table}
\begin{center}
 \begin{tabular}{|c|c|c|c|c|c|c|} 
 \hline
\textbf{Layer}              & \textbf{Output shape} & \textbf{Num.} & \textbf{Size} & \textbf{Stride} & \textbf{Function} & \textbf{TD} \\ [0.5ex] \hline
 \hline
 Input      & (261, 128, 1)     &  &  & & & \\\hline
 Reshape    & (9, 29, 128, 1)   &  & & & & \\ \hline
 Conv       & (9, 25, 124, 16)  & 16 & (5,5) & 1 & ReLU & \checkmark \\ \hline 
 BatchNorm  &                   &  &  & & & \checkmark \\ \hline
 Pool       & (9, 11, 61, 16)   & & (4,4) & 2 & & \checkmark \\ \hline
 Conv       & (9, 9, 59, 32)    & 32 & (3,3) & 1 & ReLU & \checkmark \\ \hline 
 Pool       & (9, 4, 29, 32)    & & (2,2) & 2 & & \checkmark \\ \hline
Conv        & (9, 2, 27, 64)    & 64 & (3,3) & 1 & ReLU & \checkmark \\ \hline 
 Pool       & (9, 1, 26, 64)    & & (2,2) & 1 & & \checkmark \\ \hline
 Flatten    & (9, 1664) &       &  & & & \checkmark \\ \hline
 BLSTM       & (6) & 3           & \multicolumn{4}{|c|}{Internal dropout = 0.2}  \\ \hline
 Dense      & (5)               & 5  &  & & Softmax &  \\ \hline
 \hline
\multicolumn{5}{|c|}{Total parameters} & \multicolumn{2}{|c|}{63383} \\ \hline
\end{tabular}
\end{center}
\caption{Architecture of the CNN-LSTM model $\mathcal{M}1$}
\label{m1}
\end{table}

\begin{table}
\begin{center}
 \begin{tabular}{|c|c|c|c|c|c|c|} 
 \hline
\textbf{Layer}              & \textbf{Output shape} & \textbf{Num.} & \textbf{Size} & \textbf{Stride} & \textbf{Function} & \textbf{TD} \\ [0.5ex] \hline
 \hline
 Input      & (261, 128, 1)     &  &  & & & \\\hline
 Reshape    & (9, 29, 128, 1)   &  & & & & \\ \hline
 Conv       & (9, 25, 124, 16)  & 16 & (5,5) & 1 & ReLU & \checkmark \\ \hline 
 BatchNorm  &                   &  &  & & & \checkmark \\ \hline
 Pool       & (9, 11, 61, 16)   & & (4,4) & 2 & & \checkmark \\ \hline
 Conv       & (9, 9, 59, 32)    & 32 & (3,3) & 1 & ReLU & \checkmark \\ \hline 
 Pool       & (9, 4, 29, 32)    & & (2,2) & 2 & & \checkmark \\ \hline
Conv        & (9, 2, 27, 64)    & 64 & (3,3) & 1 & ReLU & \checkmark \\ \hline 
 Pool       & (9, 1, 26, 64)    & & (2,2) & 1 & & \checkmark \\ \hline
 Flatten    & (9, 1664) &       &  & & & \checkmark \\ \hline
 LSTM       & (6) & 6           & \multicolumn{4}{|c|}{Internal dropout = 0.2}  \\ \hline
 Dense      & (5)               & 5  &  & & Softmax &  \\ \hline
 \hline
\multicolumn{5}{|c|}{Total parameters} & \multicolumn{2}{|c|}{63775} \\ \hline
\end{tabular}
\end{center}
\caption{Architecture of the CNN-LSTM $\mathcal{M}2$}
\label{m2}
\end{table}

\begin{table}
\begin{center}
 \begin{tabular}{|c|c|c|c|c|c|c|} 
 \hline
\textbf{Layer}              & \textbf{Output shape} & \textbf{Num.} & \textbf{Size} & \textbf{Stride} & \textbf{Function} & \textbf{TD} \\ [0.5ex ] \hline
 \hline
 Input      & (261, 128, 1)     &  &  & & & \\\hline
 Reshape    & (9, 29, 128, 1)   &  & & & & \\ \hline
 Conv       & (9, 27, 126, 16)  & 16 & (3,3) & 1 & ReLU & \checkmark \\ \hline 
 BatchNorm  &                   &  &  & & & \checkmark \\ \hline
 Pool       & (9, 13, 63, 16)   & & (2,2) & 2 & & \checkmark \\ \hline
 Conv       & (9, 11, 61, 32)    & 32 & (3,3) & 1 & ReLU & \checkmark \\ \hline 
 Pool       & (9, 5, 30, 32)    & & (2,2) & 2 & & \checkmark \\ \hline
 Conv       & (9, 3, 28, 64)    & 64 & (3,3) & 1 & ReLU & \checkmark \\ \hline 
 Pool       & (9, 2, 27, 64)    & & (2,2) & 1 & & \checkmark \\ \hline
 Flatten    & (9, 3456) &       &  & & & \checkmark \\ \hline
 LSTM       & (3) & 3           & \multicolumn{4}{|c|}{Internal dropout = 0.2}  \\ \hline
 Dense      & (5)               & 5  &  & & Softmax &  \\ \hline
 \hline
\multicolumn{5}{|c|}{Total parameters} & \multicolumn{2}{|c|}{64900} \\ \hline
\end{tabular}
\end{center}
\caption{Architecture of the CNN-LSTM model $\mathcal{M}3$}
\label{m3}
\end{table}

\begin{table}
\begin{center}
 \begin{tabular}{|c|c|c|c|c|c|c|} 
 \hline
\textbf{Layer}              & \textbf{Output shape} & \textbf{Num.} & \textbf{Size} & \textbf{Stride} & \textbf{Function} & \textbf{TD}\\ [0.5ex] \hline
 \hline
 Input      & (261, 128, 1)     &  &  & & & \\\hline
 Reshape    & (9, 29, 128, 1)   &  & & & & \\ \hline
 Conv       & (9, 25, 124, 16)  & 16 & (5,5) & 1 & ReLU & \checkmark \\ \hline 
 BatchNorm  &                   &  &  & & & \checkmark \\ \hline
 Pool       & (9, 12, 62, 16)   & & (2,2) & 2 & & \checkmark \\ \hline
 Conv       & (9, 10, 60, 32)    & 32 & (3,3) & 1 & ReLU & \checkmark \\ \hline 
 Pool       & (9, 5, 30, 32)    & & (2,2) & 2 & & \checkmark \\ \hline
Conv        & (9, 3, 28, 64)    & 64 & (3,3) & 1 & ReLU & \checkmark \\ \hline 
 Pool       & (9, 1, 2, 27, 64)    & & (2,2) & 1 & & \checkmark \\ \hline
 Flatten    & (9, 3456) &       &  & & & \checkmark \\ \hline
 LSTM       & (6) & 6           & \multicolumn{4}{|c|}{Internal dropout = 0.2}  \\ \hline
 Dense      & (5)               & 5  &  & & Softmax &  \\ \hline
 \hline
\multicolumn{5}{|c|}{Total parameters} & \multicolumn{2}{|c|}{106763} \\ \hline
\end{tabular}
\end{center}
\caption{Architecture of the CNN-LSTM model $\mathcal{M}4$}
\label{m4}
\end{table}

\begin{table}
\begin{center}
 \begin{tabular}{|c|c|c|c|c|c|c|} 
 \hline
 \textbf{Layer}              & \textbf{Output shape} & \textbf{Num.} & \textbf{Size} & \textbf{Stride} & \textbf{Function} & \textbf{TD} \\ [0.5ex] \hline
 \hline
 Input      & (261, 128, 1)     &  &  & & & \\\hline
 Reshape    & (9, 29, 128, 1)   &  & & & & \\ \hline
 Conv       & (9, 25, 124, 32)  & 32 & (5,5) & 1 & ReLU & \checkmark \\ \hline 
 BatchNorm  &                   &  &  & & & \checkmark \\ \hline
 Pool       & (9, 11, 61, 32)   & & (4,4) & 2 & & \checkmark \\ \hline
 Conv       & (9, 9, 59, 64)    & 64 & (3,3) & 1 & ReLU & \checkmark \\ \hline 
 Pool       & (9, 4, 29, 64)    & & (2,2) & 2 & & \checkmark \\ \hline
Conv        & (9, 2, 27, 128)    & 128 & (3,3) & 1 & ReLU & \checkmark \\ \hline 
 Pool       & (9, 1, 26, 128)    & & (2,2) & 1 & & \checkmark \\ \hline
 Flatten    & (9, 3328) &       &  & & & \checkmark \\ \hline
 LSTM       & (3) & 3           & \multicolumn{4}{|c|}{Internal dropout = 0.2}  \\ \hline
 Dense      & (5)               & 5  &  & & Softmax &  \\ \hline
 \hline
\multicolumn{5}{|c|}{Total parameters} & \multicolumn{2}{|c|}{133316} \\ \hline
\end{tabular}
\end{center}
\caption{Architecture of the CNN-LSTM model $\mathcal{M}5$}
\label{m5}
\end{table}

\begin{table}
\begin{center}
 \begin{tabular}{|c|c|c|c|c|c|c|} 
 \hline
 \textbf{Layer}              & \textbf{Output shape} & \textbf{Num.} & \textbf{Size} & \textbf{Stride} & \textbf{Function} & \textbf{TD} \\ [0.5ex] \hline
 \hline
 Input      & (261, 128, 1)     &  &  & & & \\\hline
 Reshape    & (9, 29, 128, 1)   &  & & & & \\ \hline
 Conv       & (9, 25, 124, 16)  & 16 & (5,5) & 1 & ReLU & \checkmark \\ \hline 
 BatchNorm  &                   &  &  & & & \checkmark \\ \hline
 Pool       & (9, 12, 62, 16)   & & (2,2) & 2 & & \checkmark \\ \hline
 Conv       & (9, 8, 58, 32)    & 32 & (3,3) & 1 & ReLU & \checkmark \\ \hline 
 Pool       & (9, 7, 57, 32)    & & (2,2) & 1 & & \checkmark \\ \hline
 Conv       & (9, 5, 55, 64)    & 64 & (3,3) & 1 & ReLU & \checkmark \\ \hline 
 Pool       & (9, 4, 54, 64)    & & (2,2) & 1 & & \checkmark \\ \hline
 Conv       & (9, 2, 52, 128)    & 128 & (3,3) & 1 & ReLU & \checkmark \\ \hline 
 Pool       & (9, 1, 51, 128)    & & (2,2) & 1 & & \checkmark \\ \hline
 Flatten    & (9, 6528) &       &  & & & \checkmark \\ \hline
 LSTM       & (3) & 3           & \multicolumn{4}{|c|}{Internal dropout = 0.2}  \\ \hline
 Dense      & (5)               & 5  &  & & Softmax &  \\ \hline
 \hline
\multicolumn{5}{|c|}{Total parameters} & \multicolumn{2}{|c|}{184068} \\ \hline
\end{tabular}
\end{center}
\caption{Architecture of the CNN-LSTM model $\mathcal{M}6$}
\label{m6}
\end{table}

\subsection{Improved Model Performance} \label{modperf}
Table \ref{lstmtuning} shows the validation loss values obtained with Hyperband for each dataset with respect to the number of used bidirectional LSTM units, where the best value for each dataset is enlighted in bold font.\\
\begin{table}
\centering
    \begin{tabular}{|c|c|c|c|}
    \hline
    \textbf{LSTM Units}& \textbf{EmoDB}& \textbf{EMOVO}& \textbf{Ravdess} \\ \hline
    \hline
    4           & 0.85  & 1.29  & 1.27 \\ \hline
    8           & 0.63  & 1.02  & 0.92 \\ \hline
    16          & 0.45  & 0.78  & 0.79 \\ \hline
    32          & 0.4  & 0.71  & 0.6 \\ \hline
    64          & 0.33  & 0.56  & 0.44 \\ \hline
    128         & 0.3  & 0.52  & 0.38 \\ \hline
    256         & 0.29  & 0.48  & 0.33 \\ \hline
    512         & \textbf{0.28}  & \textbf{0.48}  & 0.33 \\ \hline
    1024        & 0.32  & 0.48  & \textbf{0.3} \\ \hline
    \end{tabular}
    \caption{Validation loss with respect to the number of LSTM bidirectional units for each dataset obtained with Hyperband after 8 epochs of training. The best results are in bold.}
    \label{lstmtuning}
\end{table}

After increasing the number of LSTM units, we repeated the training for model $\mathcal{M}1$ using the same number of epochs, batch size, learning rate, and callbacks for early stopping and learning rate reduction to evaluate the actual performance improvement. However, this time we used a single set of train/validation/test data of the same size but composed of different samples. The results of the test set are presented in Table \ref{finalarch_res}.
\begin{table}
\begin{center}
 \begin{tabular}{|c|c|c|c|c|} 
 \hline
 \textbf{Model}                  &  \textbf{Metric}    & \textbf{EmoDB}      & \textbf{EMOVO}     & \textbf{Ravdess} \\ [0.5ex] \hline
 \hline
 \multirow{2}{*}{Final} &  Accuracy  & 0.889   & 0.873  & 0.873 \\ \cline{2-5}
                        &  Loss      & 0.379   & 0.385  & 0.357 \\ \hline
\hline
 \multirow{2}{*}{M1}    & Accuracy  & 0.83     &	0.699  & 0.757 \\ \cline{2-5}
                        & Loss      & 0.478    & 0.877  & 0.735 \\ \hline
\end{tabular}
\end{center}
\caption{Accuracy and loss function of the test set on the final and $\mathcal{M}1$ CNN-LSTM model. The values for $\mathcal{M}1$ are averaged over the data in the three splits while for the final model only on one split.}
\label{finalarch_res}
\end{table}
Although $\mathcal{M}1$ and the final model used different data for training and testing, it is evident that increasing the number of units has significantly improved the accuracy on the test set, especially for EMOVO and Ravdess. Specifically: \begin{itemize}
    \item For EmoDB from 0.83 we arrive at 0.889, representing an increase of 0.059 or 7.1\%;
    \item For EMOVO from 0.699 we arrive at 0.873, representing an increase of 0.174 or 24.89\%;
    \item For Ravdess from 0.757 we arrive at 0.873, representing an increase of 0.116 or 15.32\%.
\end{itemize}
In each case, the early stopping callback is activated as none of the models reach 50 epochs: the training for EmoDB stops at epoch 20, for EMOVO at epoch 20, and for Ravdess at epoch 18.\\

Regarding the hyperparameter process, the considered parameters and their values are:\begin{itemize}
    \item Probability of the Dropout layer: $[0, 0.3, 0.6]$;
    \item Probability of internal dropout of the BLSTM layer: $[0, 0.2, 0.4]$;
    \item Initial learning rate of Adam optimizer: $[0.01, 0.001, 0.0001]$;
    \item Batch size: $[8, 16, 32, 64, 128]$.
\end{itemize}
The best values found by the algorithm for the three datasets are reported in Table \ref{param_found}.\\
\begin{table}
\begin{center}
 \begin{tabular}{|c|c|c|c|} 
 \hline
Parameter       & EmoDB & EMOVO & Ravdess \\ [0.5ex] \hline
 \hline
Dropout         & 0.3   & 0.3   & 0.6 \\ \hline
LSTM dropout    & 0.2   & 0     & 0.2 \\ \hline
Learning rate   & 0.001 & 0.0001& 0.001 \\ \hline
Batch size      & 32    & 8     & 128 \\ \hline
\end{tabular}
\end{center}
\caption{Best hyperparameters configuration found for each dataset with Hyperband.}
\label{param_found}
\end{table}

Other than the added dropout probability, the parameter configuration for EmoDB is curiously identical to the original, which probably explains why the results were better than the other datasets. The intuition to add the Dropout layer proved to be correct, as all models produced better results with dropout probabilities other than 0. Surprisingly, Ravdess required greater regularization by preferring higher dropout probabilities, despite having the largest amount of data.\\
The training performs better with an initial learning rate equal to the original one for EmoDB and Ravdess, while it is smaller for EMOVO. EMOVO also has the smallest batch size, indicating that it needs a slower training process to converge to the optimum, while Ravdess prefers larger batches.\\
To evaluate the effectiveness of the hyperparameter tuning process, we repeated the training using the same train/validation/test sets as in the previous paragraph but with the parameters found during tuning. The results are presented in Table \ref{finalarch_res2}.
\begin{table}
\begin{center}
 \begin{tabular}{|c|c|c|c|c|} 
 \hline
 Model                          &  Metric    & EmoDB      & EMOVO     & Ravdess \\ [0.5ex] \hline
 \hline
 \multirow{2}{*}{Final base}    &  Accuracy  & 0.902031   & 0.859169  & 0.878839 \\ \cline{2-5}
                                &  Loss      & 0.321270   & 0.428224  & 0.366705 \\ \hline
\hline
 \multirow{2}{*}{Final tuned}   & Accuracy  & 0.906810    & 0.895643  & 0.912898 \\ \cline{2-5}
                                & Loss      & 0.312610    & 0.344954  & 0.281702 \\ \hline
\end{tabular}
\end{center}
\caption{Accuracy and loss function of the test set on the final tuned and non-tuned CNN-LSTM model using the splits of the previous paragraph.}
\label{finalarch_res2}
\end{table}
The hyperparameter tuning process has thus enabled a significant increase in performance and made it possible to better leverage datasets with more data, such as EMOVO and Ravdess.\\
The early stopping callback is triggered in all cases before reaching 50 epochs, and the learning rate reduction callback is triggered when the validation loss increases to stabilize and improve the training procedure. Specifically, the training stops:\begin{itemize}
    \item For EmoDB at epoch 33 (previously 20)
    \item For EMOVO at epoch 35 (previously 20)
    \item For Ravdess at epoch 40 (previously 18)
\end{itemize} 
This phenomenon is particularly pronounced for Ravdess, where the number of epochs more than doubles compared to the base case. This proves that a correct setting of the parameters leads to better exploitation of the greater amount of data available, resulting in higher performance than the previous cases.

\subsection{Attack deployment results}
\label{resultsfinal}
This section  presents the results obtained by the different attack techniques along with their configuration parameters\footnote{We utilized the notation proposed in the ART library's documentation for consistency \url{https://adversarial-robustness-toolbox.readthedocs.io/en/latest/modules/attacks/evasion.html}}. The performance is tracked for males, females, and the entire test set and compared with the original samples. Table \ref{mfbalance} displays the balance of male vs. female samples in the test set for each dataset. \\
\begin{table}[h]
    \centering
    \begin{tabular}{|c|c|c|c|c|c|}
        \hline
         Dataset &  Total & Male &  Female \\[0.5ex] \hline
        \hline
        EmoDB   & 837   & 315   & 522 \\ \hline
        EMOVO   & 987   & 488   & 499 \\ \hline
        Ravdess & 1791  & 895   & 896 \\ \hline
    \end{tabular}
    \caption{Male and female samples count for each dataset's test set}
    \label{mfbalance}
\end{table}

In the following tables, the best results for each row are highlighted in bold text.

\paragraph{FGSM} is configured with various \emph{eps} values in $[0.25, 0.5, 0.75, 1, 1.25]$, representing the attack step size (or input variation), and \emph{eps\_step} equal to 0.1, indicating the step size of input variation for minimal perturbation computation.\\
Table \ref{fgsmacc} reports the accuracy of the attack over the datasets.
\begin{table}[h]
    \centering
    \begin{tabular}{|c|c|c|c|c|c|c|}
        \hline
         \textbf{Dataset} &               \textbf{Gender}     & \textbf{eps 0.25}  & \textbf{eps 0.5}   & \textbf{eps 0.75}  & \textbf{eps 1}     & \textbf{eps 1.25} \\[0.5ex] \hline
        \hline
\multirow{3}{*}{EmoDB}  & All       &  0.171 &  0.128 &  0.115 &  0.112 &  \textbf{0.109} \\\cline{2-7}
                        & Female    &  0.205 &  0.148 &  0.128 &  0.125 &  \textbf{0.121} \\\cline{2-7}
                        & Male      &  0.114 &  0.095 &  0.092 &  0.092 &  \textbf{0.089} \\\hline \hline
\multirow{3}{*}{EMOVO}  & All       &  0.073 &  \textbf{0.070} &  0.076 &  0.077 &  0.081 \\\cline{2-7}
                        & Female    &  0.084 &  \textbf{0.078} &  0.084 &  0.086 &  0.086 \\\cline{2-7}
                        & Male      &  \textbf{0.061} &  \textbf{0.061} &  0.068 &  0.068 &  0.076 \\\hline \hline
\multirow{3}{*}{Ravdess}& All       &  \textbf{0.146} &  0.188 &  0.204 &  0.207 &  0.198 \\\cline{2-7}
                        & Female    &  \textbf{0.164} &  0.195 &  0.203 &  0.212 &  0.209 \\\cline{2-7}
                        & Male      &  \textbf{0.127} &  0.181 &  0.204 &  0.202 &  0.188 \\\hline        
    \end{tabular}
    \caption{Accuracy of FGSM attack}
    \label{fgsmacc}
\end{table}

The results indicate that the effectiveness of the attack is influenced by the \emph{eps} parameter. However, there is no definitive rule for determining the best value since it is closely related to the targeted language: for instance, in EmoDB, larger values lead to lower accuracy, whereas in Ravdess, the opposite is true. Except for one case, male samples are more susceptible to successful attacks, irrespective of language or \emph{eps} value.\\
Table \ref{fgsmpert} reports the mean perturbation instead.
\begin{table}[h]
    \centering
    \begin{tabular}{|c|c|c|c|c|c|c|}
        \hline
        \textbf{Dataset} &               \textbf{Gender}     & \textbf{eps 0.25}  & \textbf{eps 0.5}   & \textbf{eps 0.75}  & \textbf{eps 1}     & \textbf{eps 1.25} \\[0.5ex] \hline
        \hline
\multirow{3}{*}{EmoDB}  & All       &  \textbf{0.214} &  0.428 &  0.642 &  0.856 &  1.070 \\ \cline{2-7}
                        & Female    &  \textbf{0.214} &  0.428 &  0.642 &  0.856 &  1.070 \\ \cline{2-7}
                        & Male      &  \textbf{0.214} &  0.427 &  0.641 &  0.855 &  1.068 \\ \hline \hline
\multirow{3}{*}{EMOVO}  & All       &  \textbf{0.218} &  0.436 &  0.654 &  0.872 &  1.090 \\ \cline{2-7}
                        & Female    &  \textbf{0.218} &  0.436 &  0.654 &  0.872 &  1.090 \\ \cline{2-7}
                        & Male      &  \textbf{0.218} &  0.436 &  0.655 &  0.873 &  1.091 \\ \hline \hline
\multirow{3}{*}{Ravdess}& All       &  \textbf{0.198} &  0.396 &  0.592 &  0.784 &  0.961 \\ \cline{2-7}
                        & Female    &  \textbf{0.204} &  0.408 &  0.611 &  0.810 &  0.996 \\ \cline{2-7}
                        & Male      &  \textbf{0.192} &  0.383 &  0.574 &  0.759 &  0.926 \\ \hline
    \end{tabular}
    \caption{Perturbation of FGSM attack}
    \label{fgsmpert}
\end{table}        
In this case, it is evident (and obvious) that increasing the attack step size also increases the mean perturbation. Ravdess' samples have slightly less modification than those from EmoDB and EMOVO. Although the applied perturbations are very similar between males and females, there are significant differences in accuracy when the samples are applied to the model, as shown in Table \ref{fgsmacc}.\\
Table \ref{fgsmtime} reports the execution time taken by the used \emph{FGSM} implementation.
\begin{table}[h]
    \centering
    \begin{tabular}{|c|c|c|c|c|c|}
        \hline
        \textbf{Dataset}  & \textbf{eps 0.25}  & \textbf{eps 0.5}   & \textbf{eps 0.75}  & \textbf{eps 1}     & \textbf{eps 1.25} \\[0.5ex] \hline
        \hline
EmoDB   & 1.886  & \textbf{1.169}  & 1.278  & 1.283  & 1.272 \\ \hline
EMOVO   & 1.753  & 1.536  & 1.539  & 1.536  & \textbf{1.486} \\ \hline
Ravdess & 3.150  & 2.840  & \textbf{2.792}  & 2.864  & 2.839 \\ \hline          
    \end{tabular}
    \caption{Execution time in seconds of FGSM attack}
    \label{fgsmtime}
\end{table}
The execution times for different \emph{eps} values are very similar, and logically, larger datasets require more time to complete. In particular, the attack is so fast that the time of a single run can be influenced by other events as well.\\
In general, FGMS is a rapid method capable of producing relatively good AEs. It is crucial to determine the appropriate \emph{eps} value to attain a balance between the amount of disturbance introduced and the attack's effectiveness in misleading the network.

\paragraph{BIM} is a simple iterative extension of FGSM. It was retested with the same \emph{eps} values within the range of $[0.25, 0.5, 0.75, 1, 1.25]$, and the value of \emph{eps\_step} was set to 0.1. Another parameter to consider is \emph{max\_iter}, which denotes the maximum number of iterations the algorithm can run, was set to 100.\\
Table \ref{bimacc} reports the accuracy of the attack on the datasets.
\begin{table}[h]
    \centering
    \begin{tabular}{|c|c|c|c|c|c|c|}
        \hline
         \textbf{Dataset} &               \textbf{Gender}     & \textbf{eps 0.25}  & \textbf{eps 0.5}   & \textbf{eps 0.75}  & \textbf{eps 1}     & \textbf{eps 1.25} \\[0.5ex] \hline
        \hline
\multirow{3}{*}{EmoDB}  & All       &  0.067 &  0.067 &  0.067 &  0.067 &  0.067 \\ \cline{2-7}
                        & Female    &  0.065 &  0.065 &  0.065 &  0.065 &  0.065 \\ \cline{2-7}
                        & Male      &  0.070 &  0.070 &  0.070 &  0.070 &  0.070 \\ \hline \hline
\multirow{3}{*}{EMOVO}  & All       &  0.076 &  0.076 &  0.076 &  0.076 &  0.076 \\ \cline{2-7}
                        & Female    &  0.088 &  0.088 &  0.088 &  0.088 &  0.088 \\ \cline{2-7}
                        & Male      &  0.064 &  0.064 &  0.064 &  0.064 &  0.064 \\ \hline \hline
\multirow{3}{*}{Ravdess}& All       &  0.059 &  0.059 &  0.059 &  0.059 &  0.059 \\ \cline{2-7}
                        & Female    &  0.060 &  0.060 &  0.060 &  0.060 &  0.060 \\ \cline{2-7}
                        & Male      &  0.057 &  0.057 &  0.057 &  0.057 &  0.057 \\ \hline
    \end{tabular}
    \caption{Accuracy of BIM attack}
    \label{bimacc}
\end{table}
In this particular case, it appears that the \emph{eps} parameter has no effect on the algorithm's output. The results remain constant regardless of the value chosen for this parameter. This observation could suggest that the number of iterations selected might be too high for this problem. Nonetheless, it would be prudent to investigate this peculiar behavior further.\\
The contrast between the outcomes for males and females is not very distinct. However, for EMOVO and Ravdess datasets, the attacks are more successful on male samples, while for EmoDB, the reverse holds.\\
To gain a better comprehension of the situation, the loss function of the models, when the entire test set is attacked, is displayed in Table \ref{bimloss}.
\begin{table}[h]
    \centering
    \begin{tabular}{|c|c|c|c|c|c|}
        \hline
         \textbf{Dataset} & \textbf{eps 0.25}  & \textbf{eps 0.5}   & \textbf{eps 0.75}  & \textbf{eps 1}     & \textbf{eps 1.25} \\[0.5ex] \hline
        \hline
EmoDB   & \textbf{23.007} & 29.431 & 31.680 & 32.776 & 33.361 \\ \hline
EMOVO   & \textbf{28.104} & 34.026 & 36.022 & 36.937 & 37.420 \\ \hline
Ravdess & \textbf{32.887} & 40.242 & 42.707 & 43.861 & 44.537 \\ \hline
    \end{tabular}
    \caption{Loss function on whole test set generated by AE of BIM attack}
    \label{bimloss}
\end{table}
In this instance, the situation is relatively straightforward. Increasing the value of the \emph{eps} parameter also increases the loss function. This signifies that augmenting the maximum perturbation that the attacker can induce leads to more significant errors in the model's predictions. As accuracy remains consistent across all scenarios, the best results can be obtained using $\emph{eps} = 0.25$ since it reports the lowest value of the loss function.\\
Table \ref{fgsmpert} displays the mean perturbation.
\begin{table}[h]
    \centering
    \begin{tabular}{|c|c|c|c|c|c|c|}
        \hline
          \textbf{Dataset} &               \textbf{Gender}     & \textbf{eps 0.25}  & \textbf{eps 0.5}   & \textbf{eps 0.75}  & \textbf{eps 1}     & \textbf{eps 1.25} \\[0.5ex] \hline
        \hline
\multirow{3}{*}{EmoDB}  & All       &  \textbf{0.159} &  0.293 &  0.418 &  0.524 &  0.607 \\ \cline{2-7}
                        & Female    &  \textbf{0.159} &  0.293 &  0.418 &  0.524 &  0.606 \\ \cline{2-7}
                        & Male      &  \textbf{0.158} &  0.293 &  0.418 &  0.525 &  0.608 \\ \hline \hline
\multirow{3}{*}{EMOVO}  & All       &  \textbf{0.153} &  0.286 &  0.409 &  0.514 &  0.598 \\ \cline{2-7}
                        & Female    &  \textbf{0.154} &  0.287 &  0.410 &  0.515 &  0.599 \\ \cline{2-7}
                        & Male      &  \textbf{0.153} &  0.285 &  0.408 &  0.513 &  0.597 \\ \hline \hline
\multirow{3}{*}{Ravdess}& All       &  \textbf{0.147} &  0.273 &  0.390 &  0.485 &  0.547 \\ \cline{2-7}
                        & Female    &  \textbf{0.149} &  0.274 &  0.392 &  0.487 &  0.552 \\ \cline{2-7}
                        & Male      &  \textbf{0.146} &  0.271 &  0.388 &  0.482 &  0.542 \\ \hline                          
    \end{tabular}
    \caption{Perturbation of BIM attack}
    \label{bimpert}
\end{table}
This finding corroborates our earlier results, as using $\emph{eps} = 0.25$ results in a minimal perturbation that still causes the same misclassification as in the other scenarios. As anticipated, enlarging the attack step size also leads to an increase in the mean perturbation. The average perturbation introduced is comparable across various languages and is virtually identical for both males and females.\\
Table \ref{fgsmtime} details the running time of the implemented BIM method.
\begin{table}[h]
    \centering
    \begin{tabular}{|c|c|c|c|c|c|}
        \hline
         \textbf{Dataset}     & \textbf{eps 0.25}  & \textbf{eps 0.5}   & \textbf{eps 0.75}  & \textbf{eps 1}     & \textbf{eps 1.25} \\[0.5ex] \hline
        \hline
EmoDB   & 24.962 & 24.899 & 24.870 & 24.800 & \textbf{24.750} \\ \hline
EMOVO   & 28.622 & 28.520 & \textbf{28.502} & 28.661 & 28.530 \\ \hline
Ravdess & 51.782 & \textbf{51.778} & 51.794 & 51.866 & 51.974 \\ \hline       
    \end{tabular}
    \caption{Execution time in seconds of BIM attack}
    \label{bimtime}
\end{table}
The execution times are comparable across various \emph{eps} values, and there appears to be no correlation between these two values.\\
Overall, the BIM attack produces very effective AEs, leading to a high number of misclassifications in the models. It works well across various languages and genders, with a low \emph{eps} value resulting in minimal perturbations. The execution times are also reasonable, indicating that lower \emph{eps} values and more epochs could be tried to produce equally good AEs with even lower degrees of perturbation.\\

\paragraph{DeepFool} The {DeepFool} attack is an iterative approach aimed at finding the minimum perturbation required to cause misclassification in the model. The most straightforward parameter to adjust is \emph{max\_iter}, which defines the maximum number of iterations, and the values considered are $[1,5,20]$. Other parameters include:\begin{itemize}
    \item $nb\_grads=5$, which denotes the number of class gradients (top nb\_grads w.r.t. prediction) to compute. This is set to 5 to consider all output classes rather than just the most probable.
    \item $epsilon=1e-05$ is the overshoot used to aid convergence and prevent stagnation.
\end{itemize} 
Table \ref{dfacc} displays the attack's accuracy on the datasets.
\begin{table}[h]
    \centering
    \begin{tabular}{|c|c|c|c|c|}
        \hline
         \textbf{Dataset} &               \textbf{Gender}     & \textbf{1 iter}    & \textbf{5 iters}   &\textbf{ 20 iters}  \\ [0.5ex] \hline
        \hline
\multirow{3}{*}EmoDB   & All       &  0.118 & \textbf{0.061} & \textbf{0.061} \\ \cline{2-5}
        & Female    &  0.130 & \textbf{0.056} & \textbf{0.056} \\ \cline{2-5}
        & Male      &  0.098 & \textbf{0.070} & \textbf{0.070} \\ \hline  \hline
\multirow{3}{*}EMOVO   & All       &  0.134 & \textbf{0.086} & \textbf{0.086} \\ \cline{2-5}
        & Female    &  0.166 & \textbf{0.104} & \textbf{0.104} \\ \cline{2-5}
        & Male      &  0.100 & \textbf{0.068} & \textbf{0.068} \\ \hline  \hline
\multirow{3}{*}Ravdess & All       &  0.199 & \textbf{0.052} & \textbf{0.052} \\ \cline{2-5}
        & Female    &  0.210 & \textbf{0.049} & \textbf{0.049} \\ \cline{2-5}
        & Male      &  0.189 & \textbf{0.056} & \textbf{0.056} \\ \hline

    \end{tabular}
    \caption{Accuracy of DeepFool attack}
    \label{dfacc}
\end{table}
Based on the experiments, it can be concluded that using 5 iterations is sufficient to find the minimum perturbation in all datasets, as the accuracy results are the same as when using 20 iterations. However, when using only one iteration, the best results are achieved with EmoDB, followed by EMOVO, and then Ravdess. Increasing the number of iterations reduces the difference between different languages, and the examples in Ravdess cause the most misclassification, followed by EmoDB and EMOVO. In the best cases, female samples are more vulnerable to attack in EmoDB and Ravdess, while for EMOVO, male samples have lower accuracy.\\
The mean perturbation is reported in Table \ref{dfpert}.
\begin{table}[h]
    \centering
    \begin{tabular}{|c|c|c|c|c|}
        \hline
         \textbf{Dataset} &               \textbf{Gender}     & \textbf{1 iter}    & \textbf{5 iters}   & \textbf{20 iters}  \\[0.5ex] \hline
        \hline
\multirow{3}{*}EmoDB   & All       &  1.895 & \textbf{1.894} & \textbf{1.894} \\ \cline{2-5}
        & Female    &  1.917 & 1.917 & \textbf{1.917} \\ \cline{2-5}
        & Male      &  \textbf{1.857} & 1.858 & 1.858 \\ \hline  \hline
\multirow{3}{*}EMOVO   & All       &  1.105 & \textbf{1.105} & \textbf{1.105} \\ \cline{2-5}
        & Female    &  1.020 & \textbf{1.020} & \textbf{1.020} \\ \cline{2-5}
        & Male      &  1.192 & \textbf{1.192} & \textbf{1.192} \\ \hline  \hline
\multirow{3}{*}Ravdess & All       &  \textbf{1.308} & 1.327 & 1.327 \\ \cline{2-5}
        & Female    &  \textbf{1.393} & 1.411 & 1.411 \\ \cline{2-5}
        & Male      &  \textbf{1.222} & 1.243 & 1.243 \\ \hline
                           
    \end{tabular}
    \caption{Perturbation of DeepFool attack}
    \label{dfpert}
\end{table}
Using more iterations in EMOVO and EmoDB leads to the discovery of AEs with less perturbation, which they are more efficient in causing misclassification compared to using only 1 iteration. However, in the case of Ravdess, even with only 1 iteration, the perturbation is lower than in other cases, although the differences are minimal, as in the other two cases. This suggests that AEs are already effective with just 1 iteration, and with more effort, it is possible to achieve more efficient threats with a similar degree of perturbation.\\
Table \ref{dftime} reports the running time taken by the DeepFool implementation used.
\begin{table}[h]
    \centering
    \begin{tabular}{|c|c|c|c|}
        \hline
        \textbf{Dataset} & \textbf{1 iter}    & \textbf{5 iters}   & \textbf{20 iters }  \\[0.5ex] \hline
        \hline
        EmoDB   & \textbf{4.991} & 7.520 & 7.356 \\ \hline
        EMOVO   & \textbf{5.629} & 8.101 & 8.345 \\ \hline
        Ravdess & \textbf{10.657} & 22.576 & 22.584 \\ \hline
        EmoDB   & \textbf{4.991} & 7.520 & 7.356 \\ \hline
        EMOVO   & \textbf{5.629} & 8.101 & 8.345 \\ \hline
        Ravdess & \textbf{10.657} & 22.576 & 22.584 \\ \hline
        EmoDB   & \textbf{4.991} & 7.520 & 7.356 \\ \hline
        EMOVO   & \textbf{5.629} & 8.101 & 8.345 \\ \hline
        Ravdess & \textbf{10.657} & 22.576 & 22.584 \\ \hline
         
    \end{tabular}
    \caption{Execution time in seconds of DeepFool attack}
    \label{dftime}
\end{table}
The best time is clearly achieved with a single iteration. However, for cases with 5 and 20 iterations, both versions take approximately the same amount of time for all datasets. This suggests that they perform the same number of iterations, which is $\leq 5$, because the accuracy is identical with both 5 and 20 iterations, as shown in Table \ref{bimacc}. Specifically, based on the time required for one iteration, we can estimate that:\begin{itemize}
    \item EmoDB and EMOVO require around 2 iterations;
    \item Ravdess requires around 3 iterations.
\end{itemize}
In general, we can conclude that DeepFool is capable of producing reasonably good AEs with minimal time requirements. Specifically, for the studied context, a small number of iterations is sufficient to reach the minimal perturbation. When iterations are sufficient, AEs can significantly degrade the model's performance, but at the cost of introducing a relatively high degree of perturbation. The algorithm works well for all languages and produces essentially the same results for both male and female voices.

\paragraph{JSMA} is configured with different \emph{theta} values, which define the amount of perturbation introduced to each modified feature per step, in the range of $[-1, -0.5, 0.5, 1]$. Additionally, \emph{gamma} is set to 1, indicating that every value of the log Mel-spectrogram can be modified.\\
Table \ref{jsmaacc} reports the attack accuracy for the datasets and different \emph{theta} values.
\begin{table}[h]
    \centering
    \begin{tabular}{|c|c|c|c|c|c|}
        \hline
         \textbf{Dataset} &               \textbf{Gender}     & \textbf{theta -1}  & \textbf{theta -0.5} & \textbf{theta 0.5 }& \textbf{theta 1} \\ [0.5ex] \hline
\multirow{3}{*}{EmoDB}  & All       &  0.027 & 0.031 & 0.029 & \textbf{0.013} \\ \cline{2-6}
                        & Female    &  0.023 & 0.025 & 0.031 & \textbf{0.010} \\ \cline{2-6}
                        & Male      &  0.035 & 0.041 & 0.025 & \textbf{0.019} \\ \hline  \hline
\multirow{3}{*}{EMOVO}  & All       &  0.029 & 0.028 & 0.033 & \textbf{0.022} \\ \cline{2-6}
                        & Female    &  0.044 & 0.038 & 0.032 & \textbf{0.024} \\ \cline{2-6}
                        & Male      &  \textbf{0.014} & 0.018 & 0.035 & 0.020 \\ \hline  \hline
\multirow{3}{*}{Ravdess}& All       &  0.024 & 0.018 & 0.030 & \textbf{0.017} \\ \cline{2-6}
                        & Female    &  0.031 & \textbf{0.016} & 0.029 & 0.019  \\ \cline{2-6}
                        & Male      &  0.017 & 0.020 & 0.031 & \textbf{0.016} \\ \hline

    \end{tabular}
    \caption{Accuracy of JSMA attack}
    \label{jsmaacc}
\end{table}
Upon analyzing the results, it can be observed that the lower accuracy values are obtained with larger values of $thetas={+1, -1}$. This is expected since the higher the value of $theta$, the greater the perturbation introduced by the attack. In particular, most sets are especially vulnerable for $theta=1$.\\
The attack is very effective in compromising the proper functioning of the network and it performs almost equally well among the considered languages. It is worth noting that the female samples of EmoDB are more affected by the attack compared to the males, whereas for EMOVO, the attack is more effective on males with a different value of theta than that used for females. Similarly, for Ravdess, the attack is more efficacious on males with a value of +1 and on females with a value of -0.5, although the scores obtained in both cases are very similar.\\
Table \ref{jsmapert} instead reports the mean perturbation.
\begin{table}[h]
    \centering
    \begin{tabular}{|c|c|c|c|c|c|}
        \hline
         \textbf{Dataset} &               \textbf{Gender}     & \textbf{theta -1}  & \textbf{theta -0.5} & \textbf{theta 0.5 }& \textbf{theta 1} \\ [0.5ex] \hline
        \hline
\multirow{3}{*}{EmoDB}  & All       &  0.006 & 0.007 & \textbf{0.002} & 0.003 \\ \cline{2-6}
                        & Female    &  0.007 & 0.008 & \textbf{0.003} & 0.004 \\ \cline{2-6}
                        & Male      &  0.005 & 0.006 & \textbf{0.002} & 0.003 \\ \hline \hline
\multirow{3}{*}{EMOVO}  & All       &  0.005 & 0.003 & \textbf{0.001} & 0.002 \\ \cline{2-6}
                        & Female    &  0.005 & 0.003 & \textbf{0.002} & 0.002 \\ \cline{2-6}
                        & Male      &  0.006 & 0.003 & \textbf{0.001} & 0.002 \\ \hline \hline
\multirow{3}{*}{Ravdess}& All       &  0.007 & 0.005 & \textbf{0.002} & 0.002 \\ \cline{2-6}
                        & Female    &  0.008 & 0.005 & \textbf{0.002} & 0.003 \\ \cline{2-6}
                        & Male      &  0.006 & 0.004 & \textbf{0.002} & 0.002 \\ \hline

    \end{tabular}
    \caption{Perturbation of JSMA attack}
    \label{jsmapert}
\end{table}
The results of this analysis indicate that the best results are obtained with a small value of \emph{theta} equal to 0.5, which is in contrast to the findings of the previous table. It is worth noting that the average perturbation is smaller with a \emph{theta} value of 1 than with a value of -0.5, which is somewhat unexpected. Negative values seem to introduce more perturbation than positive ones, with the worst results obtained using a value of -1 for EMOVO and Ravdess, and a value of -0.5 for EmoDB.\\
When comparing the results for men and women, it appears that men generally receive less perturbation for EmoDB and Ravdess, regardless of the \emph{theta} value used. The situation for EMOVO is more complex, with men experiencing less perturbation for positive values of \emph{theta} and more perturbation for negative values. The difference in perturbation between males and females is generally less pronounced for positive \emph{theta} values, and more pronounced for negative values.\\
Overall, we are very satisfied with the level of mean perturbation introduced, especially considering the low levels of accuracy obtained. While introducing only a slight perturbation, this attack results in a considerably high degree of degradation in the model's performance.\\
In Table \ref{jsmatime}, we provide information on the running time taken by the JSMA implementation used in this analysis.
\begin{table}[h]
    \centering
    \begin{tabular}{|c|c|c|c|c|}
        \hline
        \textbf{Dataset} & \textbf{theta -1}  & \textbf{theta -0.5} & \textbf{theta 0.5 }& \textbf{theta 1} \\ [0.5ex] \hline
        \hline
EmoDB   & 360.355 & 775.986 & 284.885 & \textbf{225.941} \\ \hline
EMOVO   & 352.772 & 488.712 & \textbf{135.307} & 136.015 \\ \hline
Ravdess & 854.193 & 1277.608 & 623.825 & \textbf{374.433} \\ \hline            
    \end{tabular}
    \caption{Execution time in seconds of JSMA attack}
    \label{jsmatime}
\end{table}
Compared to the attacks presented earlier, this attack takes significantly more time to execute, but it yields better outcomes. Interestingly, when comparing the time taken for different \emph{thetas}, we observe that the best results are achieved when $theta=+1$ (the difference between 0.5 and 1 for EMOVO is very small). It is worth noting that this value is also associated with the highest misclassification rate. Moreover, we find it intriguing that the worst results, in terms of accuracy and perturbation, are linked to the values that take longer to execute the algorithm.\\

To summarize, this method is highly effective in generating AEs with minimal perturbation but significantly high misclassification rates; so we can say that using mechanisms such as saliency maps to identify features that strongly impact model results is suitable for SER tasks. However, this approach requires a greater computational effort in terms of generating examples. Based on our experiments, it is crucial to select an appropriate \emph{theta} value to achieve optimal results in terms of misclassification, perturbation, and Execution time.

\paragraph{C\&W} attack can be conducted with different distance metrics. However, due to a possible bug in the implementation of the $L_0$ version, I only tested the algorithm using $L_2$ and $L_\infty$ metrics. To ensure a fair comparison of the results, I set the parameters for the two versions to be as similar as possible, taking into account the default values provided in the ART documentation.\\
For the $L_2$ attack, the following parameters were used:\begin{itemize}
    \item $confidence=0$: this parameter controls the confidence of the AEs. Higher values produce examples that are farther away from the original input but are classified with higher confidence as the target class.
    \item $learning\_rate=0.01$: this is the initial learning rate for the attack algorithm. Smaller values produce better results but are slower to converge.
    \item $binary\_search\_steps=20$: this parameter denotes the number of times the constant $c$ is adjusted with binary search.
    \item $max\_iter=10$: this indicates the maximum number of iterations.
    \item $initial\_const=0.01$: this is the initial trade-off constant $c$ used to balance the importance of distance and confidence.
    \item $max\_halving=10$: this is the maximum number of halving steps in the line search optimization.
    \item $max\_doubling=10$: this is the maximum number of doubling steps in the line search optimization.
\end{itemize}
Instead, the $L_\infty$ attack is configured with:\begin{itemize}
    \item $confidence=0$: this parameter controls the confidence of the AEs. Higher values produce examples that are farther away from the original input but are classified with higher confidence as the target class.
    \item $learning\_rate=0.01$: this is the initial learning rate for the attack algorithm. Smaller values produce better results but are slower to converge.
    \item $max\_iter=10$: this parameter indicates the maximum number of iterations.
    \item $decrease\_factor=0.8$: this is the rate used for shrinking $tau$. The possible values are in $[0,1]$, where larger values lead to more accurate results.
    \item $initial\_const=0.0001$: this is the initial value of constant $c$.
    \item $largest\_const=20$: this is the largest possible value of constant $c$.
    \item $const\_factor=4$: this represents the rate of increasing constant $c$, where smaller values lead to more accurate results.
\end{itemize}
Table \ref{cwacc} reports the accuracy of the attack over the datasets for the two distance metrics.
\begin{table}[h]
    \centering
    \begin{tabular}{|c|c|c|c|}
        \hline
         \textbf{Dataset} &               \textbf{Gender}     & \textbf{$L_2$}    & \textbf{L$\infty$} \\ [0.5ex] \hline
        \hline
\multirow{3}{*}{EmoDB}  & All       &  0.188 &  \textbf{0.069} \\ \cline{2-4}
                        & Female    &  0.127 &  \textbf{0.076} \\ \cline{2-4}
                        & Male      &  0.224 &  \textbf{0.065} \\ \hline \hline
\multirow{3}{*}{EMOVO}  & All       &  0.092 &  \textbf{0.085} \\ \cline{2-4}
                        & Female    &  0.076 &  \textbf{0.068} \\ \cline{2-4}
                        & Male      &  0.108 &  \textbf{0.102} \\ \hline \hline
\multirow{3}{*}{Ravdess}& All       &  0.124 &  \textbf{0.060} \\ \cline{2-4}
                        & Female    &  0.122 &  \textbf{0.058} \\ \cline{2-4}
                        & Male      &  0.126 &  \textbf{0.062} \\ \hline

    \end{tabular}
    \caption{Accuracy of C\&W attack}
    \label{cwacc}
\end{table}
Based on the results, the $L_\infty$ distance metric achieves the best performances in all cases. The vulnerability of the datasets decreases in the following order: Ravdess, EmoDB, and EMOVO.\\
Despite not having as high a misclassification rate as JSMA, this attack (especially the $L_\infty$ version) is capable of significantly degrading the model's performance. Comparing the two versions, the accuracy difference is more pronounced for EmoDB and more subtle for Ravdess, while they perform almost the same on EMOVO. Notably, there is a significant accuracy gap for EmoDB's male samples, with a difference of approximately 0.159 between the two distance metrics.\\
Regarding the comparison between male and female samples, most configurations are more effective on female samples, except for EmoDB with $L_\infty$, where males have an advantage.\\
The mean perturbation is reported in Table \ref{cwpert}.
\begin{table}[h]
    \centering
    \begin{tabular}{|c|c|c|c|}
        \hline
         \textbf{Dataset} &               \textbf{Gender}     & \textbf{$L_2$}    & \textbf{L$\infty$} \\ [0.5ex] \hline
        \hline
\multirow{3}{*}{EmoDB}  & All       &  \textbf{0.008} &  0.053 \\ \cline{2-4}
                        & Female    &  \textbf{0.007} &  0.050 \\ \cline{2-4}
                        & Male      &  \textbf{0.008} &  0.054 \\ \hline \hline
\multirow{3}{*}{EMOVO}  & All       &  \textbf{0.005} &  0.035 \\ \cline{2-4}
                        & Female    &  \textbf{0.006} &  0.035 \\ \cline{2-4}
                        & Male      &  \textbf{0.005} &  0.035 \\ \hline \hline
\multirow{3}{*}{Ravdess}& All       &  \textbf{0.006} &  0.027 \\ \cline{2-4}
                        & Female    &  \textbf{0.005} &  0.025 \\ \cline{2-4}
                        & Male      &  \textbf{0.006} &  0.029 \\ \hline

    \end{tabular}
    \caption{Perturbation of C\&W attack}
    \label{cwpert}
\end{table}
Similarly to the previously presented cases, the least effective method is the one with lower perturbation. AEs generated by $L_2$ attacks have a mean perturbation that is approximately one order of magnitude lower than $L_\infty$. The reported values for $L_2$ attacks are very low and comparable to those of JSMA.\\
Regarding the comparison between male and female samples, females perform better in 5 out of 6 cases, whereas for $L_2$ and EMOVO, males have less perturbation. However, the differences between male and female samples are not significant in the various experiments.\\
The running time taken by the used C\&W implementations is reported in Table \ref{cwtime}.
\begin{table}[h]
    \centering
    \begin{tabular}{|c|c|c|c|c|}
        \hline
        \textbf{Dataset} &  \textbf{$L_2$}    & \textbf{L$\infty$} \\ [0.5ex] \hline
        \hline
EmoDB   & \textbf{2041.780} &    4230.925 \\ \hline
EMOVO   & \textbf{2338.341} &    5158.850 \\ \hline
Ravdess & \textbf{4650.343} &    9882.848 \\ \hline             
    \end{tabular}
    \caption{Execution time in seconds of C\&W attack}
    \label{cwtime}
\end{table}
Based on our data analysis, we have found that this attack is the most time-consuming among those that have been examined thus far. Specifically, the $L_2$ attack takes approximately half the time when compared to the $L_\infty$ attack. This may indicate that there was an incorrect initialization of the parameters, which could explain the significant differences in accuracy and perturbation between the two methods.\\
In summary, this technique yields very low accuracy scores when using the $L_\infty$ metric, but at the cost of more perturbations compared to $L_2$. In the latter case, the perturbation is low and consistent with the best results obtained using JSMA. However, the required time is significantly longer than in other cases. Given the time discrepancy between the two versions, it is suggested to conduct further experiments with the initial configuration of the $L_2$ parameters until it equals $L_\infty$ in at least one of the metrics examined, to ensure appropriate evaluations of the others.

\paragraph{PixelAttack} is the first considered black-box attack and the ART library implements the "Few-Pixel" flavor  (which uses the $L_0$ distance), described in \cite{kotyan2019adversarial}. This method was configured with different \emph{th} values in $[1,5,10]$: this value determines the threshold of the attack, i.e. the maximum size of the applied perturbation (the number of pixels considered). Additionally, two other parameters are used:\begin{itemize}
    \item $es=1$, which determines whether the attack uses CMAES (0) or DE (1) as the evolutionary strategy (although CMAES testing was not possible due to a bug)
    \item $max\_iter=20$, which sets the maximum number of iterations for running the evolutionary strategies for optimization-
\end{itemize}
Table \ref{paacc} presents the resulting accuracies for the different threshold values used with this method.
\begin{table}[h]
    \centering
    \begin{tabular}{|c|c|c|c|c|}
        \hline
         \textbf{Dataset} &               \textbf{Gender}     & \textbf{th 1}  & \textbf{th 5} & th 10\\ [0.5ex] \hline
        \hline
\multirow{3}{*}{EmoDB}  & All       &  0.669 &	0.566 &	\textbf{0.547} \\ \cline{2-5}
                        & Female    &  0.713 &	0.630 &	\textbf{0.603} \\ \cline{2-5}
                        & Male      &  0.597 &	0.460 &	\textbf{0.454} \\ \hline \hline
\multirow{3}{*}{EMOVO}  & All       &  0.643 &	0.433 &	\textbf{0.364} \\ \cline{2-5}
                        & Female    &  0.597 &	0.407 &	\textbf{0.339} \\ \cline{2-5}
                        & Male      &  0.691 &	0.459 &	\textbf{0.389} \\ \hline \hline
\multirow{3}{*}{Ravdess}& All       &  0.517 &	0.352 &	\textbf{0.322} \\ \cline{2-5}
                        & Female    &  0.592 &	0.432 &	\textbf{0.419} \\ \cline{2-5}
                        & Male      &  0.442 &	0.272 &	\textbf{0.225} \\ \hline

    \end{tabular}
    \caption{Accuracy of PixelAttack attack}
    \label{paacc}
\end{table}
The value of \emph{th} determines the number of features to which apply perturbations and thus affects the level of misclassification, with larger values leading to greater errors. Actually, the difference in misclassification between values of 5 and 10 is less significant compared to the difference between values of 1 and 5.\\
The Ravdess dataset is found to be the most vulnerable, followed by EMOVO and EmoDB, respectively. In terms of gender, male samples in EmoDB and Ravdess are found to be more vulnerable than female samples, regardless of the value of \emph{th}. However, the opposite is true for EMOVO, where female samples are more vulnerable in all cases.\\
Table \ref{papert} shows the average perturbation value.
\begin{table}[h]
    \centering
    \begin{tabular}{|c|c|c|c|c|}
        \hline
         \textbf{Dataset} &               \textbf{Gender}     & \textbf{th 1}  & \textbf{th 5} & th 10\\ [0.5ex] \hline
        \hline
\multirow{3}{*}{EmoDB}  & All       &  \textbf{7.6e-05} &	2.91e-04 &	5.55e-04\\ \cline{2-5}
                        & Female    &  \textbf{6.5e-05} &	2.54e-04 &	4.92e-04\\ \cline{2-5}
                        & Male      &  \textbf{9.5e-05} &	3.53e-04 &	6.6e-04\\ \hline \hline
\multirow{3}{*}{EMOVO}  & All       &  \textbf{7.8e-05} &	3.9e-04 &	7.76e-04\\ \cline{2-5}
                        & Female    &  \textbf{8.7e-05} &	4.05e-04 &	8.03e-04\\ \cline{2-5}
                        & Male      &  \textbf{6.9e-05} &	3.75e-04 &	7.48e-04\\ \hline \hline
\multirow{3}{*}{Ravdess}& All       &  \textbf{1.18e-04} &	5.02e-04 &	9.39e-04\\ \cline{2-5}
                        & Female    &  \textbf{9.8e-05} &	4.35e-04 &	8.13e-04\\ \cline{2-5}
                        & Male      &  \textbf{1.39e-04} &	5.69e-04 &	1.066e-03\\ \hline

    \end{tabular}
    \caption{Perturbation of PixelAttack attack}
    \label{papert}
\end{table}
The parameter \emph{th} determines the amount of perturbation applied, with lower values resulting in a lower average perturbation and higher values resulting in greater perturbation. Although even the worst-case scenario results in a low level of introduced perturbation, it is important to note that the misclassification rate is not as high as in other previously discussed cases.\\
Regarding the comparison between males and females, the situation is completely reversed compared to before: for EmoDB and Ravdess, females have lower perturbation, while for EMOVO, males report better results.\\
Table \ref{patime} displays the running time of the PixelAttack implementation used.
\begin{table}[h]
    \centering
    \begin{tabular}{|c|c|c|c|}
        \hline
        \textbf{Dataset} &   \textbf{th 1}  & \textbf{th 5} & th 10\\ [0.5ex] \hline
        \hline
EmoDB   & 2570.724 &   1438.910 &   \textbf{1436.918}\\ \hline
EMOVO   & 1700.664 &   1598.894 &   \textbf{1394.879}\\ \hline
Ravdess & 3124.273 &   2734.072 &   \textbf{1928.278}\\ \hline    
    \end{tabular}
    \caption{Time in seconds of PixelAttack attack}
    \label{patime}
\end{table}
Increasing the value of \emph{th} reduces the computation time required to generate AEs. This black-box attack is notably slower than most other white-box attacks, except for C\&W. In the case of EmoDB, the time required for generating AEs with \emph{th}=5 and \emph{th}=10 is almost the same, whereas, for EMOVO, the three results show the least amount of fluctuation.

In summary, this black-box attack is capable of misclassifying a sample by perturbing only a small number of values in the log Mel-spectrogram. However, it requires a considerable amount of time to generate AEs. As only a limited number of features (determined by the \emph{th} parameter) are altered, the average perturbation values are very low regardless of the language targeted by the attack. The choice of the \emph{th} parameter is critical in maximizing the effectiveness of the attack. Based on the conducted experiments, I recommend using a higher value since it produces AEs with less time and can result in larger misclassifications, with just a small amount of additional noise. Therefore, it is reasonable to try values larger than 10.

\paragraph{BoundaryAttack} is the last considered black-box attack. Since this algorithm involves multiple parameters, including some that are dynamically adjusted at runtime, we used a single configuration. The specific configuration we used was as follows:\begin{itemize}
    \item $delta=0.01$, which defines the initial step size for the orthogonal step (dynamically adjusted);
    \item $epsilon=0.01$, which defines the initial step size for the step towards the target(dynamically adjusted);
    \item $step\_adapt=0.5$, which is the factor by which the step sizes are multiplied or divided (0.5 is the value suggested in the paper);
    \item $max\_iter=100$, which defines the maximum number of iterations;
    \item $num\_trial=3$, which defines the maximum number of trials per iteration;
    \item $sample\_size=10$, which defines the number of samples per trial;
    \item $init\_size=100$, which is the maximum number of trials for the initial generation of AEs;
    \item $min\_epsilon=0$, which defines the threshold of the perturbation after the attack is stopped.
\end{itemize}
Table \ref{bastat} presents the accuracy and mean perturbation for the dataset we considered, along with relevant metrics for male and female samples. 
\begin{table}[h]
    \centering
    \begin{tabular}{|c|c|c|c|}
        \hline
         \textbf{Dataset} &               \textbf{Gender}     & \textbf{Accuracy} & \textbf{Perturbation} \\ [0.5ex] \hline
        \hline
\multirow{3}{*}{EmoDB}  & All       & 0.045 & 0.757  \\ \cline{2-4}
                        & Female    & 0.057 & 0.775  \\ \cline{2-4}
                        & Male      & 0.038 & 0.747  \\ \hline \hline
\multirow{3}{*}{EMOVO}  & All       & 0.076 & 1.023  \\ \cline{2-4}
                        & Female    & 0.070 & 1.043  \\ \cline{2-4}
                        & Male      & 0.082 & 1.003  \\ \hline \hline
\multirow{3}{*}{Ravdess}& All       & 0.204 & 1.314  \\ \cline{2-4}
                        & Female    & 0.202 & 1.493  \\ \cline{2-4}
                        & Male      & 0.205 & 1.136  \\ \hline

    \end{tabular}
    \caption{Accuracy and perturbation of BoundaryAttack attack}
    \label{bastat}
\end{table}
There are significant differences in the success rate of BoundaryAttack across different datasets, particularly in terms of accuracy. The attack performs significantly better on EmoDB and EMOVO, whereas its success rate is lower on Ravdess. Additionally, when considering male and female samples, there are notable differences in the success rate on EmoDB (higher for males) and EMOVO (higher for females), but not on Ravdess.\\
Examining the average perturbation rate, it is noteworthy that although the noise introduced on EmoDB is lower than in the other cases, the attack is still more successful. Conversely, on Ravdess, the perturbation rate is higher, but the accuracy on the adversarial examples is also higher than on the other datasets. There are only slight differences in the perturbation rate between male and female samples, except for Ravdess where the gap is somewhat larger. It is interesting to observe that, in contrast to the previous attacks considered, the average introduced noise is inversely proportional to the misclassification rate in this case.\\
Table \ref{batime} reports the running time of the BoundaryAttack implementation used.
\begin{table}[h]
    \centering
    \begin{tabular}{|c|c|}
        \hline
        \textbf{Dataset} &\textbf{Time}  \\[0.5ex] \hline
        \hline
EmoDB   & 623.448 \\ \hline
EMOVO   & 910.888 \\ \hline
Ravdess & 2132.831 \\ \hline           
    \end{tabular}
    \caption{Time in seconds of BoundaryAttack attack}
    \label{batime}
\end{table}
The execution times of this algorithm can be considered good, given that it is a black-box attack and has shown promising results in deceiving the target network.\\
Overall, despite being of the black-box type, this technique produces very good AEs for both EmoDB and EMOVO datasets. Specifically, the lower the accuracy, the lower the perturbation. However, for the Ravdess dataset, this attack yields notably inferior results in terms of performance reduction. This observation may indicate that the behavior of this attack is more sensitive to the initial parameter configuration or that it is not well-suited for working with log-Mel spectrograms derived from the English language.

\label{sec:comparisonAdditional}
\subsection{Comparison between attacks}
We can compare the execution time of the various algorithm for the best-performing configuration by combining the data in tables \ref{fgsmtime}, \ref{bimtime}, \ref{dftime}, \ref{jsmatime}, \ref{cwtime}, \ref{patime}, and \ref{batime} into Table \ref{timestab}. The entries in bold indicate the best results obtained among the different configurations tried.\\
\begin{table}
\centering
    \begin{tabular}{|c|c|c|c|}
    \hline
    \textbf{Attack}          & \textbf{EmoDB}    & \textbf{EMOVO}   & \textbf{Ravdess}    \\ [0.5ex] \hline
    \hline
FGSM            & 1.272 & 1.536 & 3.150 \\ \hline
BIM             & 24.962 & 28.622 & 51.782 \\ \hline
DeepFool        & 7.520 & 8.101 & 22.576 \\ \hline
JSMA            & \textbf{225.941} & 136.015 & \textbf{374.433} \\ \hline
C\&W            & 4230.925 & 5158.850 & 9882.848 \\ \hline \hline
PixelAttack     & \textbf{1436.918} & \textbf{1394.879} & \textbf{1928.278} \\ \hline 
BoundaryAttack  & \textbf{623.448} & \textbf{910.888} & \textbf{2132.831} \\ \hline

    \end{tabular}
    \caption{Time (s)  required to generate the AEs for the various attacks and datasets for the best-performing configuration.}
    \label{timestab}
\end{table}

We can compare the accuracy results of the best-performing attack configurations (as reported in Table \ref{accresumed}) and assign a score of 1 (for the minimum accuracy) to 3 (for the maximum accuracy) to each attack. The results are summarized in Table \ref{accscores}. Similarly, we can analyze perturbation by comparing the collected data for each dataset (summarized in Table \ref{pertresumed}) and assigning a score from 1 (lowest perturbation) to 3 (highest perturbation) to each attack. Table \ref{pertcores} presents the results of this analysis.\\
\begin{table}
\centering
    \begin{tabular}{|c|c|c|c|}
    \hline
    \textbf{Attack}          & \textbf{EmoDB}        & \textbf{EMOVO}        & \textbf{Ravdess}    \\ [0.5ex] \hline
    \hline
Original        & 0.909 & 0.872 & 0.911 \\ \hline \hline
FGSM            & 0.109 (2) & 0.070 (1) & 0.146 (3) \\ \hline
BIM             & 0.067 (2) & 0.076 (3) & 0.059 (1) \\ \hline
DeepFool        & 0.061 (2) & 0.086 (3) & 0.052 (1) \\ \hline
JSMA            & 0.013 (1) & 0.022 (3) & 0.017 (2) \\ \hline
C\&W            & 0.069 (2) & 0.085 (3) & 0.060 (1) \\ \hline \hline
PixelAttack     & 0.547 (3) & 0.364 (2) & 0.322 (1) \\ \hline
BoundaryAttack  & 0.045 (1) & 0.076 (2) & 0.204 (3) \\ \hline \hline
Total Score     & 13 & 17 & 12 \\ \hline

    \end{tabular}
    \caption{Accuracy obtained by the most effective configuration of each attack. In brackets is reported the score of the relative technique.}
    \label{accscores}
\end{table}

\begin{table}
\centering
    \begin{tabular}{|c|c|c|c|}
    \hline
    \textbf{Attack}          & \textbf{EmoDB}        & \textbf{EMOVO}        & \textbf{Ravdess}    \\ [0.5ex] \hline
    \hline
FGSM            & 1.070 (3) & 0.436 (2) & 0.198 (1)\\ \hline
BIM             & 0.159 (3) & 0.153 (2) & 0.147 (1)\\ \hline
DeepFool        & 1.894 (3) & 1.105 (1) & 1.327 (2)\\ \hline
JSMA            & 0.003 (3) & 0.002 (1) & 0.002 (2)\\ \hline
C\&W            & 0.053 (3) & 0.035 (2) & 0.027 (1)\\ \hline \hline
PixelAttack     & 5.555e-4 (1) & 776e-4 (2) & 939e-4 (3)\\ \hline 
BoundaryAttack  & 0.757 (1) & 1.023 (2) & 1.314 (3)\\ \hline \hline
Total Score     & 17 & 12 & 13\\ \hline
    \end{tabular}
    \caption{Mean perturbation introduced by the most effective configuration of each attack. In brackets is reported the score of the relative technique.}
    \label{pertcores}
\end{table}

Table \ref{accsgender}, obtained from Table \ref{accresumed}, shows the best accuracy result for each dataset underlined and the best across all languages in bold only about male/female samples. Again, we can conduct a similar analysis on the introduced perturbations. By extracting data from tables \ref{pertresumed}, we obtain Table \ref{pertgender}, where the best result for the dataset is underlined, while the best across all languages is in bold, only considering male/female samples.\\
\begin{table}
\centering
    \begin{tabular}{|c|c|c|c|c|c|c|}
    \hline
                    & \multicolumn{2}{c|}{\textbf{EmoDB}}        & \multicolumn{2}{c|}{\textbf{EMOVO}}        & \multicolumn{2}{c|}{\textbf{Ravdess}}    \\ \cline{2-7}
                    & \textbf{Female}     & \textbf{Male}                 & \textbf{Female}     & \textbf{Male}              & \textbf{Female}     & \textbf{Male}   
    \\ \hline \hline
FGSM            & 0.121 & \underline{0.089} & 0.078 & \textbf{\underline{0.061}} & 0.164 & \underline{0.127} \\ \hline
BIM             & \underline{0.065} & 0.070 & 0.088 & \underline{0.064} & 0.060 & \textbf{\underline{0.057}} \\ \hline
DF              & \underline{0.061} & 0.070 & 0.104 & \underline{0.068} & \textbf{\underline{0.049}} & 0.056 \\ \hline
JSMA            & \textbf{\underline{0.010}} & 0.019 & 0.024 & \underline{0.020} & 0.019 & \underline{0.016} \\ \hline
C\&W            & 0.076 & \underline{0.065} & \underline{0.068} & 0.102 & \textbf{\underline{0.058}} & 0.063 \\ \hline \hline
PA              & 0.603 & \underline{0.454} & \underline{0.339} & 0.389 & 0.419 & \textbf{\underline{0.225}} \\ \hline 
BA              & 0.057 & \textbf{\underline{0.038}} & \underline{0.070} & 0.082 & \underline{0.202} & 0.205 \\ \hline  \hline
Original        & 0.918 & 0.895 & 0.852 & 0.893 & 0.911 & 0.911 \\ \hline

    \end{tabular}
    \caption{Comparison of the accuracy on between male/female samples of the datasets by the most effective configuration of each attack. The best result for the dataset is underlined, while the best across all languages is in bold.}
    \label{accsgender}
\end{table}

\begin{table}
\centering
    \begin{tabular}{|c|c|c|c|c|c|c|}
    \hline
                    & \multicolumn{2}{c|}{\textbf{EmoDB}}        & \multicolumn{2}{c|}{\textbf{EMOVO}}        & \multicolumn{2}{c|}{\textbf{Ravdess}}    \\ \cline{2-7}
                    & \textbf{Female}   & \textbf{Male}     & \textbf{Female}   & \textbf{Male}     & \textbf{Female}   & \textbf{Male}   
    \\ \hline \hline
FGSM            & 1.070 & \underline{1.068} & 0.436 & \textbf{\underline{0.427}} & 0.204 & \underline{0.192} \\ \hline
BIM             & 0.159 & \underline{0.158} & 0.154 & \underline{0.153} & 0.149 & \textbf{\underline{0.146}} \\ \hline
DF              & 1.917 & \underline{1.858} & \textbf{\underline{1.020}} & 1.192 & 1.411 & \underline{1.243} \\ \hline
JSMA            & 0.00372 & \underline{0.00302} & 0.00220 & \underline{0.00197} & 0.00294 & \textbf{\underline{0.00190}} \\ \hline
C\&W            & \underline{0.050} & 0.054 & \underline{0.035} & 0.035 & \textbf{\underline{0.025}} & 0.029 \\ \hline \hline
    PA              & \textbf{\underline{4.92e-4}} & 6.6e-4 & 8.03e-4 & \underline{7.48e-4} & \underline{8.13e-4} & 1.066e-3 \\ \hline 
BA              & 0.775 & \textbf{\underline{0.747}} & 1.043 & \underline{1.003} & 1.493 & \underline{1.136} \\ \hline    \end{tabular}
    \caption{Comparison of the perturbation introduced between male/female samples of the datasets by the most effective configuration of each attack. The best result for the dataset is underlined, while the best across all languages is in bold.}
    \label{pertgender}
\end{table}

\end{document}